\documentclass[review,11pt]{elsarticle}

\usepackage{graphicx,xcolor}
\usepackage{hyperref}
\usepackage[relative]{overpic}
\usepackage[margin=2.5cm]{geometry}
\usepackage{amsmath,amssymb}
\usepackage{nccmath}
\usepackage{bm}
\usepackage{physics}
\usepackage{enumerate}
\usepackage{booktabs}
\usepackage{multirow}
\usepackage{tabularx}
\usepackage{tabularray}
\usepackage{subcaption}
\usepackage[geometry]{ifsym}
\usepackage{textcomp}

\usepackage{lineno}
\usepackage{mathrsfs}

\newcommand{\drv}[2]{\frac{\mathrm{d} #1}{\mathrm{d} #2}}

\newcommand{\pd}[2]{\frac{\partial #1}{\partial #2}}
\newcommand{\pdd}[2]{\frac{\partial^2 #1}{\partial #2^2}}

\begin{document}
\begin{frontmatter}


\title{Immersed Boundary Projection Method for Navier Slip Boundaries}



\author[osaka]{Takehiro Fujii\fnref{pa}}
\fntext[pa]{Present address: Industrial Technology Center of Okayama Prefecture, 5301 Haga, kita-ku, Okayama, Okayama 701-1296, Japan}
\ead{takehiro_fujii@okakogi.jp}
\author[omu,osaka]{Takeshi Omori\fnref{note}}
\fntext[note]{Corresponding author}
\ead{t.omori@omu.ac.jp}
\address[osaka]{Department of Mechanical Engineering, Osaka University, 2-1 Yamadaoka, Suita, Osaka 565-0871, Japan}
\address[omu]{Department of Mechanical Engineering, Osaka Metropolitan University, 1-1 Gakuen-tyo, Sakai-ku, Sakai, Osaka 599-8531, Japan}

\begin{abstract}
A formulation of the immersed boundary method for incompressible flow over bodies with surface slip described by the Navier boundary condition is presented. In the present method, the wall slip velocity and the boundary force are determined implicitly through a projection to satisfy the boundary conditions and the divergence-free condition of the velocity field as constraints. The present method is first-order accurate in space and fourth-order accurate in time, overcoming the difficulty of the conventional continuous forcing approaches to accurately evaluate the velocity gradient on the boundary. Results from the simulation of the flow past stationary and moving circular cylinders are in good agreement with previous experimental and numerical results for a wide range of slip length on the surface, including the no-slip case.
\end{abstract}


\end{frontmatter}


\section{Introduction}\label{S:1}
Since the pioneering work by \citet{Peskin1972}, the immersed boundary (IB) methods have been largely successful for the prediction of a wide variety of flows in complex geometries \cite{Peskin1982,Mittal2005a,Griffith2020} and one of the most popular methods to impose boundary conditions to the Navier-Stokes equation on surfaces/lines that are not aligned to the geometry of the spacial discretization \cite{Kajishima2017a}. In the formulation by \citet{Peskin1972}, which is now categorized as a continuous forcing approach, the boundary force to enforce the no-slip condition was calculated assuming a constitutive relation with some imaginary mechanical property. Such \textit{ad hoc} constitutive relations were later eliminated by \citet{Taira2007} from the continuous forcing IB approach, where the no-slip condition was enforced through the projection in a similar manner as the fractional step method enforcing the solenoidal condition to the velocity field.

Regarding the boundary condition to the Navier-Stokes equation, the fast development in the micro- and nano-fabrication in the recent decade has pushed the community to recognize the velocity slip on the solid surfaces \cite{Bocquet2010}, and for micro- and nano-fluidics it is now a common practice to employ the Navier boundary condition \cite{Bocquet2007,Matthews2008}, which relates the wall shear stress to the slip velocity on the wall. 

In contrast to the IB methods for the no-slip boundary, just a few examples \cite{He2018,WANG2021} can be found for the Navier slip boundary. For the Navier slip boundary, the tangential and normal velocities need to satisfy different types of boundary conditions (BCs): 
\begin{equation}\label{eq:navier-tau}
    \bm{\tau}\cdot(\bm{u}-\bm{U})=l_s\bm{\tau}\cdot[\nabla\bm{u}+(\nabla\bm{u})^T]\cdot\bm{n}
\end{equation}
for the tangential velocity and 
\begin{equation}
    \bm{n}\cdot(\bm{u}-\bm{U})=0
\end{equation}
for the normal velocity, where $\bm{u}-\bm{U}$ is the velocity slip between the fluid and solid on the wall, $l_s$ is the slip length, $\bm{\tau}$ and $\bm{n}$ denote the unit tangential and normal vectors on the wall. This sets the main difficulty to embed the boundary condition as the immersed boundary. It is especially challenging to evaluate the velocity gradient in Eq.~\eqref{eq:navier-tau} on the immersed boundary with a reasonable accuracy: the conventional scheme only provides the zeroth-order accuracy as we show in \ref{app:PoiCtt}.

Our formulation is the extension of the immersed boundary projection method (IBPM) by Taira and Colonius \cite{Taira2007} proposed for the no-slip boundaries, inheriting the basic structure of the method. The forces on the solid walls are calculated implicitly in the same manner as the pressure, and the method is free from the tuning parameters. The main contribution of the present study is the derivation of the interpolation operator to the velocity and the regularization operator to the boundary forces for the Navier slip boundaries that provide first-order spacial accuracy. Although the method presented here is designed for the continuous forcing approach, our interpolation principle involving the velocity gradient can also be useful for the discrete forcing approaches.

This paper is organized as follows. After the brief summary of the governing equations in Section \ref{S:gov_eq}, we review in Section \ref{S:IBPM} the IBPM for the no-slip boundaries to show the basic framework of the present method. The specific notations used in the present work is also introduced in this section. In Section \ref{S:IBPM-Navier}, we introduce the IBPM for the Navier slip boundaries: we show why the conventional operators fail for the slip boundaries and how this issue can be resolved. In Section \ref{S:results}, the method is validated on both steady and unsteady flow problems. We also show the temporal and spacial accuracy of the method by theoretical consideration and numerical results. Section \ref{S:conclusion} summarizes our formulations. Additionally, all the detailed derivations are given in the Appendices. 

\section{Governing equations}
\label{S:gov_eq}
\begin{figure}[t]
	\begin{center}
		\begin{overpic}[bb=0 0 422 302,width=8cm,clip,keepaspectratio]
				{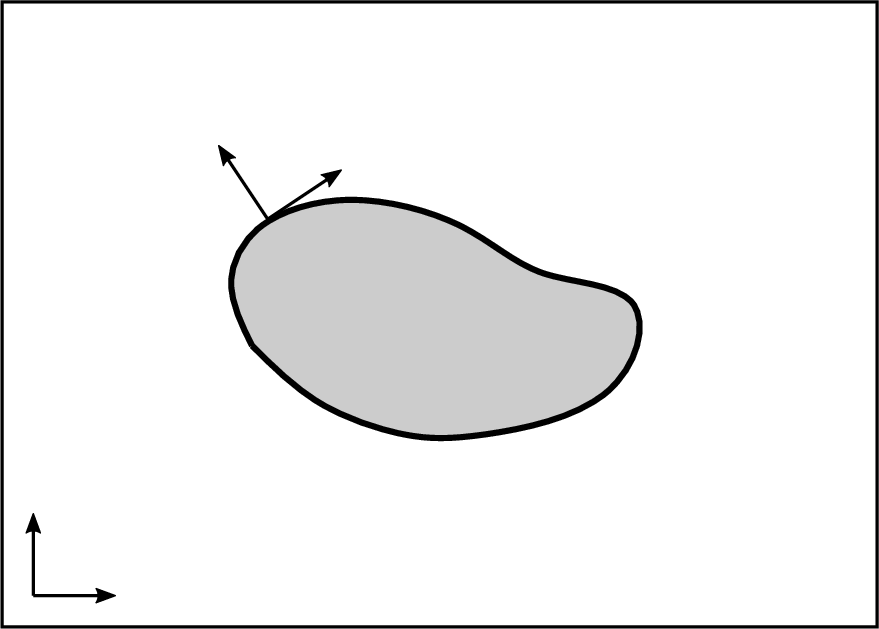}
				\put(3,64){whole domain $\Omega$}
				\put(35,32){solid body $\Omega_s$}
				\put(70,55){fluid $\Omega_f$}
				\put(35,16){body surface $\Gamma$}
				\put(38,53){$\bm{\tau}$}
				\put(25,56){$\bm{n}$}
                \put(13,5){$x$}
				\put(5,13){$y$}
		\end{overpic}
		\caption{Schematics of the computational domain. The fluid equations are solved for the whole domain $\Omega$. We consider that the immersed surface $\Gamma$ of the solid $\Omega_s$ is inside the computational domain $\Omega$.}
	 \label{fig:comp_domain}
	\end{center}
\end{figure}
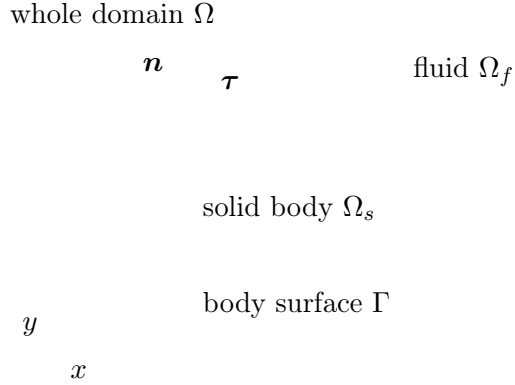

In the following discussion, we consider two dimensional incompressible flows of a Newtonian fluid around an arbitrarily shaped solid body in the computational domain (Fig.~\ref{fig:comp_domain}). The fluid is allowed to have velocity slip on the solid surface. The position and velocity of the solid body are assumed to be known. The fluid domain is denoted by $\Omega_f$, the solid domain by $\Omega_s$, the whole computational domain by $\Omega$, and the solid body surface by $\Gamma\ (=\partial\Omega_s)$, as illustrated in Fig.~\ref{fig:comp_domain}.

The governing equations for the fluid motion are the following non-dimensional 
Navier-Stokes equation and the continuity equation:
\begin{align}
	\pd{\bm{u}}{t}+\bm{u}\cdot\nabla\bm{u}&=-\nabla P
	+\frac{1}{\mathrm{Re}}\nabla^2\bm{u}+\bm{f}_{\mathrm{ext}}\ &\mathrm{in}\ \Omega_f,
\label{eq:NS}\\
	\nabla\cdot\bm{u}&=0\ &\mathrm{in}\ \Omega_f.
\label{eq:continuity}
\end{align}
Here, $ \bm{u} $, $ t $, $ P $ and $ \bm{f}_{\mathrm{ext}} $
are the non-dimensional velocity, time, pressure and external force, respectively,
and $ \mathrm{Re} $ is the Reynolds number.
The BCs on the body surface $ \Gamma $
are the Navier boundary condition \cite{Chen2014} and the impermeability condition:
\begin{align}
	\bm{\tau}\cdot(\bm{u}-\bm{U})
	&=\mathcal{L}_s\bm{\tau}\cdot[\nabla\bm{u}+(\nabla\bm{u})^T]\cdot\bm{n}\ &\mathrm{on}\ \Gamma,
\label{eq:NBC}\\
	\bm{n}\cdot(\bm{u}-\bm{U})
	&=0\ &\mathrm{on}\ \Gamma,
\label{eq:PBC}
\end{align}
where $ \bm{U} $ and $ \mathcal{L}_s $ are 
the non-dimensional fluid velocity on the surface $\Gamma$ and the slip length of the surface $\Gamma$, respectively. The unit vectors $\bm{\tau}$ and $\bm{n}$ are defined locally on $\Gamma$: $\bm{n}$ is the unit normal vector oriented outwards from the solid surface and $\bm{\tau}$ is the unit tangential vector whose direction is determined 
so that $ (\bm{\tau},\bm{n}) $ forms the right-handed system. For the no-slip condition ($\mathcal{L}_s=0$), the boundary condition is given all together by the single equation 
\begin{align}
	&\bm{u}-\bm{U}=0 & \text{on}\ \Gamma.
	\label{eq:no-slip_PBC}
\end{align}

\section{Immersed boundary projection method for the no-slip boundaries}
\label{S:IBPM}
Before proposing the IBPM for the Navier slip boundary condition, we first give a brief overview of the IBPM for the no-slip and impermeability condition, Eq.~\eqref{eq:no-slip_PBC}. In the IB formulation of the continuous forcing approach, which includes the IBPM, the fluid is also filled in the solid body domain $\Omega_s$. The governing equations Eqs.~\eqref{eq:NS} and \eqref{eq:continuity}
are solved in the whole computational domain $\Omega$,
and the solid body is replaced by a singular body force
acting only on the body surface $\Gamma$,
which constrains the flow to satisfy the boundary conditions on the surface $\Gamma$.
The governing equations in the IB formulation are given by
\begin{align}
\pd{\bm{u}}{t}+\bm{u}\cdot\nabla\bm{u}
&=
-\nabla P
+\frac{1}{\mathrm{Re}}\nabla^2\bm{u}
+\bm{f}_{\mathrm{ext}} \notag\\
&{\hphantom{===}}
+\int_{\Gamma} \bm{F}(\bm{\xi}(s)) \delta(\bm{\xi}(s)-\bm{x})\dd{s}
\ &\mathrm{in}\ \Omega,
\label{eq:NS_IB}
\\
\nabla\cdot\bm{u}&=0\ &\mathrm{in}\ \Omega,
\label{eq:continuity_IB}
\end{align}
under the no-slip BC
\begin{equation}
\bm{u}(\bm{\xi}(s))
=\int_{\Omega}\bm{u}(\bm{x}) \delta(\bm{x}-\bm{\xi}(s))\dd{V}
=\bm{U}(\bm{\xi}(s))
\label{eq:no-slip_IB}
\end{equation}
where $\bm{x}=[x,y]^T$ and $\bm{\xi}(s)=[\xi(s),\eta(s)]^T$ are the position vectors in the domain $\Omega$ and on the body surface $\Gamma$, respectively, and $s$ is the one-dimensional index on the surface $\Gamma$. $\bm{F}=[F_x,F_y]^T$ is the boundary force defined on the surface $\Gamma$. The Dirac delta function $ \delta $ in Eqs.~\eqref{eq:NS_IB} and \eqref{eq:no-slip_IB}
is used for exchanging the information between $ \Omega $ and $ \Gamma $. That is, by convolution with the delta function $ \delta $, the boundary force $\bm{F}$ on $\Gamma$ is transformed into the singular body forcing $\bm{f}(\bm{x})=\int_{\Gamma} \bm{F}(\bm{\xi}(s)) \delta(\bm{\xi}(s)-\bm{x})\dd{s}$ in $ \Omega $,
and similarly, the velocity field $\bm{u}(\bm{x})$ in $\Omega$ is transformed into the fluid velocity $\bm{u}(\bm{\xi}(s))$ on $\Gamma$. Note that by using the delta function $ \delta $, the no-slip condition $ \bm{u}(\bm{\xi}(s)) = \bm{U}(\bm{\xi}(s))$ becomes a constraint on the velocity field $ \bm{u}(\bm{x})$ defined on $\Omega$, as in the continuity equation Eq.~\eqref{eq:continuity_IB}.

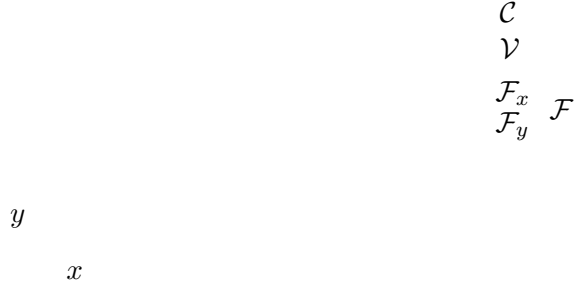
\begin{figure}[t]
	\begin{center}
		\begin{overpic}[bb=0 0 185 120, width=6cm]
				{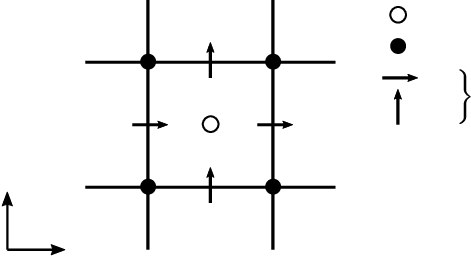}
				\put(112,60.5){$\mathcal{C}$}
				\put(112,52.5){$\mathcal{V}$}
				\put(111,43){$\mathcal{F}_x$}
				\put(111,36){$\mathcal{F}_y$}
				\put(123,39){$\mathcal{F}$}
                \put(16,3.5){$x$}
				\put(3.5,16){$y$}
		\end{overpic}
		\caption{Variable arrangements on the staggered mesh. The pressure $P$ is defined on the cell centers $\mathcal{C}$, the velocities $u$ and $v$ are defined on the cell faces $\mathcal{F}_x$ and $\mathcal{F}_y$, respectively. The velocity gradients are defined either on the cell centers $\mathcal{C}$ or the cell vertices $\mathcal{V}$.}
	 \label{fig:staggered_grid}
	\end{center}
\end{figure}

These equations \eqref{eq:NS_IB}--\eqref{eq:no-slip_IB}
are discretized on a staggered Cartesian Eulerian mesh, and the surface $ \Gamma $ is represented by a set of Lagrangian points, at which the boundary force $ \bm{F} $ is applied. The staggered mesh $\mathcal{M}=\{\mathcal{C},\mathcal{V},\mathcal{F}\}$
consists of cell centers $\mathcal{C}$, vertices $\mathcal{V}$ and faces $\mathcal{F}$,
which is further decomposed of $\mathcal{F}_x$ in the $x$-direction and $\mathcal{F}_y$ in the $y$-direction, as shown in Fig.~\ref{fig:staggered_grid}.
Let $B^A$ denote the set of all functions from a set $A$ to a set $B$, then the set of grid functions from
$\mathcal{C}$, $\mathcal{V}$, $\mathcal{F}$, $\mathcal{F}_x$, $\mathcal{F}_y$ and $\Gamma$ to $\mathbb{R}$ are denoted by
$\mathbb{R}^\mathcal{C}$, $\mathbb{R}^\mathcal{V}$,
$\mathbb{R}^\mathcal{F}$, $\mathbb{R}^{\mathcal{F}_x}$, $\mathbb{R}^{\mathcal{F}_y}$
and $\mathbb{R}^\Gamma$, respectively.
The discrete quantities that can be regarded as the grid functions are represented by the column vectors whose components are the value of those on the grid points,
so $P=[P_1,\cdots,P_{N_\Omega}]^T\in\mathbb{R}^\mathcal{C}$,
$\bm{u}=[u_1,\cdots,u_{N_\Omega},v_1,\cdots,v_{N_\Omega}]^T=[(u\in\mathbb{R}^{\mathcal{F}_x})^T,(v\in\mathbb{R}^{\mathcal{F}_y})^T]^T\in\mathbb{R}^\mathcal{F}$ and 
$\bm{F}=[F_{x,1},\cdots,F_{x,N_\Gamma},F_{y,1},\cdots,F_{y,N_\Gamma}]^T=[(F_x\in\mathbb{R}^\Gamma)^T,(F_y\in\mathbb{R}^\Gamma)^T]^T
$, where $N_\Omega$ and $N_\Gamma$ are the number of Eulerian grid points and of Lagrangian points, respectively.
The other quantities, $\bm{f}_\mathrm{ext}\in\mathbb{R}^\mathcal{F}$ and $\bm{f} \in \mathbb{R}^\mathcal{F}$,
are also represented by the column vectors.
The Dirac delta function $ \delta $ 
is approximated by a discrete delta function $ \delta_h $,
which is a continuous function with finite support
only near the surface $ \Gamma $.
Among several types of the function \cite{Kajishima2017a,Peskin2002,Beyer_LeVeque_1992}, 
we use a 3-cell discrete delta function $\delta_h$ by Roma et al.~\cite{Roma1999}, which is known to be a good choice for the staggered grids. The discrete delta function $\delta_h$ is given as follows,
\begin{align}
	\delta_h(x)=
	\left\{
		\begin{aligned}
			&\frac{1}{3\Delta x}\left\{1+\sqrt{-3\tilde{x}^2+1}\right\} &&\mathrm{for}\ |\tilde{x}|\le\frac{1}{2}\\
			&\frac{1}{6\Delta x}\left\{5-3|\tilde{x}|-\sqrt{-3(1-|\tilde{x}|)^2+1}\right\} &&\mathrm{for}\ \frac{1}{2}<|\tilde{x}|\le\frac{3}{2}\\
			&0 &&\mathrm{otherwise}
		\end{aligned}
	\right.
	\label{eq:disdelta},
\end{align}
where $\Delta x$ is the grid width in the $x$-direction and $\tilde{x}=x/\Delta x$. The discrete delta function $\delta_h$ satisfies the following properties,
\begin{align}
	\sum_i \delta_h(x_i-\xi)\Delta x &=1
	\label{eq:consv_force},\\
	\sum_i (x_i-\xi)\delta_h(x_i-\xi)\Delta x &=0
	\label{eq:consv_torque},\\
	\sum_i [\delta_h(x_i-\xi)\Delta x]^2&=\frac{1}{2}
	\label{eq:trns_invar},
\end{align}
where $x_i=i\Delta x$ and $\xi$ is an arbitrary constant. Using the discrete delta function $\delta_h$,
the convolution of $\delta$ and $\bm{u}$ (in $\Omega$) and $\bm{F}$ (on $\Gamma$) are discretized as follows, respectively,
\begin{align}
	\begin{bmatrix}
		(u_\Gamma)_l\\
		(v_\Gamma)_l
	\end{bmatrix}
	&=
	\begin{bmatrix}
		\sum_i u_i\delta_h(x^{\mathcal{F}_x}_i-\xi_l)\delta_h(y^{\mathcal{F}_x}_i-\eta_l)\Delta x\Delta y	\\
		\sum_i v_i\delta_h(x^{\mathcal{F}_y}_i-\xi_l)\delta_h(y^{\mathcal{F}_y}_i-\eta_l)\Delta x\Delta y
	\end{bmatrix}
	\label{eq:conv_u}, \\
	\begin{bmatrix}
		f_{x,i}\\
		f_{y,i}
	\end{bmatrix}
	&=
	\begin{bmatrix}
		\sum_l F_{x,l}\delta_h(\xi_l-x^{\mathcal{F}_x}_i)\delta_h(\eta_l-y^{\mathcal{F}_x}_i)\Delta s 	\\
		\sum_l F_{y,l}\delta_h(\xi_l-x^{\mathcal{F}_y}_i)\delta_h(\eta_l-y^{\mathcal{F}_y}_i)\Delta s
	\end{bmatrix}
	\label{eq:conv_F},
\end{align}
where $\bm{x}^{\mathcal{F}_x}_i=[x^{\mathcal{F}_x}_i,y^{\mathcal{F}_x}_i]^T\in\mathcal{F}_x$ and
$\bm{x}^{\mathcal{F}_y}_i=[x^{\mathcal{F}_y}_i,y^{\mathcal{F}_y}_i]^T\in\mathcal{F}_y$
are the $i$th staggered grid points, and
$\bm{\xi}_l=[\xi_l,\eta_l]^T\in\Gamma$ is the $l$th Lagrangian point.
$\Delta s$ is the interval between adjacent Lagrangian points.
Since the RHS of Eq.~\eqref{eq:conv_u} is nothing more than an interpolation caluculation,
the interpolation operator $\hat{E}$ represented by the matrix can be defined as follows,
\begin{align}
	(\hat{E})_{l,i}:=\delta_h(x_i-\xi_l)\delta_h(y_i-\eta_l)\Delta x\Delta y
	\label{eq:E},
\end{align}
and then, Eq.~\eqref{eq:conv_u} are represented by
\footnote{If we make different notations for the interpolation operator depending on the group of the variable positions $\mathcal{F}_x$ or $\mathcal{F}_y$, we write $u_\Gamma=\hat{E}^{\mathcal{F}_x}u$ ($\hat{E}^{\mathcal{F}_x}\colon\mathbb{R}^{\mathcal{F}_x}\to\mathbb{R}^\Gamma$) and $v_\Gamma=\hat{E}^{\mathcal{F}_y}v$ ($\hat{E}^{\mathcal{F}_y}\colon\mathbb{R}^{\mathcal{F}_y}\to\mathbb{R}^\Gamma$), and 
	\begin{align*}
		\bm{u}_\Gamma=
		\begin{bmatrix}
		u_\Gamma\\
		v_\Gamma	
		\end{bmatrix}
		=
		\begin{bmatrix}
			\hat{E}^{\mathcal{F}_x} & \\
			 & \hat{E}^{\mathcal{F}_y}
		\end{bmatrix}
		\begin{bmatrix}
			u\\
			v
		\end{bmatrix}.
	\end{align*}
We employ such simplified notation as Eq.~\eqref{eq:conv_u_simplified} in this work, unless it is ambiguous.}
\begin{equation}\label{eq:conv_u_simplified}
    \bm{u}_{\Gamma}=\hat{E}\bm{u}\in\mathbb{R}^\Gamma.
\end{equation}
Similarly, the RHS of Eq.~\eqref{eq:conv_F} means the regularization of the boundary force $\bm{F}$,
and the regularization operator 
$\hat{H}$ can be defined as follows,
\begin{align}
	(\hat{H})_{i,l}:=\delta_h(\xi_l-x_i)\delta_h(\eta_l-y_i)\Delta s 
	\label{eq:H}.
\end{align}
Then, Eq.~\eqref{eq:conv_F} is represented by 
\begin{equation}\label{eq:taira}
    \bm{f} = \hat{H}\bm{F}.
\end{equation}
These operators $\hat{E}$ and $\hat{H}$ play a significant role in imposing the boundary conditions.

Together with the temporal discretization schemes, the second-order Adams-Bashforth scheme for the advection term and the Crank-Nicolson scheme for the viscous term, the discretized form of the governing equations \eqref{eq:NS_IB}--\eqref{eq:no-slip_IB}
are the following linear algebraic equations:
\begin{align}
	\frac{\bm{u}^{n+1}-\bm{u}^n}{\Delta t}
	&=
	\frac{1}{2}\left(3\bm{A}^n-\bm{A}^{n-1}\right)
	-\hat{G} P
	+\frac{1}{2\mathrm{Re}}\left(\hat{L}\bm{u}^{n+1}+\hat{L}\bm{u}^n\right) \nonumber\\ 
	&+\bm{f}_{\mathrm{ext}}^{n+1}
	+\hat{H}\bm{F}
	+bc_1
	\label{eq:Dis_NS_IB},
	\\
	\hat{D}\bm{u}^{n+1}&=bc_2
	\label{eq:Dis_continuity_IB},
	\\
	\hat{E}\bm{u}^{n+1} &= \bm{U}^{n+1}
	\label{eq:Dis_no-slip_IB},
\end{align}
where
$\hat{G}$ is the discrete gradient operator,
$\hat{L}$ is the discrete Laplacian operator and 
$\hat{D}$ is the discrete divergence operator, and these operators can be represented by matrices.
$bc_1$ and $bc_2$ are terms derived from the boundary condition on the domain boundary $\partial\Omega$.
$\bm{A}$ is the advection term. The superscript $n$ denotes the time step. Eqs.~\eqref{eq:Dis_NS_IB}--\eqref{eq:Dis_no-slip_IB} are further expressed by the following matrix representation.
\begin{align}
	\begin{bmatrix}
		\hat{R} & \Delta t \hat{G} & -\Delta t \hat{H} \\
		\hat{D} & 0 & 0 \\
		\hat{E} & 0 & 0 \\
	\end{bmatrix}
	\begin{bmatrix}
		\bm{u}^{n+1} \\
		P \\
		\bm{F} \\
	\end{bmatrix}
	=
	\begin{bmatrix}
		\bm{r}_{NS} \\
		0 \\
		\bm{U}^{n+1} \\
	\end{bmatrix}
	+
	\begin{bmatrix}
		bc_1 \\
		bc_2 \\
		0	\\
	\end{bmatrix}
	\label{eq:Mtrx_GE}.
\end{align}
Here, with $\hat{I}$ being the identity operator,
$\hat{R}=\hat{I}-\Delta t/(2\mathrm{Re})\hat{L}$ and
$\bm{r}_{NS}=(\hat{I}+\Delta t/(2\mathrm{Re})\hat{L})\bm{u}^n+(\Delta t/2)(3\bm{A}^n-\bm{A}^{n-1})
+\Delta t\bm{f}_{\mathrm{ext}}^{n+1}$.
Introducing $\tilde{P}=-\Delta t P$ and $\tilde{\bm{F}}=-\Delta t\Delta s(\Delta x\Delta y)^{-1}\bm{F}$, Eq.~\eqref{eq:Mtrx_GE} is now 
\begin{align}
	\begin{bmatrix}
		\hat{R} & \hat{D}^T & \hat{E}^T \\
		\hat{D} & 0 & 0 \\
		\hat{E} & 0 & 0 \\
	\end{bmatrix}
	\begin{bmatrix}
		\bm{u}^{n+1} \\
		\tilde{P} \\
		\tilde{\bm{F}} \\
	\end{bmatrix}
	=
	\begin{bmatrix}
		\bm{r}_{NS} \\
		0 \\
		\bm{U}^{n+1} \\
	\end{bmatrix}
	+
	\begin{bmatrix}
		bc_1 \\
		bc_2 \\
		0	\\
	\end{bmatrix}
	\label{eq:Mtrx_GE_sym}
\end{align}
since $\hat{H}=\Delta s(\Delta x\Delta y)^{-1}\hat{E}^T$ from Eqs.~\eqref{eq:E} and \eqref{eq:H}, and $\hat{G}=-\hat{D}^T$ for the uniform staggered mesh.

Since $\tilde{P}$ and $\tilde{\bm{F}}$ have the same role as the Lagrange multiplier to impose the constraints (\ref{App:KKT}), the operators and vectors can be grouped as follows:
$\hat{W}=[\hat{D}^T\ \ \hat{E}^T]^T$,
$\lambda=[\tilde{P}\ \ \tilde{\bm{F}}]^T$,
$\bm{r}_1=\bm{r}_{NS}+bc_1$,
$\bm{r}_2=[bc_2\ \ \bm{U}^{n+1}]^T$.
Using these expressions,
the coefficient matrix in Eq.~\eqref{eq:Mtrx_GE_sym} 
can be LU-decomposed as follows:
\begin{align}
	\begin{bmatrix}
		\hat{R} & \hat{W}^T \\
		\hat{W} & 0 \\
	\end{bmatrix}
	=
	\begin{bmatrix}
		\hat{R} & 0 \\
		\hat{W} & -\hat{W}\hat{R}^{-1}\hat{W}^T \\
	\end{bmatrix}
	\begin{bmatrix}
		\hat{I} & \hat{R}^{-1}\hat{W}^T \\
		0 & \hat{I}
	\end{bmatrix}.
	\label{eq:LU}
\end{align}
$\hat{R}^{-1}$ can be expressed 
as an infinite series expansion by the Neumann series as follows,
\begin{align}
	\hat{R}^{-1}=\left(\hat{I}-\frac{\Delta t}{2\mathrm{Re}}\hat{L}\right)^{-1}
						  &= \hat{I}+\frac{\Delta t}{2\mathrm{Re}}\hat{L}
											 +\left(\frac{\Delta t}{2\mathrm{Re}}\right)^2(\hat{L})^2
											 +\cdots \nonumber \\
						  &= \sum_{k=1}^{\infty}\left(\frac{\Delta t}{2\mathrm{Re}}\right)^{k-1}(\hat{L})^{k-1}
	\label{eq:R_inv}.
\end{align}
Truncating at the $N$th term ($\hat{R}^{-1}=\hat{C}^N+\mathcal{O}(\Delta t^N)$), Eq.~\eqref{eq:Mtrx_GE} is finally written as
\begin{align}
	\begin{bmatrix}
		\hat{R} & 0 \\
		\hat{W} & -\hat{W}\hat{C}^{N}\hat{W}^T \\
	\end{bmatrix}
	\begin{bmatrix}
		\hat{I} & \hat{C}^{N}\hat{W}^T \\
		0 & \hat{I} \\
	\end{bmatrix}
	\begin{bmatrix}
		\bm{u}^{n+1} \\
		\lambda \\
	\end{bmatrix}
	=
	\begin{bmatrix}
		\bm{r}_1 \\
		\bm{r}_2 \\
	\end{bmatrix}
	+
	\begin{bmatrix}
		-\left(\frac{\Delta t}{2\mathrm{Re}}\right)^N(\hat{L})^N\hat{W}^T\lambda \\
		0\\
	\end{bmatrix}
	\label{eq:Mtrx_GE_LU},
\end{align}
where the last term on the RHS is the truncation error. The calculation procedure of the IBPM with the no-slip boundaries to obtain $\bm{u}^{n+1}$ and $\lambda$ is 
\begin{align}
	\hat{R}\bm{u}^F&=\bm{r}_1
	\label{eq:calc_proc_noslip_1},\\
	\hat{W}\hat{C}^N\hat{W}^T\lambda&=\hat{W}\bm{u}^F-\bm{r}_2
	\label{eq:calc_proc_noslip_2},\\
	\bm{u}^{n+1}&=\bm{u}^F-\hat{C}^N\hat{W}^T\lambda,
	\label{eq:calc_proc_noslip_3}
\end{align}
which is a fractional step algorithm.

\section{Immersed boundary projection method for the Navier slip boundaries}
\label{S:IBPM-Navier}
We invent the IBPM for the Navier slip boundaries so that the discretized equations and boundary conditions are expressed with linear operators as Eqs.~\eqref{eq:Dis_NS_IB}--\eqref{eq:Dis_no-slip_IB} whose solution satisfies the continuity equation and the boundary conditions simultaneously by the calculation procedure as Eqs.~\eqref{eq:calc_proc_noslip_1}--\eqref{eq:calc_proc_noslip_3}. It is not possible to impose the Navier BC appropriately by applying the conventional method that uses the discrete delta function to regularize the boundary forces. In this paper, we first discuss the difference between the no-slip BC and the Navier BC, and describe the relation between the velocity gradient term (shear stress term) in the Navier BC and the boundary force. Then, we derive the condition that the boundary force distribution regularized on the Eulerian mesh should satisfy, in order to evaluate the velocity gradient term appropriately. From this condition, we formulate the interpolation and regularization operators of the IBPM for the Navier slip boundaries.

\subsection{Differences between the no-slip BC and the Navier BC}
First, we discuss the difference between the no-slip BC and the Navier BC. First of all, their boundary condition types are different. The no-slip BC (Eq.~\ref{eq:no-slip_PBC}) is a Dirichlet type BC, where the velocity value on the surface $\Gamma$ is determined by the boundary conditions. On the other hand, the Navier BC (Eq.~\ref{eq:NBC}) is a Robin type BC, which only specifies a linear relationship between the velocity and the shear stress (velocity gradient) on the surface $\Gamma$ but does not specify their values: their values are obtained as a part of the solution. Therefore, when the Navier BC is applied, it is reasonable to devise a solution method that treats the boundary condition implicitly and obtains the flow field satisfying the continuity equation and the boundary condition simultaneously, like IBPM. For the method, it is required to evaluate both the velocity and the velocity gradient on $\Gamma$ appropriately. However, as described later, in order to evaluate the velocity gradient on $\Gamma$ appropriately, it is necessary to consider the effect of the boundary force on the shear stress field near $\Gamma$, which is not as simple as the evaluation of the velocity.

The second difference is the continuity of the velocity at $\Gamma $ when the boundary condition is imposed on the two sides of $\Gamma$. In the case of the no-slip BC, the velocity on both sides of $\Gamma $ coincides with the velocity at $\Gamma $ and is therefore continuous. However, in the case of the Navier BC, even if the slip lengths are the same on two sides of $\Gamma$, the slip velocities are different according to the flow in each region, and therefore the velocity is discontinuous at $\Gamma$ in general. In the continuous forcing approach including the IBPM, the continuity of the velocity at $\Gamma $ is prerequisite. 

In the present method, we impose the Navier BC only on the fluid side of $\Gamma$. The boundary forces are distributed continuously on the Eulerian mesh without changing the discretization method near $\Gamma$ as the conventional continous forcing approach. In the present method, the IBPM for the Navier BC, the appropriate evaluation of the shear stress (velocity gradient) term of the Navier BC is important. The shear stress near $\Gamma$ is affected by the boundary force distributed by the discrete delta function on the Eulerian mesh. For the appropriate evaluation of the shear stress on $\Gamma$, this effect of the boundary force must be considered. In the present method, the boundary shear stress is newly introduced in addition to the boundary force so that the shear stress on $\Gamma $ evaluated by the interpolation gives an appropriate value.

In this paper, first, the main principles of the method are shown in the continuous form, and then the final formulations are derived by discretizing the continuous form. The two dimensional version is shown after the one dimensional one.

\subsection{One dimensional case}
\subsubsection {Discussion in the continuous form}
\label{S:C1D}
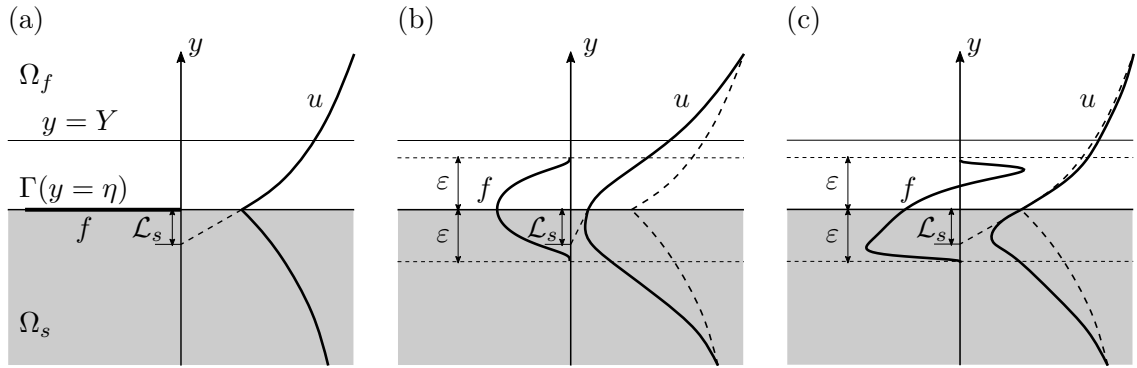
\begin{figure}[t]
	\begin{center}
		\begin{overpic}[bb=0 0 977 273,width=15cm,clip,keepaspectratio]
				{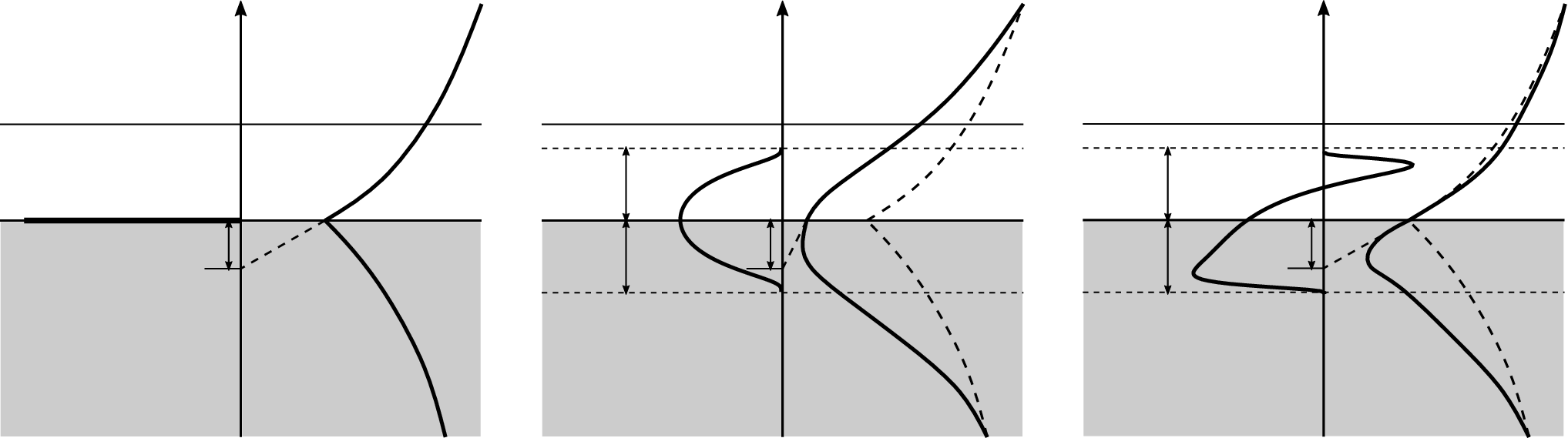}
				\put(16,28){$y$}
				\put(1,25){$\Omega_f$}
				\put(1,3){$\Omega_s$}
				\put(1,15){$\Gamma(y=\eta)$}
				\put(11,11.5){$\mathcal{L}_s$}
				\put(6,11.5){$f$}
				\put(3,21){$y=Y$}
				\put(26.5,23){$u$}
				\put(51,28){$y$}
				\put(38,15.7){$\varepsilon$}
				\put(38,11){$\varepsilon$}
				\put(46,11.5){$\mathcal{L}_s$}
				\put(41.5,15){$f$}
				\put(59,23){$u$}
				\put(85,28){$y$}
				\put(72.5,15.7){$\varepsilon$}
				\put(72.5,11){$\varepsilon$}
				\put(80.5,11.5){$\mathcal{L}_s$}
				\put(79,15){$f$}
				\put(95,23){$u$}
				\put(0,30){(a)}
				\put(34.5,30){(b)}
				\put(69,30){(c)}
		\end{overpic}
		\caption{Distributions of the fluid velocity $u$ and the regularized (distributed) force $f$ near the boundary $\Gamma$ when $f$ is regularized by (a) the Dirac delta function, (b) the conventional approximate delta function, and (c) our formulation. $\mathcal{L}_s$ is the slip length and $\varepsilon$ is the support of the approximate delta function.}
	 \label{fig:velocity-force_1D}
	\end{center}
\end{figure}

Consider a one-dimensional system with $u=u(y, t) $ and $v=0$.
The computational domain is as shown in Fig.~\ref{fig:velocity-force_1D}a: $y>\eta$ denotes the fluid domain $\Omega_f$, $y<\eta$ the object domain $\Omega_s$, and $y=\eta $ the surface $\Gamma$. The governing equation of the IB formulation for this system is
\begin{align}
	\pd{u}{t}=-\pd{P}{x}+\frac{1}{\mathrm{Re}}\pdd{u}{y}+f_\mathrm{ext}+f.
	\label{eq:GE_1D}
\end{align}
Here, the pressure gradient $\partial P/\partial x$ is a constant. A singular body force 
\begin{equation}
f(y) =F\delta (\eta-y)
\end{equation}
is imposed on the RHS of Eq.~\eqref{eq:GE_1D} with $F$ being the boundary force, to impose the Navier BC
\begin{align}\label{eq:NBC_1D}
  \lim_{y\to\eta_+}\left(u|_y-\mathcal{L}_s\left.\pd{u}{y}\right|_y\right)=U
\end{align}
on $\Gamma$. In Eq.~\eqref{eq:NBC_1D}, the one-sided limit is used because, as shown in Fig.~\ref{fig:velocity-force_1D}a, the shear stress term $(\partial u/\partial y) |_y$ is discontinuous at $y=\eta $ due to the singular body force $f$. This can also be seen from the fact that the Heaviside function $\theta (y)$ appears in Eq.~\eqref{eq:GE_1D} when it is integrated over the interval $[Y_0, Y]$, where $Y_0$ and $Y$ are any points in $\Omega_f$:
\begin{align}
  \left.\pd{u}{y}\right|_{Y_0}&=\left.\pd{u}{y}\right|_Y \nonumber\\
	&+\mathrm{Re}\int_{Y_0}^Y\left(-\pd{u}{t}-\pd{P}{x}+f_\mathrm{ext}\right) \dd{y}
	+\mathrm{Re}F\theta(\eta-Y_0)
  \label{eq:GE_1D_int}.
\end{align}
It is important to note that the last term in Eq.~\eqref{eq:GE_1D_int} is always zero, that is, for any control volume in $\Omega_f$, $f$ does not exert a net force. This is the property that the IB force should preserve for the BC involving the velocity gradient as we detail later. As an example, let us consider the steady Couette flow, where the second term on the RHS of Eq.~\eqref{eq:GE_1D_int} is zero. For the IB force to be consistent with the Navier BC ($Y_0 \to \eta_+$), the last term on the RHS of Eq.~\eqref{eq:GE_1D_int} should be zero: otherwise the velocity gradient on the boundary $(\partial u/\partial y)|_{Y_0\to\eta_+}$ shows an unphysical value even if the velocity gradient in the fluid body $(\partial u/\partial y)|_Y$ is correct. 

Next, we replace the Dirac delta function $\delta(y)$ with a $C^1$ class approximate delta function $\delta_\varepsilon (y)$ with the finite support $[-\varepsilon,\varepsilon]$ (see Fig.~\ref{fig:velocity-force_1D}b). We let $\delta_\varepsilon (y) $ have the fundamental property $\int_{-\varepsilon}^{\varepsilon}\delta_\varepsilon(y)\dd{y}=1$ and
$\delta_\varepsilon (y) =\delta_\varepsilon (-y) $. For $Y$ being any point located outside the support of $\delta_\varepsilon (y)$, Eq.~\eqref{eq:GE_1D_int} is rewitten as
\begin{align}
	\left.\pd{u}{y}\right|_{Y_0}&=\left.\pd{u}{y}\right|_Y \nonumber\\
	&+\mathrm{Re}\int_{Y_0}^Y\left(-\pd{u}{t}-\pd{P}{x}+f_\mathrm{ext}\right) \dd{y}
	+\mathrm{Re}F\theta_\varepsilon(\eta-Y_0)
	\label{eq:GE_1D_int_approx}.
\end{align}
Here, $\theta_\varepsilon (y) = \int \delta_\varepsilon (y)\dd{y}$ is the approximate Heaviside function, which changes continuously from 0 to 1 in the transition region $[-\varepsilon,\varepsilon]$. The last term of Eq.~\eqref{eq:GE_1D_int_approx} is non-zero for $\eta\le Y_0\le\eta+\varepsilon$. Therefore the conventional approximate delta functions are always inconsistent with the Navier BC ($Y_0 \to \eta$)\footnote{Since $(\partial u/\partial y) |_y$ is continuous at $y=\eta $ as shown in Fig.~\ref{fig:velocity-force_1D}b, Navier BC is evaluated at $y=\eta $ without using the one-sided limit.}. 

In the present work, we propose a regularization function (operator) that produces the regularized force distribution $f$ satisfying
\begin{equation}\label{eq:cond_f}
\lim_{Y_0\to\eta}\int_{Y_0}^Yf\dd{y}=
0.
\end{equation}
The regularized force $f$ exerts force localy in $\Omega_f$ but zero net force in the region $[\eta,Y]$. We realize this by introducing a forcing shear stress on $y=\eta$, which is distributed as $M\delta_\varepsilon(\eta-y)$. 
In terms of the force distribution, we exert 
\begin{equation}\label{eq:Fd+Md'}
f(y)=F\delta_\varepsilon(\eta-y)+\drv{}{y}\qty[M\delta_\varepsilon(\eta-y)]
\end{equation}
all together. $M$ is a parameter representing the magnitude of the forcing shear stress, which is determined by $F$ to satisty Eq.~\eqref{eq:cond_f} and $M=F/(2\delta_\varepsilon(0))$. 

For the illustration, Eq.~\eqref{eq:GE_1D_int_approx} is now rewritten as
\begin{align}
	\left.\pd{u}{y}\right|_{Y_0}+
    \mathrm{Re}M\delta_\varepsilon(\eta-Y_0)
    &=
    \left.\pd{u}{y}\right|_Y \nonumber\\
	&+\mathrm{Re}\int_{Y_0}^Y\left(-\pd{u}{t}-\pd{P}{x}+f_\mathrm{ext}\right) \dd{y}
	+\mathrm{Re}F\theta_\varepsilon(\eta-Y_0).
	\label{eq:GE_1D_int_approx_M}
\end{align}
When we take the limit $Y_0\to\eta$, the last term on the LHS cancels the last term on the RHS, and we see the present formulation is consistent with the Navier BC. 
It is important to note that $f(y)$ given by Eq.~\eqref{eq:Fd+Md'} satisfies the fundamental property of the force conservation $\int_{-\infty}^{\infty}f (y)\dd{y}=F$. On the other hand, the torque conservation is not satisfied since $\int_{-\infty}^{\infty}-(y-\eta) f (y)\dd{y}=M\neq 0$. In one-dimensional systems, however, it does not cause a problem because there is no rotational motion in the system. For two-dimensional systems, our generalized formulation for the two-dimension satisfies the torque conservation as shown later.

As the final step, we replace the limit in Eq.~\eqref{eq:cond_f} by the interpolation to evaluate the value on the boundary, as usually performed in the discretized space. We conduct the interpolation by the convolution with the approximate delta function $\delta_\varepsilon$. The condition that the regularized force should satisfy (Eq.~\ref{eq:cond_f}) is then
\begin{equation}\label{eq:cond_f_delta}
\int_{\eta-\varepsilon}^{\eta+\varepsilon}\qty(\int_y^Y f(y')\dd{y'})\delta_\varepsilon(y-\eta)\dd{y}=0,
\end{equation}
which is used to relate $M$ to $F$ in the discretized formulation. The Navier BC (Eq.~\ref{eq:NBC_1D}) is rewritten as
\begin{equation}\label{eq:NBC_1D_delta_approx}
\int_{\eta-\varepsilon}^{\eta+\varepsilon}\qty(u|_y-\mathcal{L}_s\left.\pd{u}{y}\right|_y)\delta_\varepsilon(y-\eta)\dd{y}=U.
\end{equation}

\subsubsection {IBPM for the Navier BC in one-dimension}
\label{S:IBPM-1D}
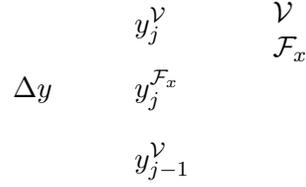
\begin{figure}[t]
	\begin{center}
		\begin{overpic}[bb=0 0 100 120,width=3cm,clip,keepaspectratio]
				{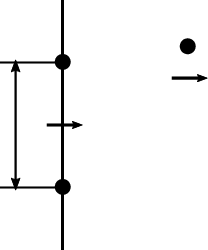}
				\put(-11,50){$\Delta y$}
				\put(34,50){$y^{\mathcal{F}_x}_j$}
				\put(34,74){$y^\mathcal{V}_j$}
				\put(34,24){$y^\mathcal{V}_{j-1}$}
				\put(85,65){$\mathcal{F}_x$}
				\put(85,78){$\mathcal{V}$}
		\end{overpic}
		\caption{Grid points on the staggered mesh in one-dimension.}
	 \label{fig:staggered_grid_1D}
	\end{center}
\end{figure}

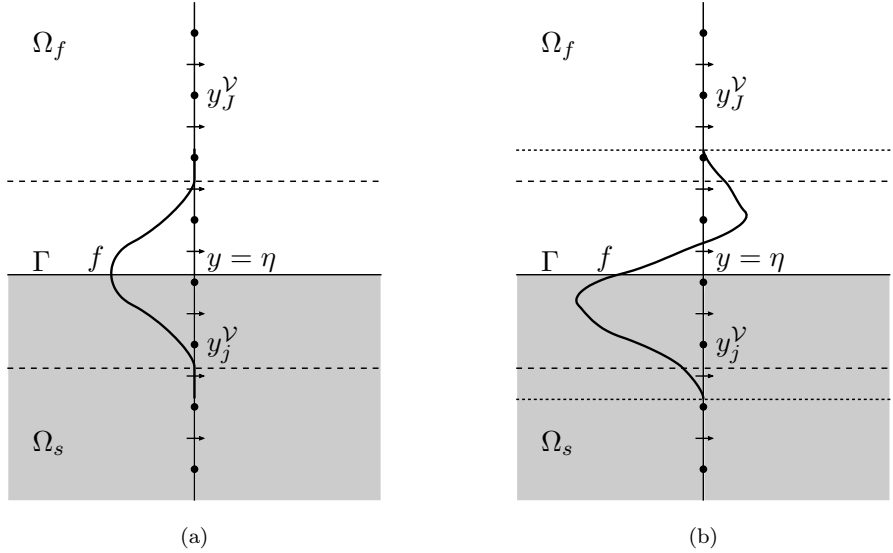
\begin{figure}[t]
    \centering
    \begin{subfigure}[t]{0.3\linewidth}
        \centering
        \begin{overpic}[width=\linewidth,clip,keepaspectratio]
            {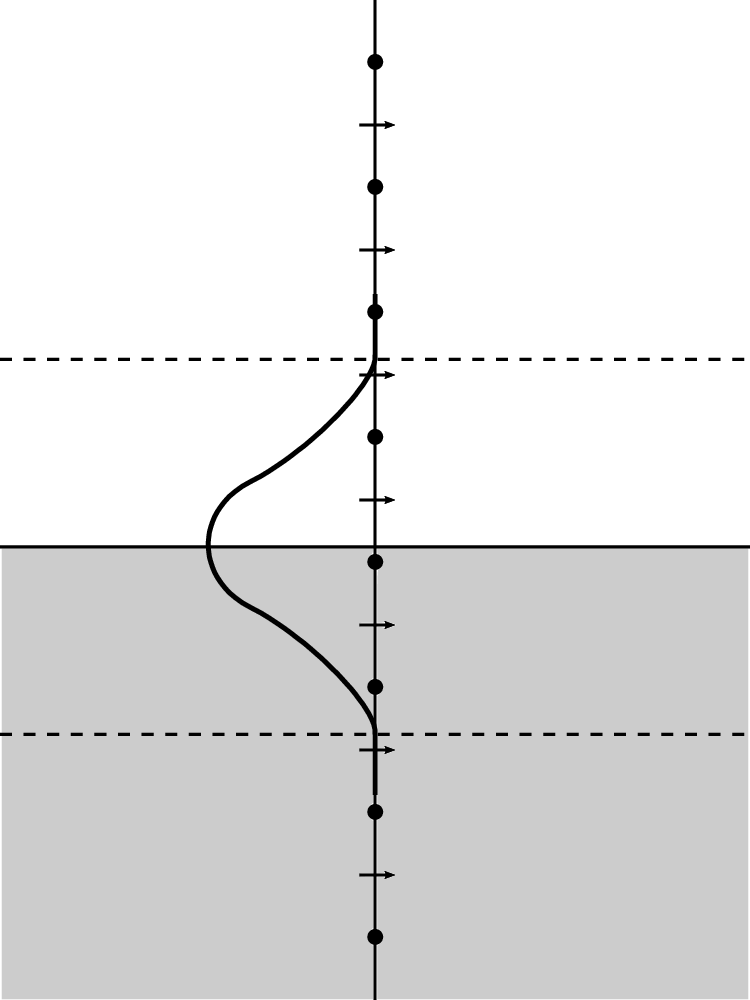}
            \put(40,80){$y^\mathcal{V}_J$}
            \put(40,30){$y^\mathcal{V}_j$}
            \put(40,47){$y=\eta$}
            \put(16,47){$f$}
            \put(5,90){$\Omega_f$}
            \put(5,10){$\Omega_s$}
            \put(5,46){$\Gamma$}
        \end{overpic}
        \caption{}
        \label{fig:force_dh_1D}
    \end{subfigure}%
    \hspace{0.1\linewidth}
    \begin{subfigure}[t]{0.3\linewidth}
        \centering
        \begin{overpic}[width=\linewidth,clip,keepaspectratio]
            {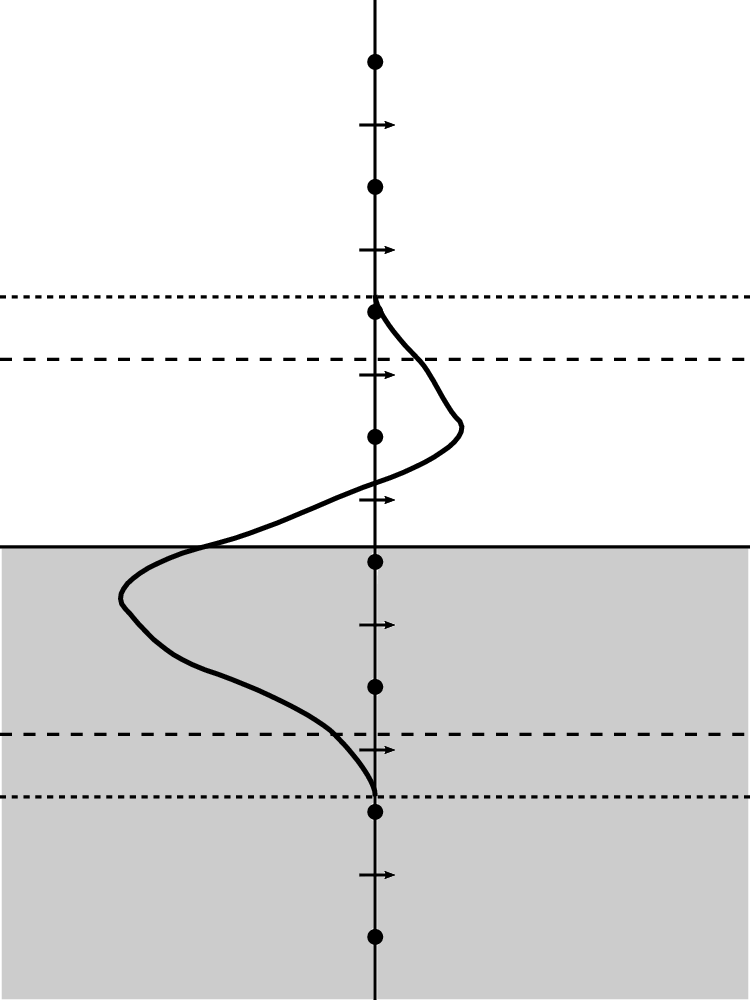}
            \put(40,80){$y^\mathcal{V}_J$}
            \put(40,30){$y^\mathcal{V}_j$}
            \put(40,47){$y=\eta$}
            \put(16,47){$f$}
            \put(5,90){$\Omega_f$}
            \put(5,10){$\Omega_s$}
            \put(5,46){$\Gamma$}
        \end{overpic}
        \caption{}
        \label{fig:force_dh+ddh_1D}
    \end{subfigure}
    \caption{Regularized force distribution by (a) the conventional method using the single approximate delta function (the dotted lines show the edge position of the support) and (b) the present method that additionally imposes shear on the boundary using three approximate delta functions (the fine dotted lines show the edge of the support).}
    \label{fig:force_1D}
\end{figure}

The principles described in \ref{S:C1D} is now expressed in the discretized form to derive the final IBPM formulations. For the derivation, only the spatial discretization matters and therefore the temporal derivative is left in the continuous form. Assume that the one-dimensional Eulerian mesh consists of a set of grid points $\mathcal{F}_x$ and $\mathcal{V}$ as shown in Fig.~\ref{fig:staggered_grid_1D}. 
The $y$ coordinates of these grid points in the $j$th cell are denoted by $y^{\mathcal{F}_x}_j$ and $y^{\mathcal{V}}_j$, respectively. If the grid width is $\Delta y$, they are related by $y^{\mathcal{V}}_j=y^{\mathcal{F}_x}_j+\Delta y/2$. The velocity $u$ is a quantity on $\mathcal{F}_x$, and the value of $u$ at $y^{\mathcal{F}_x}_j$ is denoted by $u_j$.
The finite difference of $u$ in $y$ direction $\hat{\partial}_y u$ is calculated on $y^\mathcal{V}_j\in\mathcal{V}$ as $(\hat{\partial}_y u) _j= (\Delta y) ^{-1} (-u_j+u_{j+1})$. Here, $\hat{\partial}_y $ is the difference operator in the $y $-direction. The governing equation \eqref{eq:GE_1D} is 
\begin{align}
	\pd{u_j}{t}
	=%
	-\left.\pd{P}{x}\right|_j
    +\frac{1}{\mathrm{Re}}\frac{-(\hat{\partial}_y u)_{j-1}+(\hat{\partial}_y u)_j}{\Delta y}
	+f_{\mathrm{ext},j}
	+f_j
	\label{eq:GE_1D_dis}
\end{align}
in the discretized form using the second-order finite difference. The discretized Navier BC (Eq.~\ref{eq:NBC_1D_delta_approx}) is written as
\begin{equation} \label{eq:NBC_1D_dis_j}
\sum_{y^{\mathcal{F}_x}_j\in \Omega_\delta}
u_j\delta_h(y^{\mathcal{F}_x}_j-\eta)\Delta y
-\mathcal{L}_s\sum_{y^\mathcal{V}_j\in \Omega_\delta}
(\hat{\partial}_y u)_j\delta_h(y^{\mathcal{V}}_j-\eta)\Delta y=U,
\end{equation}
employing $\delta_h$ (Eq.~\ref{eq:disdelta}) for the approximate delta function $\delta_\varepsilon$. Here, $\Omega_\delta=\mathrm{supp} (\delta_h (y-\eta))=[\eta-\varepsilon,\eta+\varepsilon]$ where $\varepsilon= (3/2)\Delta y$ for $\delta_h$. The condition (Eq.~\ref{eq:cond_f_delta}) in the discretized form is 
\begin{equation}\label{eq:cond_f_delta_dis}
\sum_{y^{\mathcal{V}}_j\in \Omega_\delta}
	\qty(
		\sum_{j'=j+1}^{J}f_{j'}\Delta y
	)\delta_h(y^{\mathcal{V}}_j-\eta)\Delta y=0,
\end{equation}
where the numerical integration is conducted for the range $[y^\mathcal{V}_j, y_J^\mathcal{V}]$ with the end point $y_J^\mathcal{V}\in\mathcal{V}$ being a point of $\Omega_f$ outside the support of $f$ (See Figs.~\ref{fig:staggered_grid_1D} and \ref{fig:force_1D}). The summation index $j'$ starts from $j+1$ since Eq.~\eqref{eq:GE_1D_int_approx} is discretized as
\begin{align}\label{eq:GE_1D_dis_sum}
	(\hat{\partial}_y u)_j
	&=
	(\hat{\partial}_y u)_J \notag\\
	&+\mathrm{Re}\sum_{j'=j+1}^{J}
	\left(
		-\pd{u_{j'}}{t}-\left.\pd{P}{x}\right|_{j'}+f_{\mathrm{ext},j'}
	\right)\Delta y\notag\\
	&+\mathrm{Re}\sum_{j'=j+1}^{J}f_{j'}\Delta y
\end{align}
and this last term should correspond to the term in the parentheses of Eq.~\eqref{eq:cond_f_delta_dis}. 

As elucidated in the continuous form, the conventional IB force $f_j=F\delta_h (\eta-y^{\mathcal{F}_x}_j)$ does not sasisfy the condition \eqref{eq:cond_f_delta_dis}. 
Our formulation of the IB force is obtained by discretizing Eq.~\eqref{eq:Fd+Md'}:
\begin{equation}\label{eq:Fd+Md'_dis}
f_j=F\delta_h(\eta-y^{\mathcal{F}_x}_j)+M(\Delta y)^{-1}(-\delta_h(\eta-y^{\mathcal{V}}_{j-1})+\delta_h(\eta-y^{\mathcal{V}}_j)).
\end{equation}
As in the continous form, $M$ is related to $F$ in order to satisfy the condition \eqref{eq:cond_f_delta_dis} and together with the  property of $\delta_h$ (Eq.~\ref{eq:trns_invar}) it is written as
\begin{equation}\label{eq:M-F_1D}
M=\qty[2\Delta y\sum_{y^{\mathcal{V}}_j\in \Omega_\delta}
\qty(\sum_{j'=j+1}^{J}\delta_h(\eta-y^{\mathcal{F}_x}_{j'})\Delta y)
\delta_h(y^{\mathcal{V}}_j-\eta)\Delta y]F.
\end{equation}
The IB force by Eq.~\eqref{eq:Fd+Md'_dis} with $M$ given by Eq.~\eqref{eq:M-F_1D} satisfies the condition \eqref{eq:cond_f_delta_dis} regardless of the value of $F$, and is expected to be consistent with the Navier BC (Eq.~\ref{eq:NBC_1D_dis_j}). Similar formulations can be derived for the boundary whose normal vector is pointing in the negative $y$ direction (See \ref{App:negative}).

Note that the present IB force (Eq.~\ref{eq:Fd+Md'_dis}) is also written as
\begin{align}\label{eq:Fd+Md'_dis'}
	f_j=F\delta_h(\eta-y^{\mathcal{F}_x}_j)+M(\Delta y)^{-1}(-\delta_h(\eta^{+}-y^{\mathcal{F}_x}_j)+\delta_h(\eta^{-}-y^{\mathcal{F}_x}_j)),
\end{align}
considering $y^{\mathcal{V}}_{j-1}=y^{\mathcal{F}_x}_j-\Delta y/2 $ and $y^{\mathcal{V}}_j=y^{\mathcal{F}_x}_j+\Delta y/2$ with $\eta^{-}$ and $\eta^{+}$ being $\eta-\Delta y/2$ and $\eta+\Delta y/2$ respectively. It can be interpreted that $F$ is imposed on $\eta$, $M (\Delta y)^{-1}$ is on $\eta^{-}$, and $-M (\Delta y)^{-1}$ is on $\eta^{+}$ in our formulation. It is reasonable for the boundary forces to act on these three locations because the discretized Navier BC (Eq.~\ref{eq:NBC_1D_dis_j}) is actually the relationship between the interpolated values on $\eta$, $\eta^{-}$ and $\eta^{+}$:
\begin{align} \label{eq:NBC_1D_dis_int}
	\hat{E}_\eta u-\mathcal{L}_s(\Delta y)^{-1}(-\hat{E}_{\eta^{-}} u+\hat{E}_{\eta^{+}} u)=U,
\end{align}
where the interpolation operator to an arbitrary point $y$ using $\delta_h$ is expressed as $\hat{E}_y$.
Figs.~\ref{fig:force_dh_1D} and \ref{fig:force_dh+ddh_1D} show the distribution of $f$ by the conventional method and by the present method, respectively. In order to satisfy Eq.~\eqref{eq:cond_f_delta_dis}, $f$ by the present method acts in the opposite direction to the conventional $f$ in the $\Omega_f$ region. It should also be noted that the support of $f$ in the present method is $[\eta-2\Delta y,\eta+2\Delta y]$ and wider than the conventional support by $\Delta y/2 $ on each side (Fig.~\ref{fig:force_dh+ddh_1D}).

The regularized force $f$ must have the following fundamental properties as well as the condition \eqref{eq:cond_f_delta_dis}. These are the conservation of force, the conservation of torque and the translational invariance: 
\begin{align}
	\sum_j f_j\Delta y&=F
	\label{eq:consv_force_1D},\\
	\sum_j -(y^{\mathcal{F}_y}_j-\eta)f_j\Delta y &= 0
	\label{eq:consv_torque_1D},\\
	\sum_j f_j\delta_h(y^{\mathcal{F}_y}_j-\eta)\Delta y&=cF/\Delta y,
	\label{eq:trns_invar_1D}
\end{align}
where $c$ is a constant. 
The force by the conventional method $f_j=F\delta_h (\eta-y^{\mathcal{F}_x}_j)$ can be shown to have these properties, considering the property of $\delta_h$ (Eqs.~\ref{eq:consv_force}--\ref{eq:trns_invar}). 
The conservation of force (Eq.~\ref{eq:consv_force_1D})  must be satisfied to guarantee the conservation of the total momentum in the system. If it is not satisfied, the action $\sum_j f_j\Delta y$ exerted by the body on the fluid does not coincide with the reaction $F $ exerted by the fluid on the body. The conservation of torque (Eq.~\ref{eq:consv_torque_1D}) is necessary for the conservation law of angular momentum, but for one-dimensional systems, there is no problem even if it is not satisfied because the rotational motion of the body is not considered. The translational invariance (Eq.~\ref{eq:trns_invar_1D}) requires that the regularized force gives a constant value when it is interpolated back on the boundary regardless of the relative position between the Eulerian mesh $\mathcal{M}$ and the boundary $\Gamma$. It is an important property in the moving boundary problems for the method to give consistent results.

Now we have to see if the present method, Eq.~\eqref{eq:Fd+Md'_dis}, possesses these three properties. The first two properties have already been discussed in the continuous form, but it is repeated here for the completeness. First, the force conservation is satisfied by virtue of Eq.~\eqref{eq:consv_force}, regardless of the value of $M$. For the torque, from Eqs.~\eqref{eq:consv_force} and \eqref{eq:consv_torque} it can be shown that $\sum_j -(y^{\mathcal{F}_y}_j-\eta)f_j\Delta y = M \neq 0$,
therefore torque is not conserved. As mentioned above, it is not a problem in one-dimensional systems, and it will be shown later that it is indeed conserved if the method is generalized for two-dimensional systems. For the translation invariance, from Eqs.~\eqref{eq:trns_invar} and \eqref{eq:M-F_1D} it is shown that 
\begin{align}
	\sum_j f_j\delta_h(y^{\mathcal{F}_y}_j-\eta)\Delta y&=(c+\phi)F/\Delta y.
	\label{eq:trns_invar_1D_M}
\end{align}
Here, $\phi$ is a quantity depending on the relative position between $\mathcal{M}$ and $\Gamma$: by some arithmetic it can be shown that $|\phi/c|<1/10$. Therefore the violation of the translational invariance is not significant although it is not strictly satisfied. 

Finally, the above discrete expressions are written down with linear operators. Navier BC (Eq.~\ref{eq:NBC_1D_dis_j}) is written with the interpolation operator $\hat{E}$ (Eq.~\ref{eq:E}) as
\footnote{Strictly speaking, $\hat{E}$ acting on $u$ and $\hat{\partial}_y u$ are different due to the difference in the definition locations of $u$ and $\hat{\partial}_y u$. Therefore, we cannot write $\hat{E}(\hat{I}-\mathcal{L}_s\hat{\partial}_y) u=U$.}
\begin{align}
	(\hat{E}-\mathcal{L}_s\hat{E}\hat{\partial}_y)u=U
	\label{eq:NBC_1D_intp}
\end{align}
and the force regularization (Eq.~\ref{eq:Fd+Md'_dis}) is written as 
\begin{equation}\label{eq:temp}
f=\hat{H}F+\hat{\partial}_y\hat{H}M
\end{equation}
using the operator $\hat{H}$ (Eq.~\ref{eq:H}). 
The condition \eqref{eq:cond_f_delta_dis} is expressed as $\hat{E}\hat{J}f=0$ when the numerical integration in the interval $[y^\mathcal{V}_j, y_J^\mathcal{V}]$ is expressed by the linear operator $\hat{J}$, and Eq.~\eqref{eq:M-F_1D} is now 
\begin{equation}\label{eq:M}
M=-(\hat{E}\hat{J}\hat{\partial}_y\hat{H}) ^{-1}\hat{E}\hat{J}\hat{H}F
\end{equation}
using these operators.
Note that $\hat{J}\hat{\partial}_y=-\hat{I}$ and 
$-(\hat{E}\hat{J}\hat{\partial}_y\hat{H}) ^{-1}= (\hat{E}\hat{H}) ^{-1}=2\Delta y$.
The final form of the force regularization is given by
\begin{align}
	f=(\hat{H}+\hat{\partial}_y\hat{H}\hat{K})F,
	\label{eq:f_1D_K}
\end{align}
substituting Eq.~\eqref{eq:M} into Eq.~\eqref{eq:temp} and writing $\hat{K}=-(\hat{E}\hat{J}\hat{\partial}_y\hat{H}) ^{-1}\hat{E}\hat{J}\hat{H}$.
The IBPM for the Navier BC is obtained by replacing $\hat{E}u=U$ and $f=\hat{H}F$ in the conventional method for the no-slip BC with Eq.~\eqref{eq:NBC_1D_intp} and Eq.~\eqref{eq:f_1D_K}, respectively. Note that the operator on the velocity (Eq.~\ref{eq:NBC_1D_intp}) reduces to the no-slip version when $\mathcal{L}_s=0$, but not the operator on the force (Eq.~\ref{eq:f_1D_K}). The operator $\hat{K}$ is independent of $\mathcal{L}_s$.
Before concluding this section, the relationship between the interpolation and the regularization operators in the present method is furthur discussed. Having in mind $\hat{H}= (\Delta y) ^{-1}\hat{E}^T$ and $\hat{\partial}_y\hat{H}=-(\Delta y) ^{-1} (\hat{E}\hat{\partial}_y) ^T$
\footnote{
	For the sake of notational simplicity, we use the same notation for $\hat{\partial}_y$ acting on a quantity on $\mathcal{F}_x$ and
	$\hat{\partial}_y$ acting on a quantity on $\mathcal{V}$ , but their matrix elements are actually different. The former ($\hat{\partial}^{\mathcal{F}_x}_y$) and the latter ($\hat{\partial}^\mathcal{V}_y$) are related by $\hat{\partial}^\mathcal{V}_y=-(\hat{\partial}^{\mathcal{F}_x}_y) ^T$.
}, the force regularization (Eq.~\ref{eq:f_1D_K}) can also be written as $f=(\Delta y)^{-1}(\hat{E}-\hat{K}^T\hat{E}\hat{\partial_y})^T F$, 
from which we see the correspondence between the shear stress term $\mathcal{L}_s\hat{E}\hat{\partial}_y$ in the Navier BC and the new term $(\hat{K}^T\hat{E}\hat{\partial_y}) ^T$ in the regularization operator. 

\subsection{Two dimensional case}
The method for one-dimensional systems is generalized to two-dimensional systems. The main generalization is in the (local) direction of the tangential/normal vector of the boundary and the integration in the consistency condition (Eq.~\ref{eq:cond_f_delta}). As in the previous section, the principles of the method is presented first in the continuous form. Then, IBPM for the Navier BC is formulated by the discretization of the expressions in the continuous form.
\subsubsection {Discussion in the continuous form}
\label{S:2DC}
We generalize the consistency condition (Eq.~\ref{eq:cond_f_delta}) for the regularized force in  two-dimension. The condition that the regularized force must satisfy in the one-dimensional system is that the boundary force term appearing in the relation (Eq.~\ref{eq:GE_1D_int}) between the shear stress at $\Gamma$ $(y=Y_0\to\eta)$ and the shear stress at an arbitrary point $y=Y\in\Omega_f$ sufficiently far from $\Gamma$ should be zero. This idea is extended to two-dimensional systems: we consider the shear stress relationship between a point $\bm{\xi}$ on the surface $\Gamma$ and a point $\bm{\xi}+\nu\bm{n}\in\Omega_f$ ($\nu>0$) sufficiently far from $\bm{\xi}$ in the normal direction of $\Gamma$. In deriving the relation of the shear stress, we use $\{\bm{\tau},\bm{n}\} $ at $\bm{\xi}$ as basis vectors, which form the local $\tau$-$n$ Cartesian coordinate system\footnote{It is not a curvilinear coordinate system along $\Gamma$.}. Then, under $\nabla\cdot\bm{u}=0$, the following equation holds:
\begin{align}
	\pd{}{n}\qty{\bm{\tau}\cdot[\nabla\bm{u}+(\nabla\bm{u})^T]\cdot\bm{n}}+\pd{}{\tau}\qty{\bm{\tau}\cdot[\nabla\bm{u}+(\nabla\bm{u})^T]\cdot\bm{\tau}}
	=\bm{\tau}\cdot\laplacian{\bm{u}}
	\label{eq:st_dndt}.
\end{align}
Integrating Eq.~\eqref{eq:st_dndt} from $\bm{x}$ to $\bm{x}+\nu\bm{n}$ in the $n$ direction and also considering the Navier-Stokes equation with the regularized force, we obtain the relation of the shear stress at two points $\bm{x} $ and $\bm{x}+\nu\bm{n}$ as
\begin{align}
	\bm{\tau}\cdot[\nabla\bm{u}+(\nabla\bm{u})^T]_{\bm{x}}\cdot\bm{n}
	&=\bm{\tau}\cdot[\nabla\bm{u}+(\nabla\bm{u})^T]_{\bm{x}+\nu\bm{n}}\cdot\bm{n}\notag\\
	&+\mathrm{Re}\int_{\bm{x}}^{\bm{x}+\nu\bm{n}}
	\bm{\tau}\cdot\left(-\pd{\bm{u}}{t}-\bm{u}\cdot\nabla\bm{u}-\nabla P+\bm{f}_\mathrm{ext}\right)\dd{n} \notag\\
&+\mathrm{Re}\int_{\bm{x}}^{\bm{x}+\nu\bm{n}}\bm{\tau}\cdot\bm{f}\dd{n} \notag\\
	&+\int_{\bm{x}}^{\bm{x}+\nu\bm{n}}\pd{}{\tau}\qty{\bm{\tau}\cdot[\nabla\bm{u}+(\nabla\bm{u})^T]\cdot\bm{\tau}}\dd{n}.
	\label{eq:st_dndt_int}
\end{align}
We require for the third term on the RHS to vanish as the consistency condition. Therefore the generalization of the condition \eqref{eq:cond_f_delta} is 
\begin{align}\label{eq:bf_cond}
	\int_{\Omega_{\delta}} \qty{\int_{\bm{x}}^{\bm{x}+\nu\bm{n}}\bm{\tau}\cdot\bm{f}\dd{n}}\delta_\varepsilon(\bm{x}-\bm{\xi})\dd{V}=0,
\end{align}
where $\delta_\varepsilon(\bm{x}) =\delta_\varepsilon (x)\delta_\varepsilon (y)$. The conventionally regularized force $\bm{f}=\int_ \Gamma \bm{F} (\bm{\xi} (s))\delta_\varepsilon (\bm{\xi} (s)-\bm{x})\dd{s} $ does not satisfy this condition. 

Our idea is to introduce the forcing shear stress tensor $M\delta_\varepsilon (\bm{\xi}-\bm{x}) (\bm{\tau n}+\bm{n\tau})$ around the points $\bm{\xi}$ on the boundary. The total body force imposed by this forcing shear stress is $\div{\bm{m}}$ with $\bm{m}=\int_\Gamma M(\bm{\xi}(s))\delta_\varepsilon(\bm{\xi}(s)-\bm{x})(\bm{\tau n}+\bm{n\tau})\dd{s}$. The regularized force is now given by
\begin{align}\label{eq:bf_FM}
	\bm{f}=\int_\Gamma \bm{F}(\bm{\xi}(s))\delta_\varepsilon(\bm{\xi}(s)-\bm{x})\dd{s}
				+\nabla\cdot\int_\Gamma M(\bm{\xi}(s))\delta_\varepsilon(\bm{\xi}(s)-\bm{x})(\bm{\tau n}+\bm{n\tau})\dd{s},
\end{align}
where $M$ is determined by $\bm{F}$ to satisfy Eq.~\eqref{eq:bf_cond}. Since $M\delta_\varepsilon (\bm{\xi}-\bm{x}) (\bm{\tau n}+\bm{n\tau})$ is a symmetric tensor and has non-zero elements only near the boundary,
it is shown that the conservations of force
\begin{align}
	\int_\Omega\bm{f}(\bm{x})\dd{V}=\int_\Gamma\bm{F}(\bm{\xi}(s))\dd{s}
	\label{eq:consv_force_2D}
\end{align}
and the torque
\begin{align}
	\int_\Omega\bm{x}\times\bm{f}(\bm{x})\dd{V}=\int_\Gamma\bm{\xi}(s)\times\bm{F}(\bm{\xi}(s))\dd{s}
	\label{eq:consv_torque_2D}
\end{align}
are always satisfied in the present method. The two-dimensional generalization of the Navier BC (Eq.~\ref{eq:NBC_1D_delta_approx} in one-dimension) is 
\begin{align}
	\bm{\tau}\cdot\int_{\Omega}\bm{u}(\bm{x})\delta_\varepsilon(\bm{x}-\bm{\xi})\dd{V}
	-\mathcal{L}_s\bm{\tau}\cdot\qty{\int_{\Omega}[\nabla\bm{u}+(\nabla\bm{u})^T](\bm{x})\delta_\varepsilon(\bm{x}-\bm{\xi})\dd{V}}
	\cdot\bm{n}
	&=\bm{\tau}\cdot\bm{U}, %
    \label{eq:NBC_delta_approx}\\
	\bm{n}\cdot\int_{\Omega}\bm{u}(\bm{x})\delta_\varepsilon(\bm{x}-\bm{\xi})\dd{V}
	&=\bm{n}\cdot\bm{U},
    \label{eq:PBC_delta_approx}
\end{align}
where the normal commponent of the velocity is additionally constrained to be impermeable.

\subsubsection {IBPM for the Navier BC in two-dimension}
By discretizing the formulations in \ref{S:2DC}, we derive the IB expressions for the Navier BC in two-dimension. As in one-dimension, we only describe the spacial discretization. The semi-discretized Navier-Stokes equation is written as
\begin{align}\label{eq:NS_spacedis}
	\pd{\bm{u}}{t}=\bm{A}-\hat{G}P+\frac{1}{\mathrm{Re}}\hat{L}\bm{u}+\bm{f}_\mathrm{ext}+\bm{f},
\end{align}
where $\bm{A} $ is the convection term. Remember the interpolated velocity $(\bm{u}_\Gamma) _l $ at the $l$th Lagrangian point $\bm{\xi}_l $ is given by Eq.~\eqref{eq:conv_u}, and the interpolated velocity gradient $[(\nabla\bm{u}) _\Gamma] _l$ is given by
\begin{align}\label{eq:gradu_l}
	[(\nabla\bm{u})_\Gamma]_l=
	\begin{bmatrix}
		\sum_i(\hat{\partial}_x u)_i\delta_h(\bm{x}^\mathcal{C}_i-\bm{\xi}_l)\Delta x\Delta y &
		\sum_i(\hat{\partial}_x v)_i\delta_h(\bm{x}^\mathcal{V}_i-\bm{\xi}_l)\Delta x\Delta y \\
		\sum_i(\hat{\partial}_y u)_i\delta_h(\bm{x}^\mathcal{V}_i-\bm{\xi}_l)\Delta x\Delta y &
		\sum_i(\hat{\partial}_y v)_i\delta_h(\bm{x}^\mathcal{C}_i-\bm{\xi}_l)\Delta x\Delta y
	\end{bmatrix},
\end{align}
where $\bm{x}^\mathcal{C}_i\in\mathcal{C} $ and $\bm{x}^\mathcal{V}_i\in\mathcal{V}$. The boundary conditions for the velocity (Eqs.~\ref{eq:NBC_delta_approx} and \ref{eq:PBC_delta_approx}) are discretized as
\begin{align}
	\bm{\tau}_l\cdot(\bm{u}_\Gamma)_l
	-\mathcal{L}_s(\bm{\tau}_l\bm{n}_l+\bm{n}_l\bm{\tau}_l):[(\nabla\bm{u})_\Gamma]_l
	&=\bm{\tau}_l\cdot\bm{U}_l
	\label{eq:NBC_l},\\
	\bm{n}_l\cdot(\bm{u}_\Gamma)_l
	&=\bm{n}_l\cdot\bm{U}_l
	\label{eq:PBC_l}.
\end{align}
Note that $\bm{\tau}\cdot [\nabla\bm{u}+ (\nabla\bm{u}) ^T]\cdot\bm{n}= (\bm{\tau n}+\bm{n\tau}):\nabla\bm{u}$. Compiling for all $l$, Eqs.~\eqref{eq:NBC_l} and \eqref{eq:PBC_l} can be expressed with the linear operators as
\begin{align}
	\hat{T}^T\hat{E}\bm{u}-\mathcal{L}_s(\hat{T}^T\hat{N}^T_2+\hat{N}^T\hat{T}^T_2)\hat{E}\hat{G}\bm{u}&=\hat{T}^T\bm{U}
	\label{eq:NBC_op},\\
	\hat{N}^T\hat{E}\bm{u}&=\hat{N}^T\bm{U}.
	\label{eq:PBC_op}
\end{align}
Here, the discrete velocity gradient $\hat{G}\bm{u}$ is a column vector, $\hat{G}\bm{u}=[(\hat{\partial}_x u\in\mathbb{R}^\mathcal{C})^T,(\hat{\partial}_x v\in\mathbb{R}^\mathcal{V})^T,(\hat{\partial}_y u\in\mathbb{R}^\mathcal{V})^T,(\hat{\partial}_y v\in\mathbb{R}^\mathcal{C})^T]^T$ and its interpolation to the Lagrangian points on the boundary is written as $\hat{E}\hat{G}\bm{u}=[(\hat{E}\hat{\partial}_x u\in\mathbb{R}^{\Gamma})^T, (\hat{E}\hat{\partial}_x v\in\mathbb{R}^{\Gamma})^T, (\hat{E}\hat{\partial}_y u\in\mathbb{R}^{\Gamma})^T, (\hat{E}\hat{\partial}_y v\in\mathbb{R}^{\Gamma})^T]^T$. $\hat{T}$ is the operator to extract the tangential component: 
\begin{equation}
\hat{T}=%
\begin{bmatrix}
    \hat{T}_x\\
    \hat{T}_y
\end{bmatrix}
\end{equation}
with $\hat{T}_x=\mathrm{diag}(\tau_{x,1},\tau_{x,2},\cdots,\tau_{x,N_\Gamma})$ and $\hat{T}_y=\mathrm{diag}(\tau_{y, 1},\tau_{y, 2},\cdots,\tau_{y, N_\Gamma})$. $\hat{T}_2 $ is defined by
\begin{align}\label{eq:T2}
	\hat{T}_2
	=&
	\begin{bmatrix}
		\hat{T} & \\
		& \hat{T}
	\end{bmatrix}.
\end{align}
For the normal component, $\hat{N}$ and $\hat{N}_2$ are defined in a similar fashion. Moreover, for the simplicity of notation, writing 
$\hat{E}_{\tau n}=[(\hat{T}^T\hat{E})^T \ \ (\hat{N}^T\hat{E})^T]^T$,
$\hat{G}_{\tau n}=(\hat{T}^T\hat{N}^T_2+\hat{N}^T\hat{T}^T_2)\hat{E}\hat{G}$, and $\hat{\mathcal{L}}_s= [\mathcal{L}_s\hat {I}\ \ 0] ^T$, the velocity boundary condition (Eqs.~\ref{eq:NBC_op} and \ref{eq:PBC_op}) can be finally expressed as
\begin{align}\label{eq:NBC+PBC_op}
	(\hat{E}_{\tau n}-\hat{\mathcal{L}_s}\hat{G}_{\tau n})\bm{u}=\bm{U}_{\tau n},
\end{align}
where $\bm{U}_{\tau n}=[(\hat{T}^T\bm{U}) ^T, (\hat{N}^T\bm{U}) ^T]^T$, which is the $\tau$-$n $ coordinate representation of $\bm{U}$.

Next, we want to descretize Eq.~\eqref{eq:st_dndt_int} to derive the condition (Eq.~\ref{eq:bf_cond}) in the discretized form. In the discretized form, the LHS of Eq.~\eqref{eq:st_dndt_int} consists of the terms either defined on $\mathcal{C}$ to give $\sigma^\mathcal{C}_n \equiv 2\tau_x n_x (\hat{\partial}_x u) +2\tau_y n_y (\hat{\partial}_y v)$ or on $\mathcal{V}$ to give $\sigma^\mathcal{V}_n \equiv (\tau_x n_y+n_x\tau_y) (\hat{\partial}_x v+\hat{\partial}_y u)$\label{pp:decomp}, where $\bm{\tau} $ and $\bm{n}$ are constant vectors specified later in Eq.~\eqref{eq:bf_cond_l_op}. They are interpolated to the Lagrangian points on the boundary to form the shear stress term in Eq.~\eqref{eq:NBC_l}. Since $\sigma^\mathcal{C}_n$ and $\sigma^\mathcal{V}_n$ are defined on $\mathcal{C}$ and $\mathcal{V}$ respectively, so should be the origins of the integration intervals that appear on the RHS of Eq.~\eqref{eq:st_dndt_int}. Therefore the two sets of the discretization are given for Eq.~\eqref{eq:st_dndt_int} by the following: 
\begin{align}
	\sigma_n^\mathcal{C}|_{\bm{x}^\mathcal{C}}
    &=\sigma_n^\mathcal{C}|_{\bm{x}^\mathcal{C}+\nu\bm{n}}\notag\\
	&+\mathrm{Re}\hat{J}^\mathcal{C}\hat{\tau}^\mathcal{C}
	\qty[-\pd{\bm{u}}{t}+\bm{A}-\hat{G}P+\bm{f}_\mathrm{ext}]\notag\\
	&+\mathrm{Re}\hat{J}^\mathcal{C}\hat{\tau}^\mathcal{C}\bm{f}\notag\\
	&+\hat{J}^\mathcal{C}\hat{\partial}_{\tau}\sigma^\mathcal{V}_{\tau}
	\label{eq:st_dndt_int_C_op},\\
	\sigma_n^\mathcal{V}|_{\bm{x}^\mathcal{V}}
	&=\sigma_n^\mathcal{V}|_{\bm{x}^\mathcal{V}+\nu\bm{n}}\notag\\
	&+\mathrm{Re}\hat{J}^\mathcal{V}\hat{\tau}^\mathcal{V}
	\qty[-\pd{\bm{u}}{t}+\bm{A}-\hat{G}P+\bm{f}_\mathrm{ext}]\notag\\
	&+\mathrm{Re}\hat{J}^\mathcal{V}\hat{\tau}^\mathcal{V}\bm{f}\notag\\
	&+\hat{J}^\mathcal{V}\hat{\partial}_{\tau}\sigma^\mathcal{C}_{\tau}
    \label{eq:st_dndt_int_V_op},
\end{align}
where the linear operator performing the line integral from $\bm{x}^\mathcal{C}\in\mathcal{C}$ and $\bm{x}^\mathcal{V}\in\mathcal{V}$ are written as $\hat{J}^\mathcal{C}$ and $\hat{J}^\mathcal{V}$ respectively, and the operators performing coordinate transformation $\hat{\tau}^\mathcal{C}$ and $\hat{\tau}^\mathcal{V}$ are defined by Eqs.~\eqref{eq:st_Lu_C} and \eqref{eq:st_Lu_V} respectively. Complete derivation of these equations are shown in \ref{app:derivation}. Summing the interpolation of the LHS of Eqs.~\eqref{eq:st_dndt_int_C_op} and \eqref{eq:st_dndt_int_V_op} to a Lagrangian point $\bm{\xi}_l$ on the boundary, 
\begin{equation}\label{eq:interpolation}
\hat{E}_l\sigma_n^\mathcal{C}|_{\bm{x}^\mathcal{C}}+\hat{E}_l\sigma_n^\mathcal{V}|_{\bm{x}^\mathcal{V}}, 
\end{equation}
gives the shear stress term in Eq.~\eqref{eq:NBC_l}. The consistency condition (Eq.~\ref{eq:bf_cond}) requires that the boundary force term vanishes in Eq.~\eqref{eq:interpolation}, and therefore is written in the discretized form as
\begin{align}
(\hat{E}_l\hat{J}^\mathcal{C}_l\hat{\tau}^\mathcal{C}_l+\hat{E}_l\hat{J}^\mathcal{V}_l\hat{\tau}^\mathcal{V}_l)\bm{f}=0
\label{eq:bf_cond_l_op}
\end{align}
for each Lagrangian point $\bm{\xi}_l$: for $\bm{\tau}$ and $\bm{n}$ the values on $\bm{\xi}_l$ are employed for all the line integration paths involved in the interpolation. Eq.~\eqref{eq:bf_cond_l_op} must hold for all $l$. Therefore, it is convenient to summarize Eq.~\eqref{eq:bf_cond_l_op} for all $l$ as
\begin{align}
	\hat{J}\bm{f}:=
	\begin{bmatrix}
		\hat{E}_1\hat{J}^\mathcal{C}_1\hat{\tau}^\mathcal{C}_1+
		\hat{E}_1\hat{J}^\mathcal{V}_1\hat{\tau}^\mathcal{V}_1\\
		\vdots\\
		\hat{E}_{N_\Gamma}\hat{J}^\mathcal{C}_{N_\Gamma}\hat{\tau}^\mathcal{C}_{N_\Gamma}+
		\hat{E}_{N_\Gamma}\hat{J}^\mathcal{V}_{N_\Gamma}\hat{\tau}^\mathcal{V}_{N_\Gamma}
	\end{bmatrix}\bm{f}=0
	\label{eq:bf_cond_op}.
\end{align}

The conventional $\bm{f}=\hat{H}\bm{F}$ does not satisfy this consistency condition \eqref{eq:bf_cond_op}. The discretization of the first term of our fomulation (Eq.~\ref{eq:bf_FM}) gives $\hat{H}\bm{F}$, which is identical to the conventional method. For the second term, $\nabla\cdot\bm{m}=[(\partial m_{xx}/\partial x) + (\partial m_{yx}/\partial y), (\partial m_{xy}/\partial x) + (\partial m_{yy}/\partial y)]^T $ is the quantity defined on $\mathcal{F}$ exactly as $\bm{f}$: $\hat{\partial}_x m_{xx}+\hat{\partial}_y m_{yx}\in\mathbb{R}^{\mathcal{F}_x}$,
$\hat{\partial}_x m_{xy}+\hat{\partial}_y m_{yy}\in\mathbb{R}^{\mathcal{F}_y}$,
Therefore, $m_{xx}, m_{yy}\in\mathbb{R}^\mathcal{C} $ and $m_{xy}=m_{yx}\in\mathbb{R}^\mathcal{V}$, and in the discretized form written as $(m_{xx})_i=\sum_l 2\tau_{x,l}n_{x,l}M_l\delta_h(\bm{\xi}_l-\bm{x}^\mathcal{C}_i)\Delta s$, $(m_{yy})_i=\sum_l 2\tau_{y,l}n_{y,l}M_l\delta_h(\bm{\xi}_l-\bm{x}^\mathcal{C}_i)\Delta s$, and 
$(m_{xy})_i=(m_{yx})_i=\sum_l (\tau_{x,l}n_{y,l}+n_{x,l}\tau_{y,l})M_l\delta_h(\bm{\xi}_l-\bm{x}^\mathcal{V}_i)\Delta s$. Similarly as $\hat{G}\bm{u}$, $\bm{m}$ is written as a column vector $\bm{m}=[m_{xx}^T, m_{xy}^T, m_{yx}^T, m_{yy}^T]^T$ and expressed with the linear operators as
\begin{align}
	\bm{m}&=
	\begin{bmatrix}
		\hat{H}(\hat{T}_x\hat{N}_x+\hat{N}_x\hat{T}_x)M\\
		\hat{H}(\hat{T}_x\hat{N}_y+\hat{N}_x\hat{T}_y)M\\
		\hat{H}(\hat{T}_y\hat{N}_x+\hat{N}_y\hat{T}_x)M\\
		\hat{H}(\hat{T}_y\hat{N}_y+\hat{N}_y\hat{T}_y)M
	\end{bmatrix}\notag\\
	&=\hat{H}(\hat{T}_2\hat{N}+\hat{N}_2\hat{T})M,
	\label{eq:m_op}
\end{align}
and $\div{\bm{m}}$ is written as $\hat{D}\hat{H}(\hat{T}_2\hat{N}+\hat{N}_2\hat{T})M$. The discretized form of the regularized force (Eq.~\ref{eq:bf_FM}) is now
\begin{align}
	\bm{f}=\hat{H}\bm{F}+\hat{D}\hat{H}(\hat{T}_2\hat{N}+\hat{N}_2\hat{T})M
	\label{eq:bf_FM_op},
\end{align}
where $M$ is determined to satisfy the condition \eqref{eq:bf_cond_op} and
\begin{align}
	M=-[\hat{J}\hat{D}\hat{H}(\hat{T}_2\hat{N}+\hat{N}_2\hat{T})]^{-1}(\hat{J}\hat{H})\bm{F}=:\hat{K}\bm{F}.
	\label{eq:M_F}
\end{align}
By substituting Eq.~\eqref{eq:M_F} in Eq.~\eqref{eq:bf_FM_op}, we obtain the final form of our regularization operator of the boundary force:
\begin{align}
	\bm{f}=[\hat{H}+\hat{D}\hat{H}(\hat{T}_2\hat{N}+\hat{N}_2\hat{T})\hat{K}]\bm{F},
	\label{eq:bf_F_op}
\end{align}
which is consistent with the Navier BC (Eqs.~\ref{eq:NBC_op} and \ref{eq:PBC_op}) by construction. 

As in the one-dimensional case, the regularization of the boundary force should satisfy the force conservation
\begin{subequations}\label{eq:consv_force_dis}
\begin{align}
	\sum_i f_{x,i}\Delta x\Delta y&=\sum_l F_{x,l}\Delta s
	\label{eq:consv_force_dis_x},\\
	\sum_i f_{y,i}\Delta x\Delta y&=\sum_l F_{y,l}\Delta s
	\label{eq:consv_force_dis_y}
\end{align}
\end{subequations}
and the torque conservation
\begin{align}
	\sum_i x^{\mathcal{F}_y}_i f_{y,i}\Delta x\Delta y-\sum_i y^{\mathcal{F}_x}_i f_{x,i}\Delta x\Delta y
	=\sum_l(\xi_lF_{y,l}-\eta_lF_{x,l})\Delta s.
	\label{eq:consv_torque_dis}
\end{align}
As shown in \ref{app:consv}, Eq.~\eqref{eq:bf_FM_op} satisfies these conditions regardless of the value of $M$. We also want $\bm{f} $ to maintain the translational invariance. Ideally, the translational invariance that $\hat{E}\bm{f} $ does not depend on the positional relation between Eulerian mesh $\mathcal{M}$ and Lagrangian points $\Gamma$ is desirable. However, it is a strict condition, and in the conventional regularization $\bm{f}=\hat{H}\bm{F} $ with $\delta_h$ (Eq.~\ref{eq:disdelta}), the weaker condition is imposed that the regularized boundary force $\bm{f}_l$ (see also Eq.~\ref{eq:f_l}) interpolated back on the Lagrangian point $\bm{\xi}_l$, that is $\hat{E}_l\bm{f}_l$, is independent of the positional relation between $\mathcal{M}$ and $\Gamma$. As shown in \ref{app:consv}, $\bm{f}$ given by Eq.~\eqref{eq:bf_F_op} does not strictly satisfy this translational invariance in the weak sense, but its violation is not significant (the error is smaller than 20 \%).

\subsection{Calculation procedure}
\label{S:procedure}
Finally, the calculation procedure of the present method is shown. The discretized expressions of the Navier BC and the impermeability condition with linear operators are given by Eq.~\eqref{eq:NBC+PBC_op}, and the corresponding regulalized force $\bm{f} $ is given by Eq.~\eqref{eq:bf_F_op}. Thus, the discretized governing equations including the discretization in time are written as
\begin{align}
	\frac{\bm{u}^{n+1}-\bm{u}^n}{\Delta t}
	&=
	\frac{1}{2}\left(3\bm{A}^n-\bm{A}^{n-1}\right)
	-\hat{G}P
	+\frac{1}{2\mathrm{Re}}\left(\hat{L}\bm{u}^{n+1}+\hat{L}\bm{u}^n\right) \notag\\
	&+\bm{f}_{\mathrm{ext}}^{n+1}
	+[\hat{H}+\hat{D}\hat{H}(\hat{T}_2\hat{N}+\hat{N}_2\hat{T})\hat{K}]\bm{F}
	+bc_1
	\label{eq:Dis_NS_IB_NBC},
	\\
	\hat{D}\bm{u}^{n+1}&=bc_2
	\label{eq:Dis_continuity_IB_NBC},
	\\
	(\hat{E}_{\tau n}-\hat{\mathcal{L}_s}\hat{G}_{\tau n})\bm{u}^{n+1} &= \bm{U}_{\tau n}^{n+1},
	\label{eq:Dis_BC_IB_NBC}
\end{align}
where for the Navier-Stokes equation the second-order Adams-Bashforth scheme and the Crank-Nicolson scheme are adopted for the advection term and the viscous term, respectively. Similarly as ~\eqref{eq:Mtrx_GE}, Eqs.~\eqref{eq:Dis_NS_IB_NBC}--\eqref{eq:Dis_BC_IB_NBC} can be expressed in the matrix form as
\begin{align}
	\begin{bmatrix}
		\hat{R} & \Delta t \hat{G} & -\Delta t [\hat{H}+\hat{D}\hat{H}(\hat{T}_2\hat{N}+\hat{N}_2\hat{T})\hat{K}] \\
		\hat{D} & 0 & 0 \\
		\hat{E}_{\tau n}-\hat{\mathcal{L}_s}\hat{G}_{\tau n} & 0 & 0 \\
	\end{bmatrix}
	\begin{bmatrix}
		\bm{u}^{n+1} \\
		P \\
		\bm{F} \\
	\end{bmatrix}
	=
	\begin{bmatrix}
		\bm{r}_{NS} \\
		0 \\
		\bm{U}_{\tau n}^{n+1} \\
	\end{bmatrix}
	+
	\begin{bmatrix}
		bc_1 \\
		bc_2 \\
		0	\\
	\end{bmatrix}.
	\label{eq:Mtrx_GE_NBC}
\end{align}
Furthermore, the deformation of the equation is carried out considering the symmetry. First, $\bm{F} $ is displayed in the $\tau$-$n$ coordinate according to $\bm{U}_{\tau n} $ as $\bm{F}_{\tau n}=[(\hat{T}^T\bm{F}) ^T, (\hat{N}^T\bm{F}) ^T]^T$, and then by introducing $\hat{H}_{\tau n}=\hat{H} [\hat {T}\ \ \hat{N}] $ and $\hat{K}_{\tau n}=\hat{K} [\hat {T}\ \ \hat{N}] $,
$\bm{f}=(\hat{H}_{\tau n}+\hat{D}_{\tau n}\hat{K}_{\tau n})\bm{F}_{\tau n}$.
For the notational simplicity, we set $\hat{D}_{\tau n}=\hat{D}\hat{H} (\hat{T}_2\hat{N}+\hat{N}_2\hat{T}) $. In addition, as $\hat{H}_{\tau n}=\Delta s (\Delta x\Delta y) ^{-1}\hat{E}_{\tau n}^T$ and
$\hat{D}_{\tau n}=-\Delta s (\Delta x\Delta y) ^{-1}\hat{G}_{\tau n}^T$, by writing 
$\tilde{P}=-\Delta t P$, $\tilde{\bm{F}}_{\tau n}=-\Delta t\Delta s (\Delta x\Delta y) ^{-1}\bm{F}_{\tau n}$,
Eq.~\eqref{eq:Mtrx_GE_NBC} becomes
\begin{align}
	\begin{bmatrix}
		\hat{R} & \hat{D}^T & \hat{E}_{\tau n}^T-(\hat{K}_{\tau n}^T\hat{G}_{\tau n})^T \\
		\hat{D} & 0 & 0 \\
		\hat{E}_{\tau n}-\hat{\mathcal{L}}_s\hat{G}_{\tau n} & 0 & 0 \\
	\end{bmatrix}
	\begin{bmatrix}
		\bm{u}^{n+1} \\
		\tilde{P} \\
		\tilde{\bm{F}}_{\tau n} \\
	\end{bmatrix}
	=
	\begin{bmatrix}
		\bm{r}_{NS} \\
		0 \\
		\bm{U}_{\tau n}^{n+1} \\
	\end{bmatrix}
	+
	\begin{bmatrix}
		bc_1 \\
		bc_2 \\
		0	\\
	\end{bmatrix}
	\label{eq:Mtrx_GE_sym_NBC}.
\end{align}
Since $\hat{K}_{\tau n}^T\ne\hat{\mathcal{L}}_s $ in general, the coefficient matrix is not symmetric unlike Eq.~\eqref{eq:Mtrx_GE_sym}. The coefficient matrix of Eq.~\eqref{eq:Mtrx_GE_sym_NBC} can be LU-decomposed as follows:
\begin{align}
	\begin{bmatrix}
		\hat{R} & \hat{Q} \\
		\hat{W} & 0 \\
	\end{bmatrix}
	=
	\begin{bmatrix}
		\hat{R} & 0 \\
		\hat{W} & -\hat{W}\hat{R}^{-1}\hat{Q}^T \\
	\end{bmatrix}
	\begin{bmatrix}
		\hat{I} & \hat{R}^{-1}\hat{Q} \\
		0 & \hat{I} \\
	\end{bmatrix}
	\label{eq:LU_NBC}.
\end{align}
Here we write $\hat{W}=[\hat{D}^T \ \ (\hat{E}_{\tau n}-\hat{\mathcal{L}}_s\hat{G}_{\tau n})^T]^T$ and $\hat{Q}=[\hat{D}^T \ \ \hat{E}_{\tau n}^T-(\hat{K}_{\tau n}^T\hat{G}_{\tau n})^T]$. Truncating Eq.~\eqref{eq:R_inv} at the $N$th term, Eq.~\eqref{eq:Mtrx_GE_sym_NBC} becomes
\begin{align}
	\begin{bmatrix}
		\hat{R} & 0 \\
		\hat{W} & -\hat{W}\hat{C}^{N}\hat{Q} \\
	\end{bmatrix}
	\begin{bmatrix}
		\hat{I} & \hat{C}^{N}\hat{Q} \\
		0 & \hat{I} \\
	\end{bmatrix}
	\begin{bmatrix}
		\bm{u}^{n+1} \\
		\lambda \\
	\end{bmatrix}
	=
	\begin{bmatrix}
		\bm{r}_1 \\
		\bm{r}_2 \\
	\end{bmatrix}
	+
	\begin{bmatrix}
		-\left(\frac{\Delta t}{2\mathrm{Re}}\right)^N(\hat{L})^N\hat{Q}\lambda \\
		0\\
	\end{bmatrix},
	\label{eq:Mtrx_GE_LU_NBC}
\end{align}
where $\lambda=[\tilde{P}^T,\tilde{\bm{F}}_{\tau n}^T]^T$,
$\bm{r}_1=\bm{r}_{NS}+bc_1$, $\bm{r}_2=[bc_2 \ \ \bm{U}_{\tau n}^{n+1}]^T$, and the second term on the RHS is the truncation error. 

There is another step forward to reach our final solution algorithm. Since $\hat{R}^{-1}$ is approximated by $\hat{C}^N$, the regularized body force is effectively 
$\bm{f}'=\bm{f}-(\Delta t/2\mathrm{Re})^N(\hat{L})^N\bm{f}$ and $\hat{J}\bm{f}'=-(\Delta t/2\mathrm{Re}) ^N\hat{J} (\hat{L}) ^N\bm{f}$. The consistency condition \eqref{eq:bf_cond_op} is not completely satisfied with the error $\order{(\Delta t) ^N}$. Since this error is ultimately related to the accuracy of the predicted slip velocity, it is important to reduce it. In order to improve the accuracy, we apply the delta form to the boundary force: $\bm{F}_{\tau n}=\bm{F}_{\tau n}^n + \delta\bm{F}_{\tau n}$. Here, $\bm{F}_{\tau n}^n $ is the boundary force obtained in the previous step. Correspondingly, $\bm{f}=\bm{f}^n+\delta\bm{f}$. If we also apply the delta form to the pressure $P$, as $\lambda=\lambda^n+\delta\lambda$,  Eq.~\eqref{eq:Mtrx_GE_LU_NBC} is written as
\begin{align}
	\begin{bmatrix}
		\hat{R} & 0 \\
		\hat{W} & -\hat{W}\hat{C}^{N}\hat{Q} \\
	\end{bmatrix}
	\begin{bmatrix}
		\hat{I} & \hat{C}^{N}\hat{Q} \\
		0 & \hat{I} \\
	\end{bmatrix}
	\begin{bmatrix}
		\bm{u}^{n+1} \\
		\delta\lambda \\
	\end{bmatrix}
	=
	\begin{bmatrix}
		\bm{r}_1-\hat{Q}\lambda^n \\
		\bm{r}_2 \\
	\end{bmatrix}
	+
	\begin{bmatrix}
		-\left(\frac{\Delta t}{2\mathrm{Re}}\right)^N(\hat{L})^N\hat{Q}\delta\lambda \\
		0\\
	\end{bmatrix}
	\label{eq:Mtrx_GE_LU_NBC_delta}.
\end{align}
In this case, $\bm{f}'=\bm{f}-(\Delta t/2\mathrm{Re})^N(\hat{L})^N\delta\bm{f}$ and $\hat{J}\bm{f}'=-(\Delta t/2\mathrm{Re})^N\hat{J}(\hat{L})^N\delta\bm{f}$. Since $\delta\bm{F}_{\tau n}=\order{\Delta t}$ and $\delta\bm{f}=\order{\Delta t}$, the error in the consistency condition \eqref{eq:bf_cond_op} becomes $\order{(\Delta t) ^{N+1}}$, improving the accuracy by one order. 

Thus, with the last term on the RHS of Eq.~\eqref{eq:Mtrx_GE_LU_NBC_delta} as the truncation error, the final form of our solution procedure for the velocity $\bm{u}^{n+1}$ and the constraint (the pressure and the boundary force) $\lambda$ is 
\begin{align}
	\hat{R}\bm{u}^F&=\bm{r}_1-\hat{Q}\lambda^n
	\label{eq:calc_proc_NBC_1},\\
	\hat{W}\hat{C}^N\hat{Q}\delta\lambda&=\hat{W}\bm{u}^F-\bm{r}_2
	\label{eq:calc_proc_NBC_2},\\
	\bm{u}^{n+1}&=\bm{u}^F-\hat{C}^N\hat{Q}\delta\lambda
	\label{eq:calc_proc_NBC_3},\\
	\lambda&=\lambda^n+\delta\lambda
	\label{eq:calc_proc_NBC_4}.
\end{align}
Since $\hat{Q}\ne\hat{W}^T$ ($\hat{K}_{\tau n}^T\ne\hat{\mathcal{L}}_s$ in general), Eq.~\eqref{eq:calc_proc_NBC_2} has a non-symmetric coefficient matrix $\hat{W}\hat{C}^N\hat{Q}$ unlike Eq.~\eqref{eq:calc_proc_noslip_2} for the no-slip boundary. For the moving boundary problems, the operators $\hat{W}$ and $\hat{Q}$ are updated every time step. 

\section{Results}
\label{S:results}
The present method is validated on various benchmark problems for 1 and 2 dimensional systems. In the following calculations, the Eulerian mesh is equally spaced with $\Delta x=\Delta y$ in the whole domain or only near the boundary $\Gamma$. The approximation of $\hat{R}^{-1}$ is truncated at $N=3$.
\subsection{One dimensional problem}\label{S:1D}
\begin{figure}[t]
	\vspace{2mm}
  \begin{center}
    \begin{overpic}[bb = 0 0 679 186,width=6.3in,clip,keepaspectratio]
        {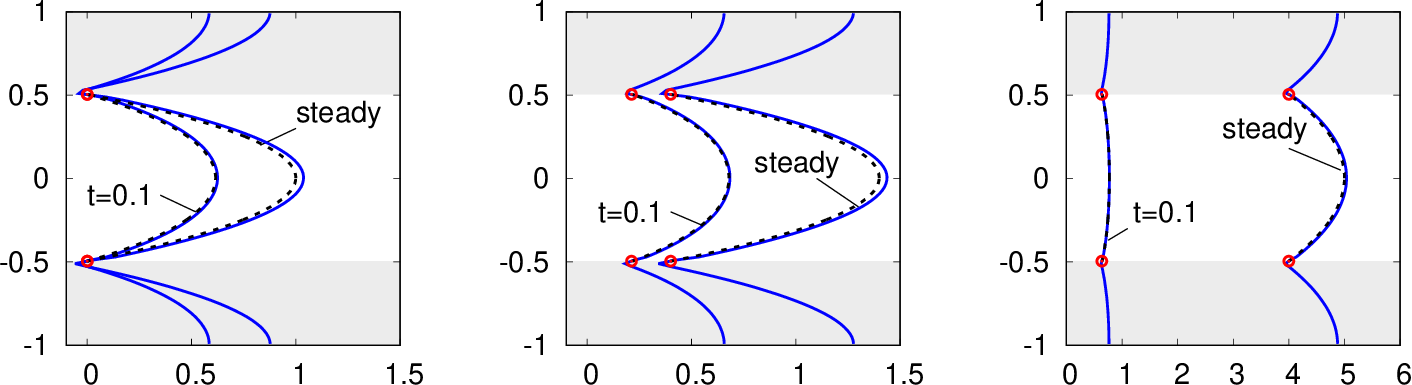}
        \put(16.5,-2){$u$}
				\put(52.5,-2){$u$}
				\put(87,-2){$u$}
				\put(-1.5,14){\rotatebox{90}{$y$}}
				\put(33.5,14){\rotatebox{90}{$y$}}
				\put(69,14){\rotatebox{90}{$y$}}
				\put(-1,27){(a)}
				\put(34,27){(b)}
				\put(69.5,27){(c)}
    \end{overpic}
    \caption{
			The velocity profiles of the Poiseuille flow at $t=0.1$ and in the steady state for (a) $\mathcal{L}_s=0$, (b) $\mathcal{L}_s=0.1$ and (c) $\mathcal{L}_s=1$.
			The shaded regions are the walls. The solid and dashed lines represent the present results and the analytical solution, respectively. The interpolated velocities to the walls using $\hat{E}^{\mathcal{F}_x}$ are represented by $\circ$.
		}
   \label{fig:vel1D_poi}
  \end{center}
\end{figure}
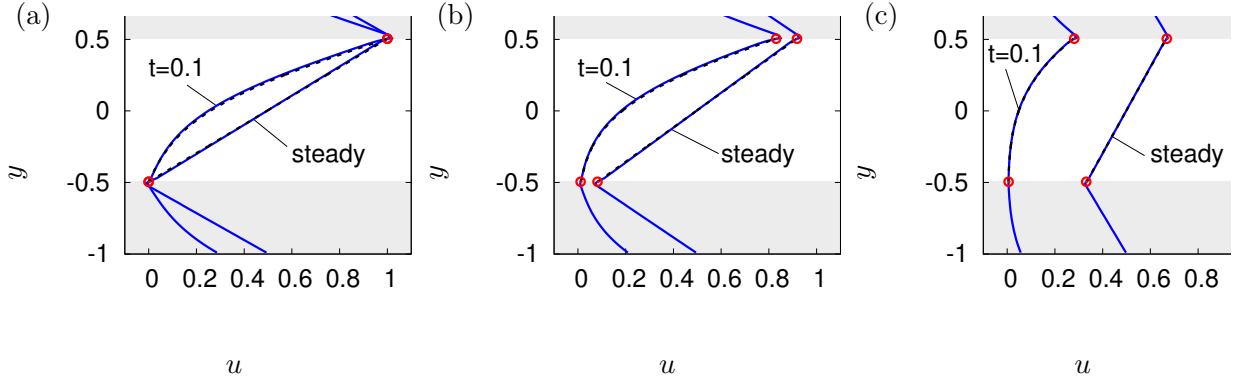
\begin{figure}[t]
  \begin{center}
    \begin{overpic}[bb = 0 0 673 186,width=6.3in,clip,keepaspectratio]
        {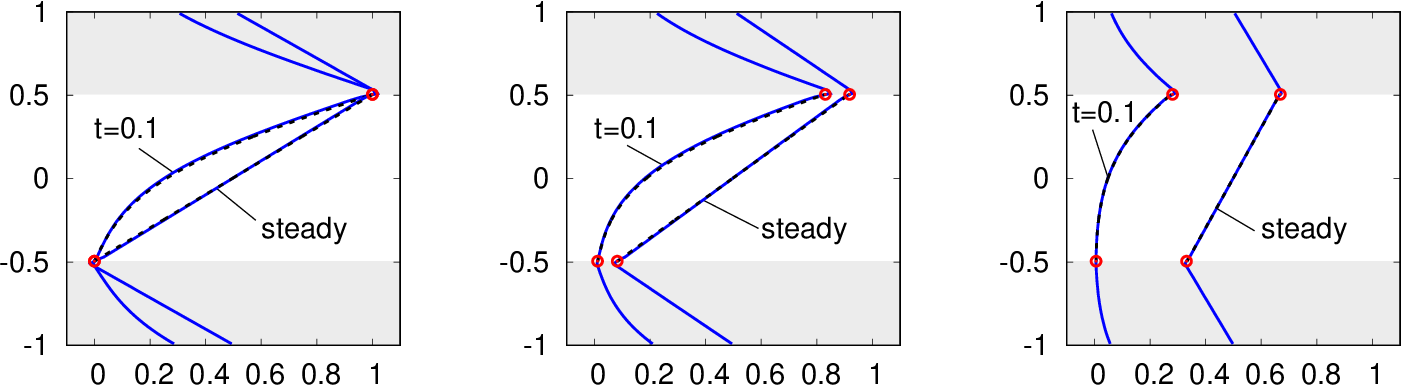}
        \put(16.5,-2){$u$}
				\put(52.5,-2){$u$}
				\put(87,-2){$u$}
				\put(-1.5,14){\rotatebox{90}{$y$}}
				\put(33.5,14){\rotatebox{90}{$y$}}
				\put(69,14){\rotatebox{90}{$y$}}
				\put(-1,27){(a)}
				\put(34,27){(b)}
				\put(69.5,27){(c)}
    \end{overpic}
    \caption{
			The velocity profiles of the Couette flow at $t=0.1$ and in the steady state for (a) $\mathcal{L}_s=0$, (b) $\mathcal{L}_s=0.1$ and (c) $\mathcal{L}_s=1$.
			The shaded regions are in the walls.
			The solid and dashed lines represent the present results and the analytical solution, respectively.
			The interpolated velocities to the walls using $\hat{E}^{\mathcal{F}_x}$ are represented by $\circ$.
		}
   \label{fig:vel1D_ctt}
  \end{center}
\end{figure}
First, we calculate Poiseuille and Couette flows. We set the flow direction in $x$ direction, and the computational domain is considered to be a periodic domain of $[-1,1]$ in $y$ direction. The wall is located at $\eta_1=-0.5+\varepsilon_y\Delta y$ and $\eta_2=\eta_1+H$, where the Navier BC is imposed. The channel width is $H=1$. Here, $\varepsilon_y\in [-0.5,0.5]$ is a constant parameter to adjust the relative position between the Euler grid and the Lagrange points. The velocity $u_j$ is defined at $y^{\mathcal{F}_x}_j= (j-1/2)\Delta y$. We choose a low Reynolds number, $\mathrm{Re}=1$, since we are interested in the microscopic flows where the velocity slip on the wall is important. The Poiseuille flow is driven by applying an external force (pressure gradient) of $f_\mathrm{ext}=8$ in the whole domain in the direction parallel with the wall. The Couette flow is driven from a stationary state by moving the wall at $\eta_2$ with a constant velocity $U=1$. 

In order to impose the Navier BCs at $\eta_1$ and $\eta_2$, the boundary forces $F_1$ and $F_2$ and the boundary shear stresses $M_1$ and $M_2$ are introduced. From Eqs.~\eqref{eq:Fd+Md'_dis} and \eqref{eq:Fd+Md'_dis_2}, they are distributed on the Eulerian mesh by the following equations.
\begin{align}
	f_j&=F_1[\delta_h(\eta_1-y^{\mathcal{F}_x}_j)+a_1(-\delta_h(\eta_1-y^\mathcal{V}_{j-1})+\delta_h(\eta_1-y^\mathcal{V}_j))]\notag\\
		 &+F_2[\delta_h(\eta_2-y^{\mathcal{F}_x}_j)-a_2(-\delta_h(\eta_2-y^\mathcal{V}_{j-1})+\delta_h(\eta_2-y^\mathcal{V}_j))]
	\label{eq:f_poictt}.
\end{align}
where $a_1=M_1/(F_1\Delta y) $and $a_2=M_2/(F_2\Delta y) $. Since $f$ must satisfy Eq.~\eqref{eq:cond_f_delta_dis} at $\eta_1$ and Eq.~\eqref{eq:cond_f_delta_dis_2} at $\eta _2$, $a_1$ and $a_2$ are determinde as follows:
\begin{align}
	a_1=2\sum_{y^{\mathcal{V}}_j\in \Omega_\delta}
	\sum_{j'=j+1}^{J}\delta_h(\eta_1-y^{\mathcal{F}_x}_{j'})
	\delta_h(y^{\mathcal{V}}_j-\eta_1)(\Delta y)^2
	\label{eq:a1},\\
	a_2=2\sum_{y^{\mathcal{V}}_j\in \Omega_\delta}
	\sum_{j'=J+1}^{j}\delta_h(\eta_2-y^{\mathcal{F}_x}_{j'})
	\delta_h(y^{\mathcal{V}}_j-\eta_2)(\Delta y)^2
	\label{eq:a2}.
\end{align}

The calculation results with $\Delta y=2\times10^{-2}$, $\Delta t=1\times10^{-6}$ and $\varepsilon_y=0.25$ are shown in Figs.~\ref{fig:vel1D_poi} and \ref{fig:vel1D_ctt}. Both for the Poiseuille and Couette flows, three slip length cases are calculated: $\mathcal{L}_s=0$, 0.1 and 1. 
These figures show good agreement of the present results with the analytical solutions in both steady and unsteady states, not only in the case without slip but also in the case with slip. However, in the steady state, the deviation from the analytical solution is larger for the Poiseuille flow than for the Couette flow. The reason for this is as follows. As described in \ref{app:PoiCtt}, the model error of the present method is proportional to the boundary force $F$. As can be seen from Eq.~\eqref{eq:GE_1D_int}, the boundary force increases linearly with the difference between the velocity gradients on both sides of $\Gamma$. Therefore, in the Poiseuille flow predictions with larger differences between the velocity gradients on $\Gamma$ the error becomes larger. 

\begin{figure}[t]
  \begin{center}
    \begin{overpic}[bb = 0 0 679 289,width=6.3in,clip,keepaspectratio]
        {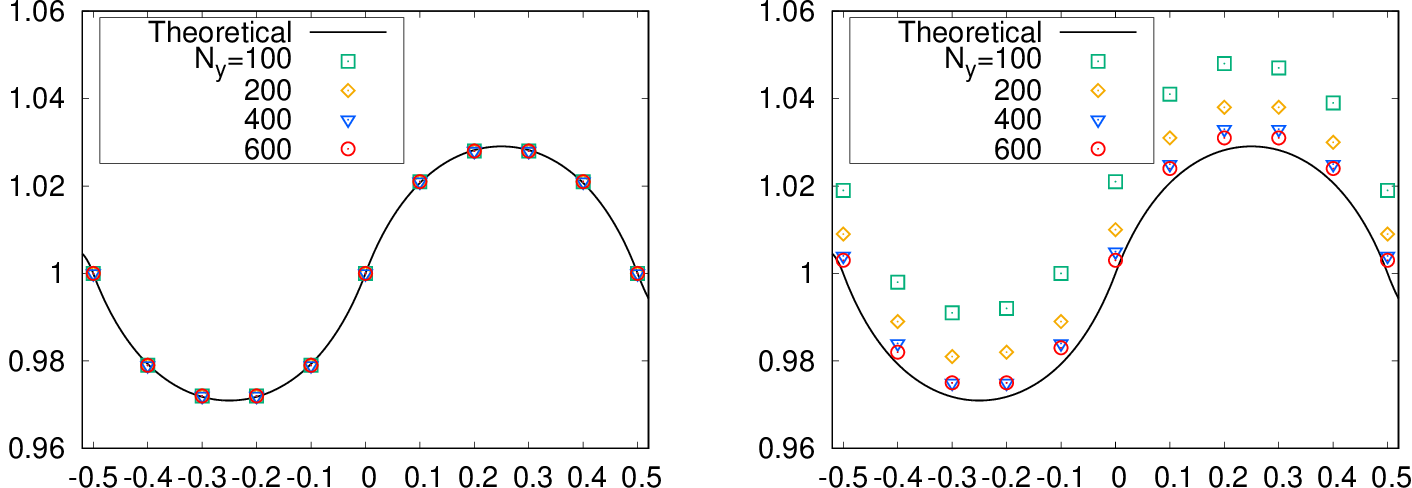}
        \put(-2,15){\rotatebox{90}{$a_1$}}
        \put(52,15){\rotatebox{90}{$a_1$}}
        \put(24,-1){$\varepsilon_y$}
        \put(77,-1){$\varepsilon_y$}
        \put(0,37){(a)}
        \put(53,37){(b)}
    \end{overpic}
    \caption{Nondimensionalized forcing shear stress $a_1$ determined by the numerical optimization (symbols) and by our method (solid lines, Eq.~\ref{eq:a1}) for (a) the Poiseuille flow and (b) the Couette flow. Note that $a_1$ obtained by Eq.~\eqref{eq:a1} is independent of the mesh resolution.}
   \label{fig:a_vs_y0_Ls1_10-1_t5}
  \end{center}
\end{figure}
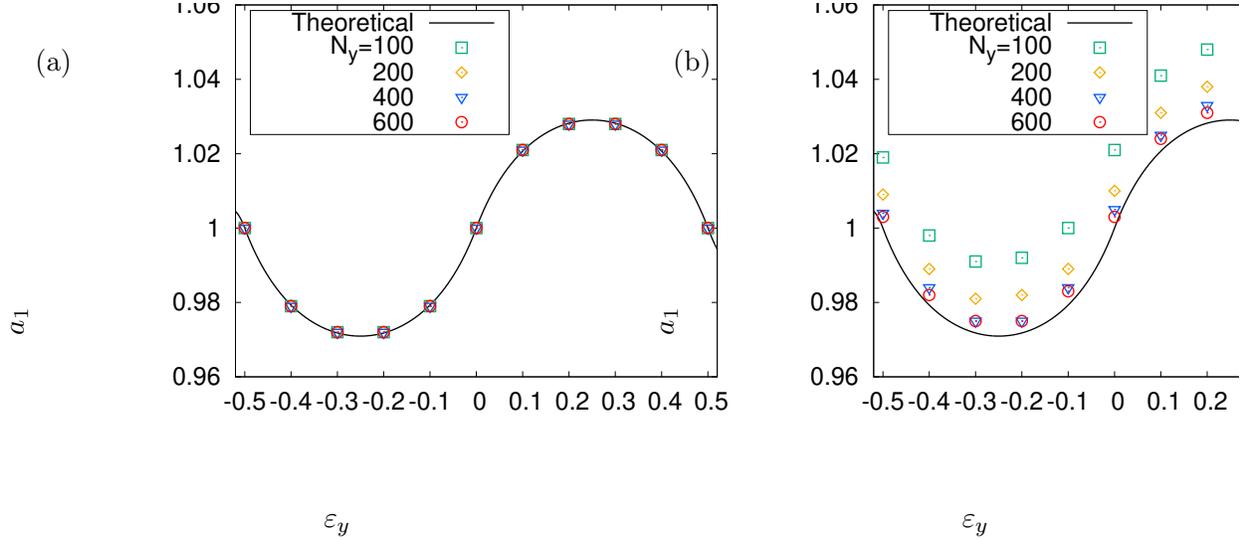

Next, the key principle in the present method to determine the forcing shear stress by the consistency condition (Eq.~\ref{eq:cond_f_delta_dis}) is verified. For this purpose, we compare numerically determined $a_1$ and $a_2$ values minimizing the error of the wall slip velocities and those values given by Eqs.~\eqref{eq:a1} and \eqref{eq:a2} derived from the consistency condition. The error of the wall slip velocities is calculated as the RMS of the deviation from the analytical solutions from $t=0$ to $t=5$, where the flow reaches the steady state. 
The theoretical values by Eqs.~\eqref{eq:a1} and \eqref{eq:a2} depend on the relative position between the Eulerian mesh $\mathcal{M}$ and the Lagrange points $\Gamma$, and are the functions of $\varepsilon_y\in [-0.5,0.5]$. Fig.~\ref{fig:a_vs_y0_Ls1_10-1_t5} shows the comparison results of $a_1$ for $\mathcal{L}_s=0.1$ with $\Delta t=1\times10^{-5}$ and different mesh resolutions $N_y$ ($\Delta y=2/N_y$). Since $a_2 (\varepsilon_y) =a_1 (-\varepsilon_y)$ from the geometrical symmetry, only the result of $a_1$ is shown. In case of the Poiseuille flow, numerically obtained $a_1$ value agrees well with the theoretical value. In case of the Couette flow, numerically obtained value asymptotes to the theoretical value with increasing the spatial resolution. These results demonstrate the validity of the principle in the present method.

\begin{figure}[tb]
  \begin{center}
    \begin{overpic}[bb = 0 0 612 239,width=6.3in,clip,keepaspectratio]
        {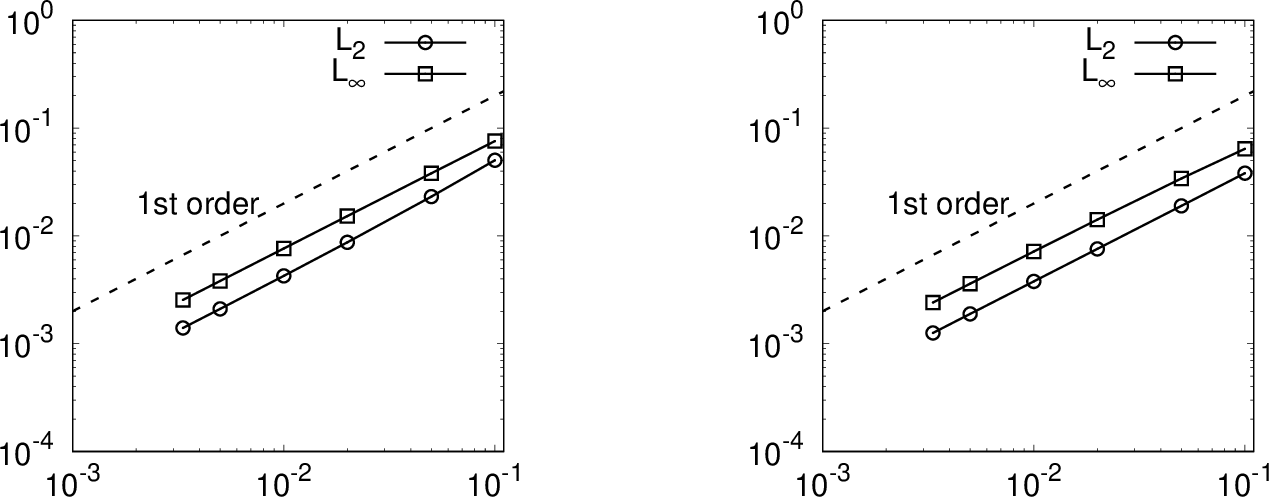}
        \put(0,40){(a)}
        \put(59,40){(b)}
        \put(30,-1){$\Delta y$}
        \put(90,-1){$\Delta y$}
        \put(-3,17){\rotatebox{90}{Error}}
        \put(56,17){\rotatebox{90}{Error}}
    \end{overpic}
    \caption{Spacial accuracy of the present method evaluated by $L_2$ and $L_\infty$ norms (Eq.~\ref{eq:L}) for (a) the Poiseuille flow and (b) the Couette flow. The slip length of the wall is $\mathcal{L}_s=0.1$.}
   \label{fig:1D_err_vel_space}
  \end{center}
\end{figure}

The spatial accuracy of the present method is evaluated by calculating the $L_2$ and $L_\infty$ errors to the analytical solution $u_a$ for $\mathcal{L}_s=0.1$ with $\Delta t=1\times10^{-6}$ by several mesh resolutions. The errors are calculated for the unsteady solution at $t=0.1$ by 
\begin{equation}\label{eq:L}
\begin{aligned}
L_2&=\sqrt{\sum_{y^{\mathcal{F}_x}_j\in\Omega_f}(u_j-u_{a,j})^2/N_{\Omega_f}}\\
L_\infty&=\max_{y^{\mathcal{F}_x}_j\in\Omega_f}(u_j-u_{a,j}),
\end{aligned}
\end{equation}
where $N_{\Omega_f}$ is the number of grid points belonging to $\mathcal{F}_x$ in $\Omega_f$. 
The spacial resolution dependence of the $L_2$ and $L_\infty$ errors for the Poiseuille and Couette flows are shown in Fig.~\ref{fig:1D_err_vel_space}. The figure shows that the order of accuracy is the first order, which is consistent with the theoretical evaluation detailed in \ref{app:PoiCtt}.

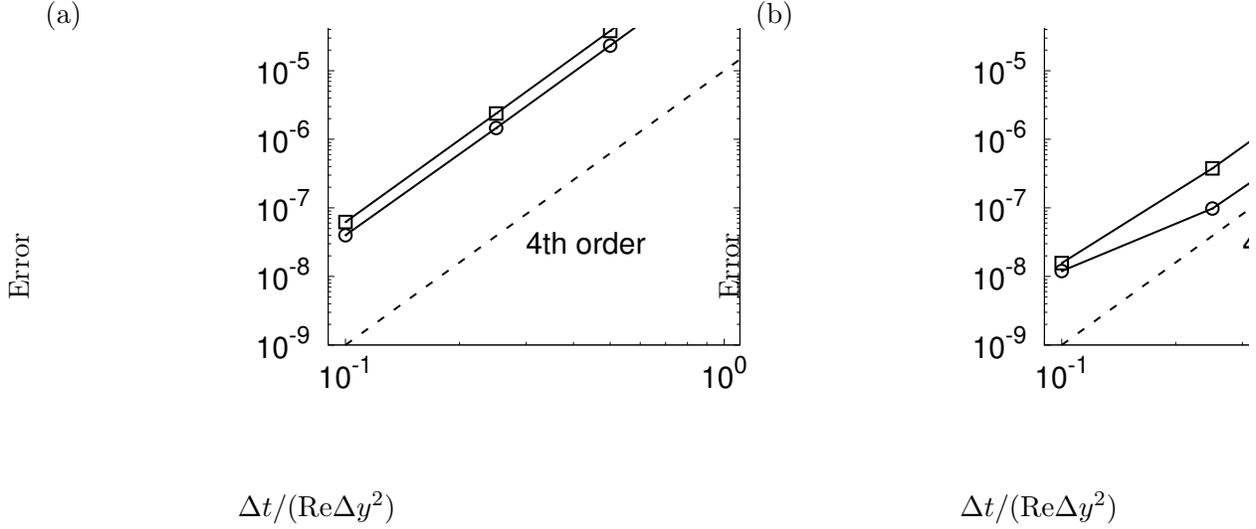
\begin{figure}[tb]
  \begin{center}
    \begin{overpic}[bb = 0 0 605 239,width=6.3in,clip,keepaspectratio]
        {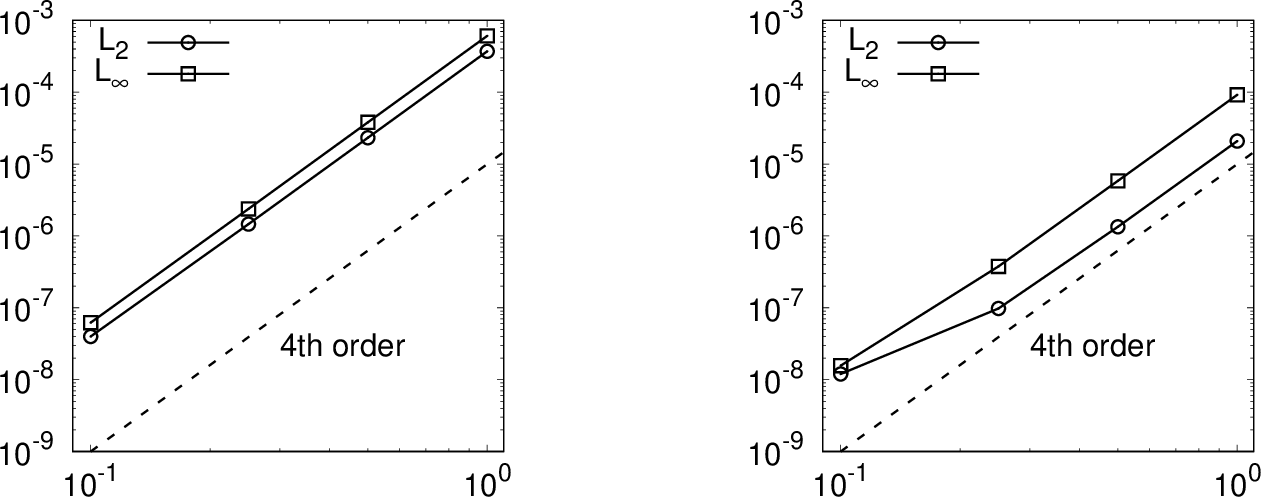}
        \put(0,40){(a)}
        \put(59,40){(b)}
        \put(16,-1){$\Delta t/(\mathrm{Re}\Delta y^2)$}
		\put(76,-1){$\Delta t/(\mathrm{Re}\Delta y^2)$}
        \put(-3,17){\rotatebox{90}{Error}}
        \put(56,17){\rotatebox{90}{Error}}
    \end{overpic}
    \caption{Temporal accuracy of the present method evaluated by $L_2$ and $L_\infty$ norms (Eq.~\ref{eq:L}) for (a) the Poiseuille flow and (b) the Couette flow. The slip length of the wall is $\mathcal{L}_s=0.1$.}
   \label{fig:1D_err_vel_time}
  \end{center}
\end{figure}

The temporal accuracy of the present method is also evaluated for $\mathcal{L}_s=0.1$ with $\Delta y=2\times10^{-2}$ by several temporal resolutions. The $L_2$ and $L_\infty$ errors are calculated for the unsteady solution at $t=0.5$ by Eq.~\eqref{eq:L}, replacing $u_a$ with the numerical solution with a very fine time step ($\Delta t=1\times10^{-6}$) to extract only the temporal discretization error. For the Couette flow, the wall velocity at $y=\eta_2$ is given by the following equation to avoid the initial discontinuity:
\begin{align}
	U(t) = \frac{1}{2}\qty{1+\tanh\qty(\frac{t-0.2}{0.05})}
	\label{eq:1D_ctt_wallvel}.
\end{align}
Fig.~\ref{fig:1D_err_vel_time} shows that $L_2$ and $L_\infty$ errors for both the Poiseuille and Couette flows follow the fourth order trend. This agrees with the splitting error for $N = 3$ predicted by Eq.~\eqref{eq:Mtrx_GE_LU_NBC_delta} based on the theory of the fractional step algorithm, which dominates over the underlying second-order error resulting from the time integration schemes \cite{Taira2007}. 

\subsection{Two dimensional problem}
\subsubsection{Flow confined between two concentric cylinders}
\begin{figure}[t]
	\begin{center}
		\begin{overpic}[bb=0 0 408 427,width=6cm,clip,keepaspectratio]
				{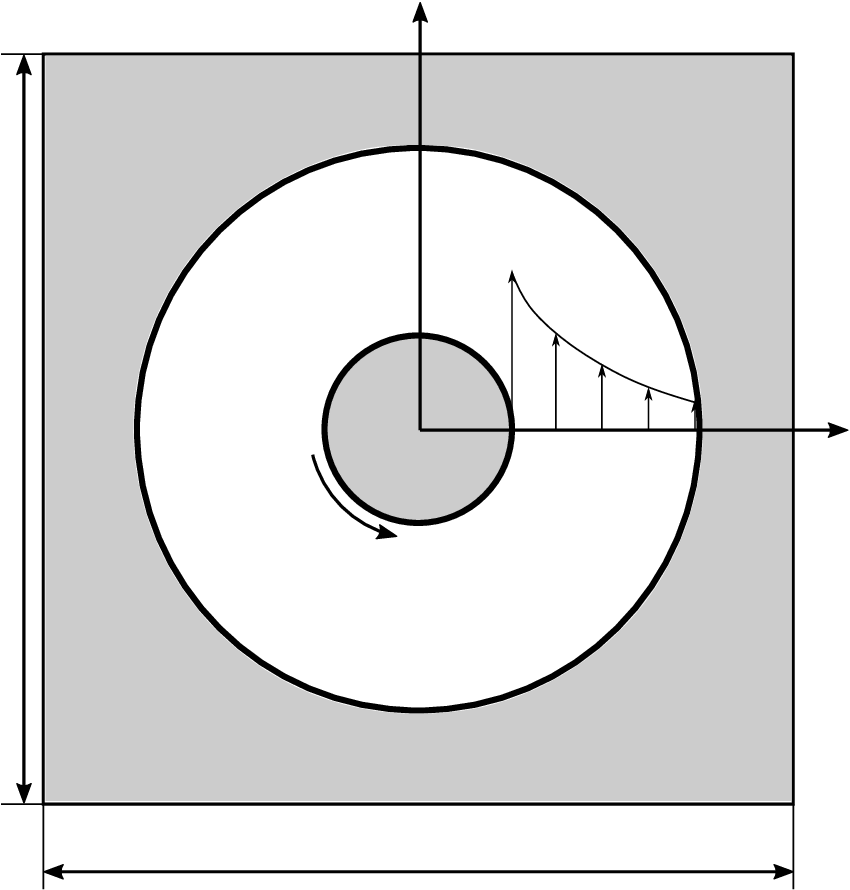}
				\put(34,39){\small $\omega$}
				\put(30,32){\small $D_1=2R_1=1$}
 				\put(67,18){\small $D_2=3$}
				\put(46,3){\small $4$}
				\put(-3,50){\small \rotatebox{90}{$4$}}
				\put(60.5,65){\small $u_\theta(r)$}
				\put(95,53){\small $x$}
				\put(48,100){\small $y$}
		\end{overpic}
		\caption{Flow between two concentric cylinders as a two-dimensional benchmark problem. The inner cylinder rotates counterclockwise with the angular velocity $\omega$ while the outer cylinder is fixed. The two cylinder surfaces have the same slip length. The boundary for the computational domain $[-2,2]\times [-2,2]$ is periodic.}
	 \label{fig:cyl_ctt_comp_domain}
	\end{center}
\end{figure}
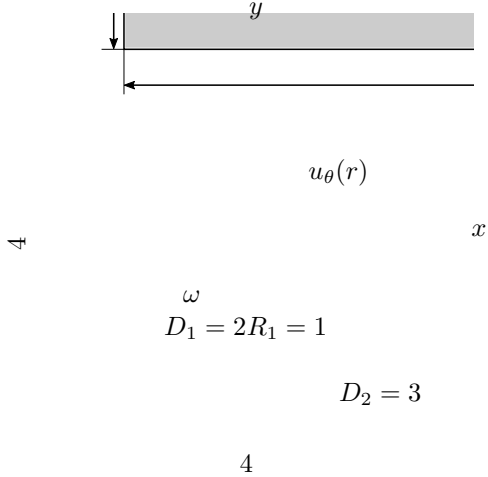

For the first test on the two-dimensional problem, the spacial and temporal accuracy of the present method is evaluated on the flow between two concentric cylinders. The schematic of the flow is shown in Fig.~\ref{fig:cyl_ctt_comp_domain}.
The outer cylinder is stationary and the inner cylinder rotates counterclockwise at an angular velocity $\omega$ given by
\begin{align}
	\omega(t)=1+\tanh\qty(\frac{t-0.2}{0.05}).
	\label{eq:2D_ctt_wallvel}
\end{align}
The two cylinder surfaces have the same slip length, and the Reynolds number is set as $\mathrm{Re}=1$. The calculation domain is a periodic domain of $[-2,2]\times [-2,2]$ with the cylinder axis at the origin, and the grid width is constant throughout the domain, $\Delta x=\Delta y=\Delta$. The surface of the cylinder is represented by the fixed Lagrangian points equally spaced with $\Delta s\simeq\Delta$. 

\begin{figure}[tb]
  \begin{center}
    \begin{overpic}[bb = 0 0 685 186,width=6.3in,clip,keepaspectratio]
        {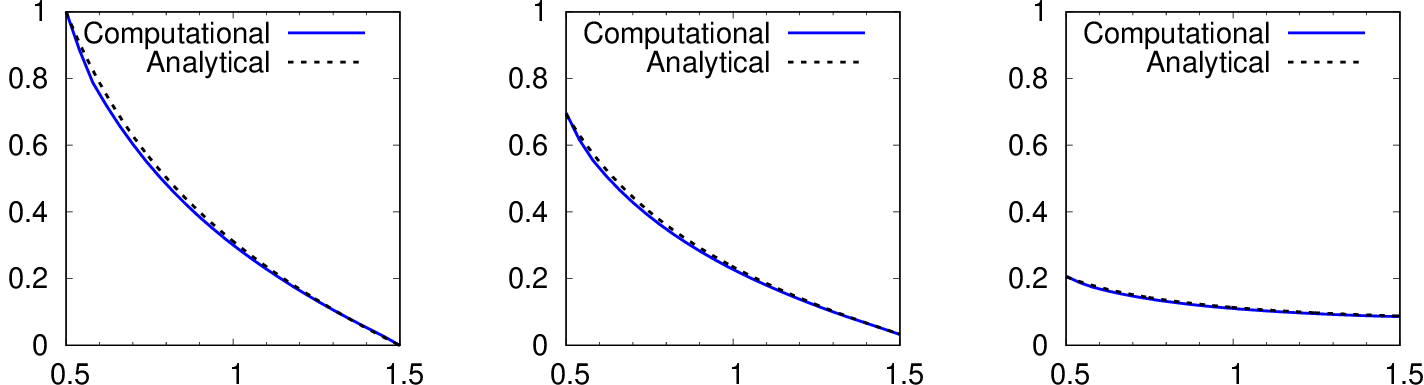}
            \put(17,-2){$r$}
			\put(52.5,-2){$r$}
			\put(87,-2){$r$}
			\put(-2,14){\rotatebox{90}{$u_\theta/(\omega R_1)$}}
			\put(33,14){\rotatebox{90}{$u_\theta/(\omega R_1)$}}
			\put(68,14){\rotatebox{90}{$u_\theta/(\omega R_1)$}}
			\put(-1,27){(a)}
			\put(34,27){(b)}
			\put(69.5,27){(c)}
    \end{overpic}
    \caption{Circumferential velocity $u_\theta$ predicted by the present method (solid lines) and the analytical solution (dotted lines) for the slip length (a) $\mathcal{L}_s=0$, (b) $\mathcal{L}_s=0.1$ and (c) $\mathcal{L}_s=1$.}
   \label{fig:vel1D_theta30_t5}
  \end{center}
\end{figure}

First, for slip lengths $\mathcal{L}_s=0$, 0.1 and 1, the computational results with $\Delta=4\times10^{-2}$ and $\Delta t=1\times10^{-4}$ are compared to the analytical result. The inner and outer cylinder surfaces are represented by 78 and 234 Lagrangian points, respectively. The circumferential velocity distributions at the steady state are shown in Fig.~\ref{fig:vel1D_theta30_t5}. The present results are evaluated on the line rotated counterclockwise by 30 degree from $y$ axis: the velocity in the bulk is given by the bilinear interpolation and the velocity on the cylinder surface is given by the interpolation operator $\hat{E}$. 
We see that the predictions by the present method agree well with the analytical solutions for both slip and no-slip cases. 

\begin{figure}[tb]
  \begin{center}
    \begin{overpic}[bb = 0 0 360 252,width=.48\linewidth]{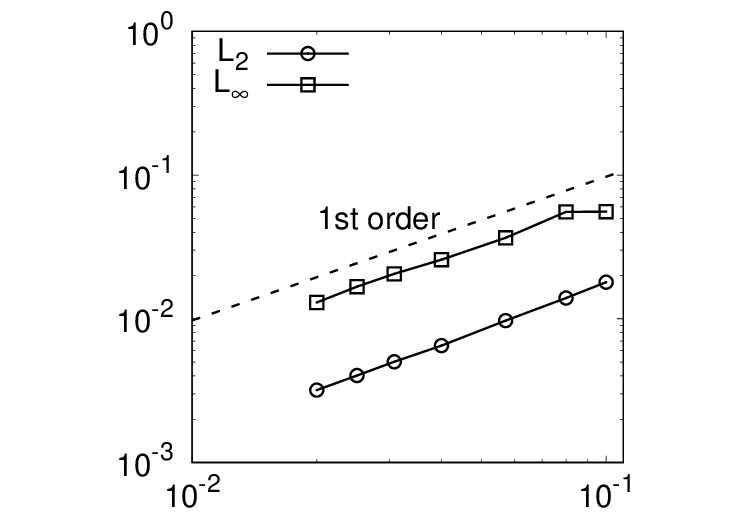}
     \put(10,29){\rotatebox{90}{Error}}
     \put(15,70){(a)}
     \put(54,-3){$\Delta$}
     \end{overpic}
    \begin{overpic}[bb = 0 0 360 252,width=.48\linewidth]{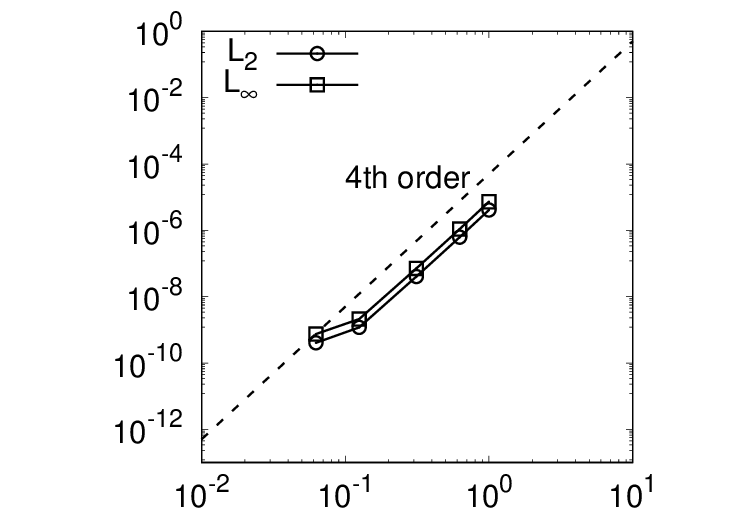}
    \put(10,29){\rotatebox{90}{Error}}
    \put(15,70){(b)}
    \put(45,-3){$\Delta t/(\mathrm{Re}\Delta^2)$}
    \end{overpic}
    \caption{The accuracy of the present method evaluated for the flow between two concentric cylinders in two-dimensions: (a) spacial accuracy and (b) temporal accuracy.}
   \label{fig:cyl_err}
  \end{center}
\end{figure}
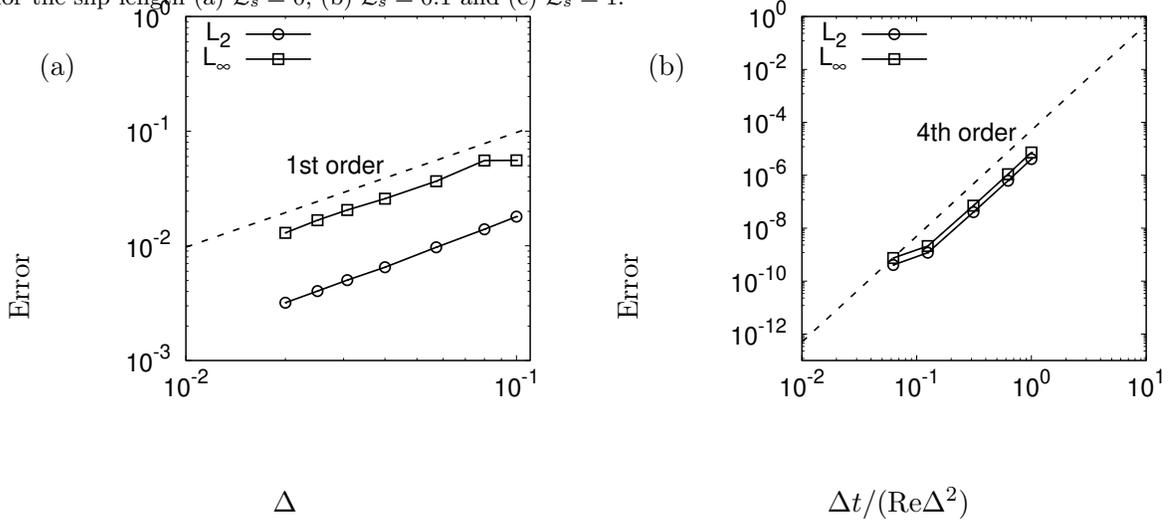

The spacial accuracy of the present method is evaluated by the error to the analytical solution for $\mathcal{L}_s=0.1$ at the steady state. The time step width is fixed at $\Delta t=1\times10^{-4}$. The $L_2$ and $L_\infty$ errors are calculated as Eq.~\eqref{eq:L} and shown for different spacial resolutions in Fig.\ref{fig:cyl_err}(a). It is found that the spacial accuracy is of the 1st order as for the one-dimensional system. The temporal accuracy of the present method is evaluated in a similar way and the result is shown in Fig.~\ref{fig:cyl_err}(b). Here the spacial resolution is fixed at $\Delta=4\times10^{-2}$ and the error is calculated at $t=1$ by the difference to the numerical solution with a very fine time step width ($\Delta t=5\times10^{-6}$) for the same reason described in \ref{S:1D}. Fig.~\ref{fig:cyl_err}(b) shows that the fourth-order accuracy is achieved as in the one-dimensional system.

\subsubsection{Flow over a stationary cylinder}
\label{subsubsec:stationary_cylinder}
As the next benchmark problem, the flow around a stationary cylinder is calculated and compared with the experimental and numerical results in the literature. A cylinder of diameter $D=1$ is placed in the fluid which is initially a uniform flow with $U_\infty=1$. 
The calculations are carried out for the cases where the cylinder surface is with or without slip. The computational domain is $[-30,30]\times [-30,30]$ with the axis of the cylinder at the origin. The boundary conditions of the domain are assumed to be $u=U_\infty$ and $v=0$ at the inlet ($x=-30$) and the top and bottom ($y=-30,30$), which are the same as the initial uniform flow. The convective outflow condition is imposed at the domain outlet ($x=30$). The calculations are performed with the resolution of the mesh and the Lagrangian points listed in Table~\ref{tab:comp_param_stationary_cylinder}, depending on the Reynolds number $\mathrm{Re}$.
\begin{table}[t]
	\centering
	\caption{Three sets of the number of computational cells and the Lagrangian points used to calculate the flow past a cylinder.}
	\label{tab:comp_param_stationary_cylinder}
		\begin{tabular}{ccccc}
			\toprule
			{}&	$N_x\times N_y$ & $\Delta x_\mathrm{min}$ & $N_\Gamma$ & $\Delta t$\\
			\midrule
			Resolution A & $150\times 150$ & $4\times 10^{-2}$ & 78 & $5\times 10^{-3}$\\
			Resolution B & $300\times 300$ & $2\times 10^{-2}$ & 156 & $5\times 10^{-3}$\\
			Resolution C & $300\times 300$ & $2\times 10^{-2}$ & 152 & $5\times 10^{-3}$\\
			\bottomrule
		\end{tabular}
\end{table}
The number of Lagrangian points $N_\Gamma$ in Resolution C is slightly smaller than Resolution B to get a better convergence in the solution of Eq.~\eqref{eq:calc_proc_NBC_2} when the slip length of the cylinder surface is equal or close to zero. 
The domain near the cylinder is discretized with equally spaced grids with a minimum grid width of $\Delta x_\mathrm{min}=\Delta y_\mathrm{min}$, and the rest of the domain is discretized with unequally spaced grids with gradually increasing grid width toward the domain boundaries. On the cylinder surface, Lagrangian points are equally spaced with $\Delta s\simeq\Delta x_\mathrm{min}$.

\begin{figure}[t]
	\begin{center}
		\begin{overpic}[bb=0 0 431 397,width=6in,clip,keepaspectratio]
				{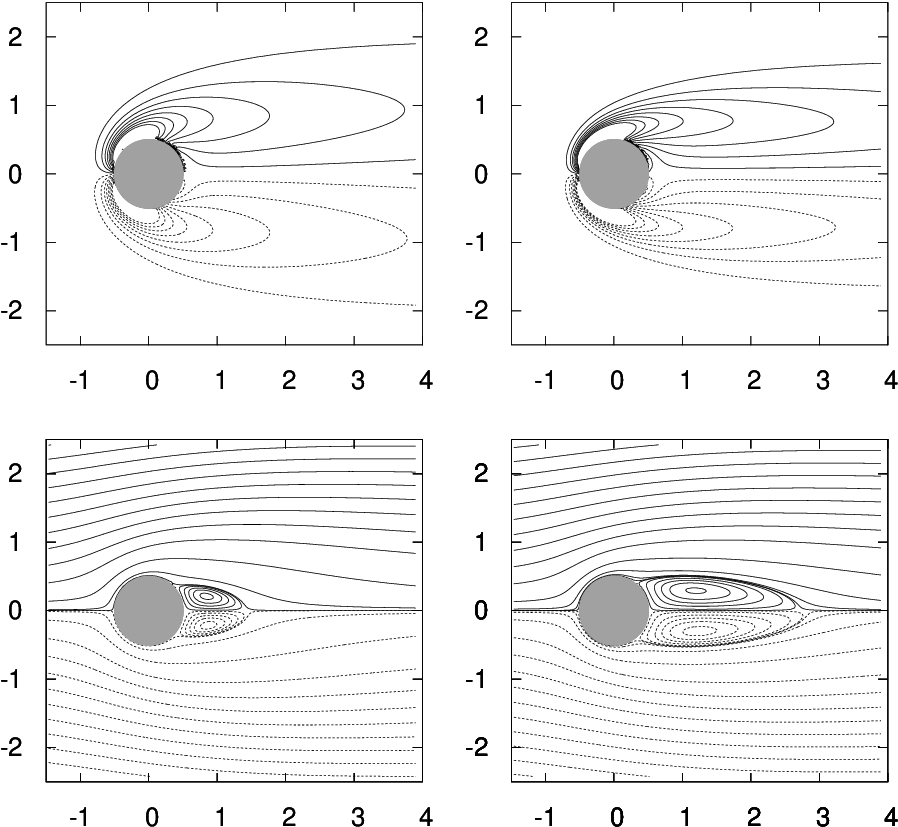}
				\put(22,93){$\mathrm{Re}=20$}
                \put(74,93){$\mathrm{Re}=40$}
		\end{overpic}
		\caption{Vorticity distribution (top) and streamline (bottom) of the flow past a no-slip cylinder: the present method (solid lines) and the method by Taira and Colonius \cite{Taira2007} (dotted lines).}
	 \label{fig:vorticity-stream_no-slip}
	\end{center}
\end{figure}

First, the steady state solution of the flow over a no-slip cylinder for $\mathrm{Re}=20$ and 40 is calculated. 
Fig.~\ref{fig:vorticity-stream_no-slip} shows the steady state vorticity distributions and streamlines for $\mathrm{Re}=20$ calculated with Resolution B and for $\mathrm{Re}=40$ with Resolution C. In the lower half of each figure in Fig.~\ref{fig:vorticity-stream_no-slip}, the result using the conventional regularization operator (Eq.~\ref{eq:taira}) is also shown for comparison. 
Both results agree reasonably well. The unsmooth vorticity profile near the cylinder surface is limited in the region where the boundary force is distributed. The present method is constructed so that the less accurate velocity distribution in the boundary force support does not affect the velocity distribution outside this support. Fig.~\ref{fig:vorticity-stream_no-slip} shows that this design philosophy is well realized. 

\begin{figure}[t]
	\begin{center}
		\begin{overpic}[bb=0 0 182 56,width=10cm,clip,keepaspectratio]{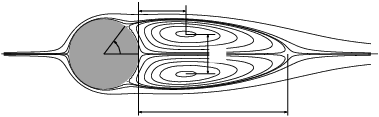}
				\put(55.5,15.5){$b_\mathrm{w}$}
				\put(53.5,2.5){$l_\mathrm{w}$}
				\put(32,18){$\theta_\mathrm{w}$}
				\put(41,29){$a_\mathrm{w}$}
		\end{overpic}
		\caption{Calculated wake dimensions:     $l_\mathrm{w}$ represents the length of the recirculation zone, $a_w$ the downstream position of the recirculation vortex center, $b_w$ the distance between the two recirculation vortex centers, and $\theta_\mathrm{w}$ the separation angle measured from the stagnation point on the downstream side.}
	 \label{fig:wakedim}
	\end{center}
\end{figure}

\begin{figure}[t]
	\centering
		\begin{overpic}[bb=0 0 360 252,width=7cm,clip,keepaspectratio]
			{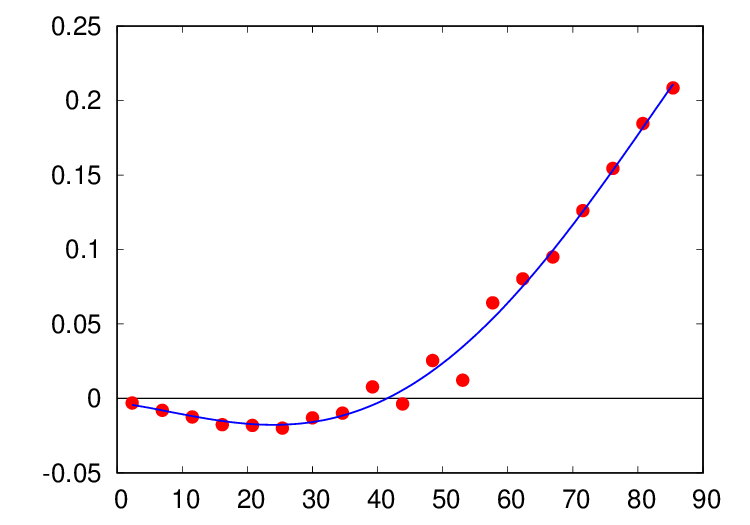}
            \put(45,70){$\mathrm{Re}=20$}
			\put(20,57){Resolution A}
			\put(53.5,-4){$\theta$}
			\put(0,20){\rotatebox{90}{Shear stress}}
		\end{overpic}
		\vspace{5mm}
		\begin{overpic}[bb=0 0 360 252,width=7cm,clip,keepaspectratio]
			{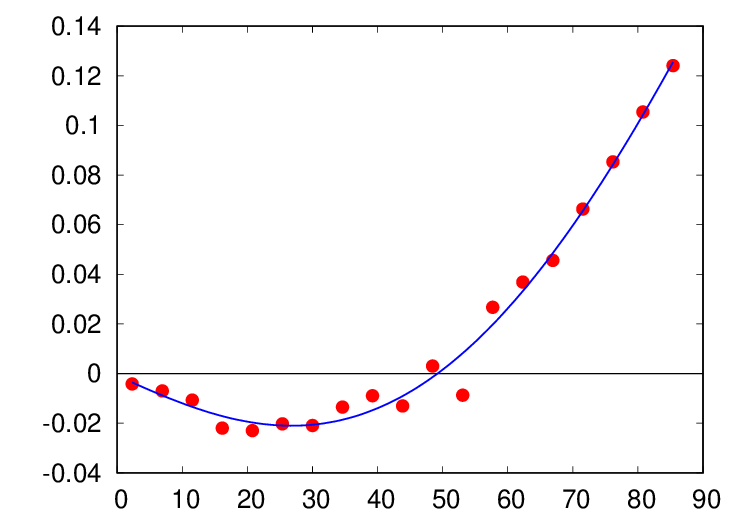}
            \put(45,70){$\mathrm{Re}=40$}
			\put(20,57){Resolution A}
			\put(53.5,-4){$\theta$}
			\put(0,20){\rotatebox{90}{Shear stress}}
		\end{overpic}
		\vspace{5mm}
		\begin{overpic}[bb=0 0 360 252,width=7cm,clip,keepaspectratio]
			{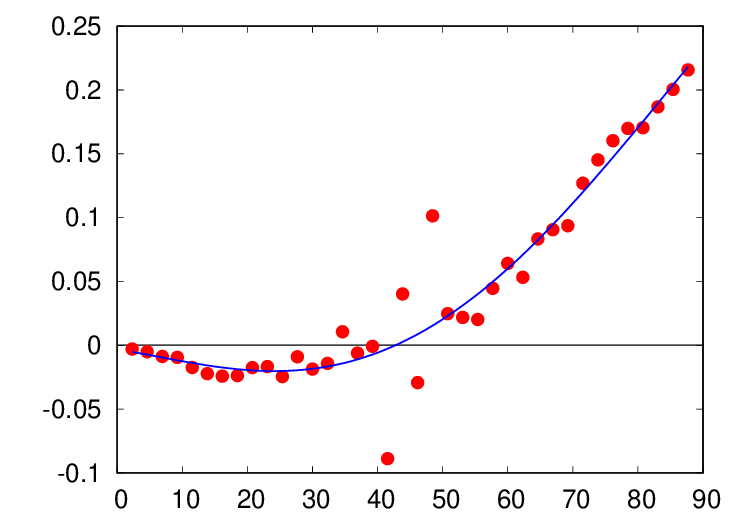}
			\put(20,57){Resolution B}
			\put(53.5,-4){$\theta$}
			\put(0,20){\rotatebox{90}{Shear stress}}
		\end{overpic}
		\begin{overpic}[bb=0 0 360 252,width=7cm,clip,keepaspectratio]
			{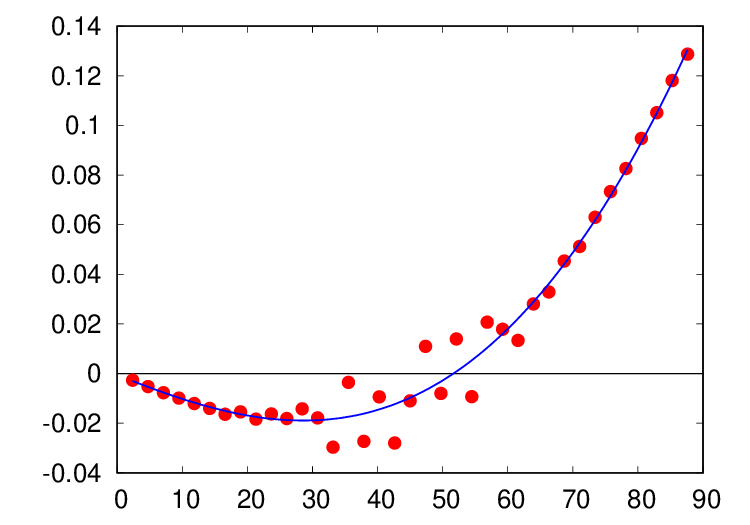}
			\put(20,57){Resolution C}
			\put(53.5,-4){$\theta$}
			\put(0,20){\rotatebox{90}{Shear stress}}
		\end{overpic}
	\caption{Shear stress distribution evaluated on the no-slip cylinder. The angular position $\theta$ is measured in the same way as the separation angle $\theta_w$ (Fig.~\ref{fig:wakedim}). The symbols are the values on the Lagrangian points calculated by the interpolation operator $\hat{G}_{\tau n}$ (Eq.~\ref{eq:NBC+PBC_op}), and the solid lines are the quartic polynomial fits to the symbols. $\theta_w$ is calculated as $\theta$ where the fit intersects the zero horizontal line.}
	\label{fig:strss-fit}
\end{figure}

For more detailed comparison, the drag coefficient $C_\mathrm{D}$ and the wake dimensions shown in Fig.~\ref{fig:wakedim} are compared with the literature values. 
The drag coefficient $C_\mathrm{D}$ is calculated by the following equation using the boundary force $\bm{F}$:
\begin{align}
	C_\mathrm{D}=-\sum_l F_{x,l}\Delta s/\qty(\frac{1}{2}U_\infty^2 D)
	\label{eq:CD}.
\end{align}
Table~\ref{tab:wake} summarizes the present results and the literature values. The separation angle is calculated from the shear stress distribution on the cylinder surface, interpolated to the Lagrangian points using the $\hat{G}_{\tau n}$ operator in Eq.~\eqref{eq:NBC+PBC_op}. 
\begin{table}[t]
	\centering
    \caption{Comparison of the present results and the literature values for the flow past a no-slip cylinder. The quantities with the subscript w is the wake dimensions shown in Fig.\ref{fig:wakedim} and $C_\mathrm{D}$ is the drag coefficient given by Eq.~\eqref{eq:CD}.}
	\label{tab:wake}
		\begin{tabular}{lllllll}
			\toprule
			{} & {} & $l_\mathrm{w}$ & $a_\mathrm{w}$ & $b_\mathrm{w}$ & $\theta_\mathrm{w}$ & $C_\mathrm{D}$\\
			\midrule
			$\mathrm{Re}=20$ 	& \citet{Coutanceau_Bouard_1977} & 0.93 	& 0.33 	& 0.46 	& 45.0\textdegree	& - 	\\
												& \citet{Tritton_1959}								& -			&	-			&	-			&	-						& 2.09\\
												& \citet{Dennis_Chang_1970}			& 0.94	&	-			&	-			&	43.7\textdegree	&	2.05\\
												& \citet{LINNICK2005157}			& 0.93	&	0.36	&	0.43	&	43.5\textdegree	&	2.06\\
												& \citet{Taira2007}		& 0.94	&	0.37	&	0.43	&	43.3\textdegree	&	2.06\\
												&	\citet{Canuto_Taira_2015}			&	0.92	&	0.36	&	0.42	&	43.7\textdegree	&	2.07\\
                                                & \citet{LEGENDRE_LAUGA_MAGNAUDET_2009}&-&-			&	-			& -						& 2.04\\
												& Present (Resolution A)							& 0.88	& 0.34	&	0.42	&	41.5\textdegree	&	2.01\\
												& Present (Resolution B)							& 0.90	& 0.34	&	0.42	&	42.7\textdegree	&	2.03\\
												&																&				&				&				&							&			\\
			$\mathrm{Re}=40$	& \citet{Coutanceau_Bouard_1977} & 2.13 	& 0.76 	& 0.59 	& 53.8\textdegree & - 	\\
												& \citet{Tritton_1959}								& -			&	-			&	-			&	-						& 1.59\\
												& \citet{Dennis_Chang_1970}			& 2.35	&	-			&	-			&	53.8\textdegree	&	1.52\\
												& \citet{LINNICK2005157}			& 2.28	&	0.72	&	0.60	&	53.6\textdegree	&	1.54\\
												& \citet{Taira2007}		& 2.30	&	0.73	&	0.60	&	53.7				&	1.54\\
												&	\citet{Canuto_Taira_2015}			&	2.24	&	0.72	&	0.59	&	53.7				&	1.54\\
												& Present (Resolution A)							& 2.12	& 0.65	&	0.58	&	49.2\textdegree	&	1.49\\
												& Present (Resolution C)							& 2.19	& 0.69	&	0.59	&	51.6\textdegree	&	1.51\\
			\bottomrule
		\end{tabular}
\end{table}
The present results are generally in good agreement with the literature values, and better predictions are obtained with increased resolution. The noticeable deviation from the literature value is observed for the separation angle $\theta_w$. As shown in Fig.~\ref{fig:strss-fit}, the shear stress distribution near the separation point fluctuates, which deteriorates the prediction of $\theta_w$. It should be noted however that this fluctuation disappears for the non-zero slip length, which is the main target of our study. 

\begin{figure}[t]
	\begin{center}
		\begin{overpic}[bb=0 0 360 252,width=8cm,clip,keepaspectratio]
				{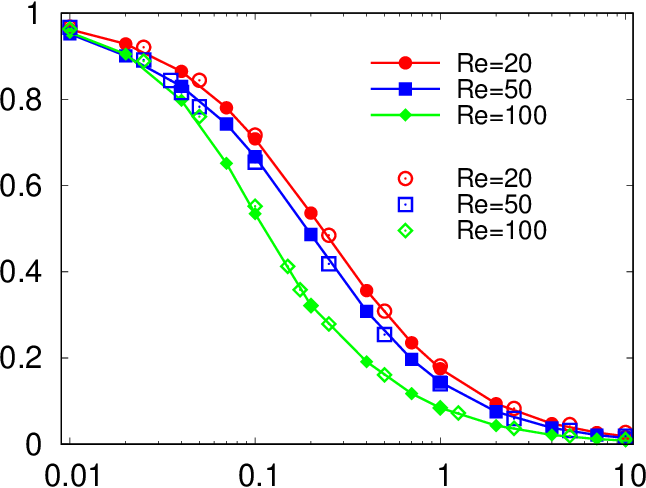}
				\put(50,59){\scriptsize \textsf{Present}}
				\put(50,44){\scriptsize \textsf{\textsf{\citet{LEGENDRE_LAUGA_MAGNAUDET_2009}}}}
				\put(45,-3){$\mathcal{L}_s$}
				\put(-9,33){\rotatebox{90}{$C^*_\mathrm{D}$}}
		\end{overpic}
		\caption{Normalized drag coefficient (Eq.~\ref{eq:CD_norm}) of a cylinder with various slip lengths. The results obtained on the body-fitted mesh by \citet{LEGENDRE_LAUGA_MAGNAUDET_2009} are also shown for comparison.}
	 \label{fig:CD_Ls}
	\end{center}
\end{figure}

As the final test on the flow past a stationary cylinder, we calculate the slip length dependence of the drag coefficient for $\mathrm{Re}=20$, 50 and 100. We employ Resolution B for $\mathrm{Re}=20$ and Resolution C for $\mathrm{Re}=50$ and 100. The normalized drag coefficient given by
\begin{align}
	C^*_\mathrm{D}(\mathcal{L}_s)=\frac{C_\mathrm{D}(\mathcal{L}_s)-C_\mathrm{D}(\mathcal{L}_s=\infty)}
 {C_\mathrm{D}(\mathcal{L}_s=0)-C_\mathrm{D}(\mathcal{L}_s=\infty)}
	\label{eq:CD_norm}
\end{align}
is compared in Fig.~\ref{fig:CD_Ls} with the results by \citet{LEGENDRE_LAUGA_MAGNAUDET_2009} obtained on the body-fitted mesh. In the present results, we employ $C_\mathrm{D} (\mathcal{L}_s=100)$ as $C_\mathrm{D}(\mathcal{L}_s=\infty)$ to calculate $C^*_\mathrm{D} (\mathcal{L}_s)$, and the drag coefficient value is either the steady state value or the time-averaged value depending on the Reynolds number.
The present results agree well with the results obtained on the body-fitted mesh for a wide range of the slip length. 

\subsubsection{Flow around a moving cylinder}
\label{S:movingcylinder}
As the final test, the flow around a circular cylinder moving at a constant velocity is calculated. In this case, the Lagrangian points on the cylinder surface move relative to the Eulerian mesh. A cylinder of diameter $D=1$ placed in a fluid at rest initially ($t=0$) moves at a constant velocity $U=1$ in the negative $x$ direction for $t>0$. The calculations are performed for various slip lengths under $\mathrm{Re}=40$. The calculation domain is $[-16.5,13.5]\times [-15,15] $ with the initial cylinder center at the origin. On the boundaries of the calculation domain the no-slip condition is imposed. The calculation is performed up to $t=3.5$ with a time step width of $\Delta t=5\times 10^{-3}$. The region of $[-4.5,1]\times [-1,1]$ that covers the region through which the cylinder passes is discretized with equally spaced grids with a minimum grid width of $\Delta x_\mathrm{min}=0.02$, and the rest of the domain is discretized with unequally spaced grids with gradually increasing grid width toward the domain boundaries. The Lagrangian points are equally spaced on the cylinder surface with $N_\Gamma=152$. For the adopted time step, grid width and moving velocity of the cylinder, the relative position between the Eulerian mesh $\mathcal{M}$ and the Lagrangian points $\Gamma$ have 4 patterns. 

\begin{figure}[t]
	\begin{center}
		\begin{overpic}[bb=0 0 677 376,width=6in,clip,keepaspectratio]
				{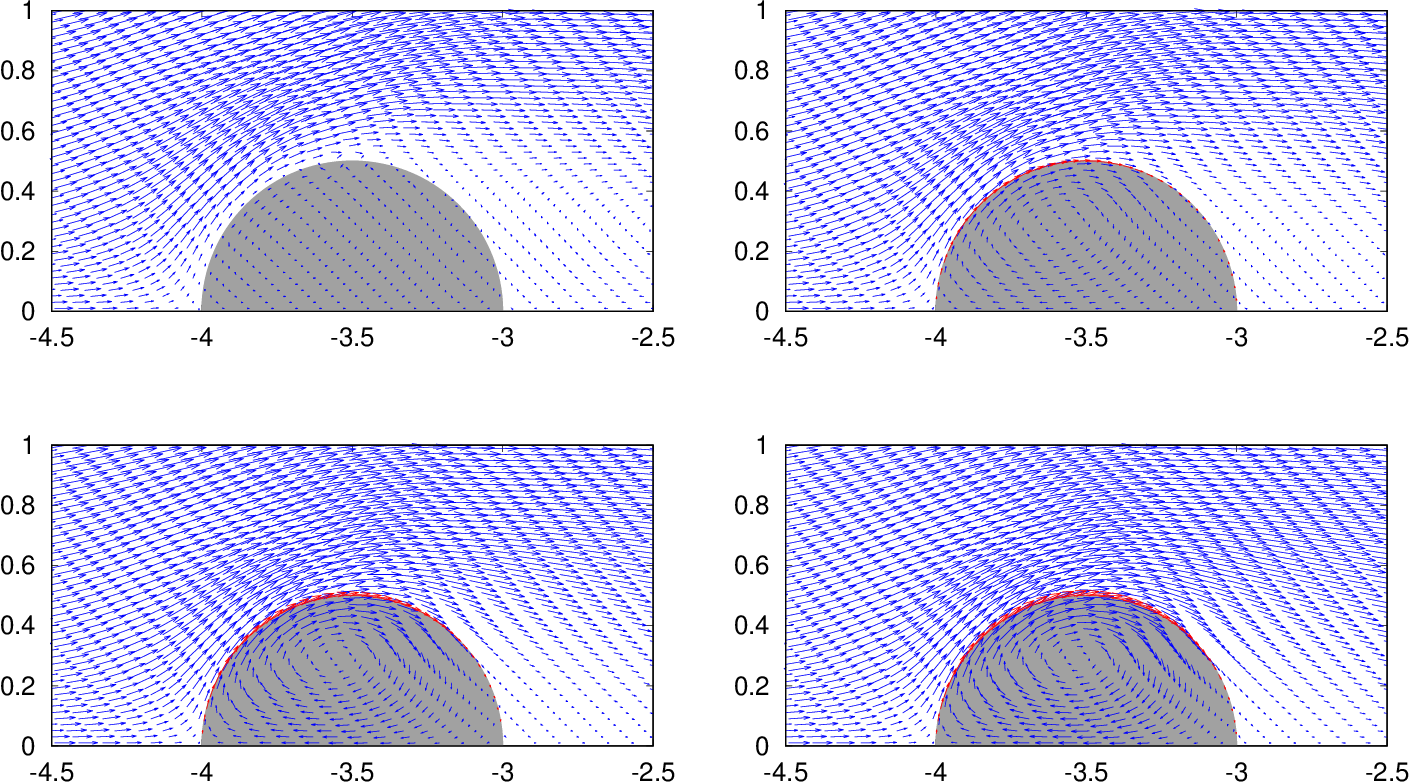}
				\put(0,57){(a)}
				\put(52,57){(b)}
				\put(0,26){(c)}
				\put(52,26){(d)}
		\end{overpic}
		\caption{Velocity field around the cylinder at $t=3.5$ measured from the moving cylinder for the slip lengths
		(a) $\mathcal{L}_s=0$, (b) $\mathcal{L}_s=0.1$, (c) $\mathcal{L}_s=0.5$, and (d) $\mathcal{L}_s=10$. The red arrows show the velocity on the cylinder surface calculated on the Lagrangian points.}
	 \label{fig:vel2D_mvcyl_Re40}
	\end{center}
\end{figure}

\begin{figure}[t]
	\begin{center}
		\begin{overpic}[bb=0 0 360 252,width=8cm,clip,keepaspectratio]
                {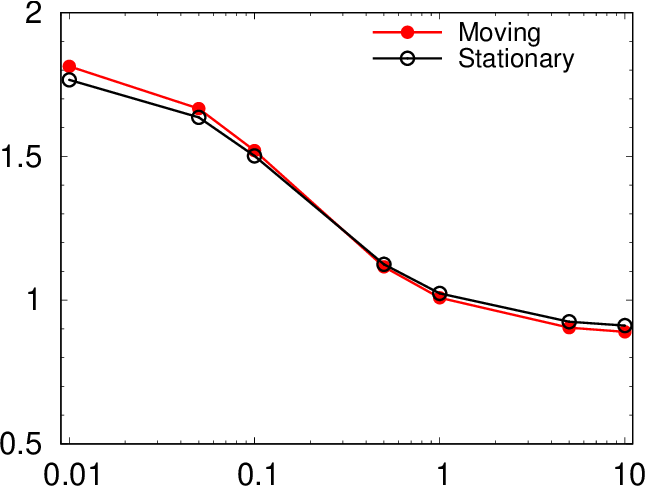}
				\put(45,-3){$\mathcal{L}_s$}
				\put(-9,33){\rotatebox{90}{$C_\mathrm{D}$}}
		\end{overpic}
		\caption{The slip length dependence of the drag coefficient for $\mathrm{Re}=40$. The results obtained for the moving (\ref{S:movingcylinder}) and the stationary cylinder (\ref{subsubsec:stationary_cylinder}) are compared.}
	 \label{fig:CD_Ls_move}
	\end{center}
\end{figure}

Fig.~\ref{fig:vel2D_mvcyl_Re40} shows the velocity field near the cylinder at $t=3.5$. The slip on the cylinder surface has a significant impact on the velocity profile near the cylinder, reducing the wake region as expected from the slip length dependence of the drag coefficient shown in Fig.~\ref{fig:CD_Ls_move}.
In Fig.~\ref{fig:CD_Ls_move}, the results obtained for the stationary cylinder (\ref{subsubsec:stationary_cylinder}) is also shown for comparison. Both results agree reasonably well considering the differences in the calculation parameters such as the boundary condition on the domain boundaries.

\begin{figure}[t]
	\begin{center}
		\begin{overpic}[bb=0 0 340 234,height=5.5cm,keepaspectratio]
				{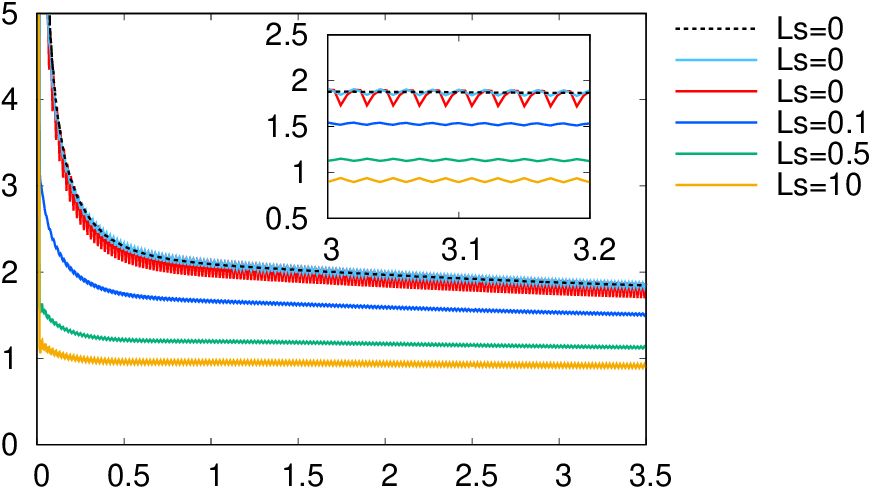}
				\put(48,-5){$t$}
				\put(-8,36){\rotatebox{90}{$C_\mathrm{D}$}}
				\put(120,63.5){\footnotesize \textsf{(Taira\cite{Taira2007}})}
				\put(120,59){\footnotesize \textsf{(Conventional)}}
		\end{overpic}
		\caption{Time variation of the drag coefficient calculated for the cylinder moving at a constant velocity ($\mathrm{Re}=40$). For the no-slip case, the results reported by \citet{Taira2007} and the results obtained by the present authors using Taira's method (denoted as ``Conventional") are also shown for comparison. In the inset from $t=3$ to $t=3.2$ there are 10 periods of oscillation for 40 calculation time steps.}
	 \label{fig:CD_time}
	\end{center}
\end{figure}

Fig.~\ref{fig:CD_time} shows the time variation of the drag coefficient for the slip lengths $\mathcal{L}_s=$ 0, 0.1, 0.5, and 10. The value at every time step is shown to examine the translational invariance property of the present method. For the no-slip case, the results reported by \citet{Taira2007} and the results obtained by the present authors using Taira's method are also shown for comparison. In \citet{Taira2007}, the calculation was performed with a time step width of $\Delta t=0.01$ and a minimum grid width of $\Delta x_\mathrm{min}=0.02$, and therefore the relative position between $\mathcal{M}$ and $\Gamma$ had 2 patterns instead of 4 patterns in the present calculations. From Fig.~\ref{fig:CD_time}, we see first that the results reported by \citet{Taira2007} and the present results using their method (denoted as ``conventional") agree well with each other. In the present calculation, a periodic fluctuation is slightly pronounced due to the doubled number of positional patterns between the Eulerian mesh and the Lagrangian points. In the prediction of the drag coefficient for the no-slip case by the present method ($\mathcal{L}_s=0$), there is a periodic oscillation whose period is 4 time steps (See the inset). This corresponds to the fact that there are 4 patterns of the relative position between the Eulerian mesh $\mathcal{M}$ and the Lagrangian points $\Gamma$ under the present calculation condition. When applied to the no-slip cylinder, the present method shows larger oscillation than the conventional method because it does not strictly satisfy the translational invariance (in the weak sense) as derived in \ref{app:consv}. However, if the slip length is non-zero, the oscillation amplitude is much smaller, and the translational invariance is not significantly violated as also theoretically shown in \ref{app:consv}. 

\section{Conclusions}
\label{S:conclusion}
We presented a formulation of the immersed boundary projection method (IBPM) for the Navier slip boundaries. The IBPM is a continuous forcing approach and treats the boundary condition as a constraint just as the solenoidal condition on the velocity field. The boundary force is determined implicitly as the pressure without any \textit{ad hoc} constitutive relations. The present method is first-order accurate in space, while the conventional IBPM designed for the no-slip boundaries is zeroth-order accurate when applied to the Navier slip boundaries. The temporal accuracy of the present method is fourth-order, which is one order higher than the truncation error in the expansion of the inverse operator to solve for the prediction velocity. These order of accuracies were shown theoretically and by numerical tests in one and two-dimensional benchmark problems. The present method was applied to predict the flow past stationary and moving circular cylinders with and without slip on the surface, and the results show excellent agreement with the experimental and numerical results in the literature.

\section*{Acknowledgements}
This work was supported by JSPS KAKENHI Grant Nos.~JP18K03929 and JP23H01346.

\appendix
\section{One-dimensional formulation for the Navier boundary whose normal points in the negative coordinate direction}\label{App:negative}
The IBPM formulation shown in \ref{S:IBPM-1D} is for the boundary ($y=\eta$) whose normal is pointing in the positive $y$ direction, where $y>\eta $ is a fluid domain. The same argument can be applied to the case where $y<\eta$ is a fluid domain. First, the Navier BC is expressed as
\begin{align}
	\sum_{y^{\mathcal{F}_x}_j\in \Omega_\delta}
	u_j\delta_h(y^{\mathcal{F}_x}_j-\eta)\Delta y
	+\mathcal{L}_s\sum_{y^\mathcal{V}_j\in \Omega_\delta}
	(\hat{\partial}_y u)_j\delta_h(y^{\mathcal{V}}_j-\eta)\Delta y=U, 
	\label{eq:NBC_1D_dis_j_2}
\end{align}
where the sign of the shear stress term is reversed. 
The condition that the regularized boundary force $f$ should satisfy is given as 
\begin{align}
	\sum_{y^{\mathcal{V}}_j\in \Omega_\delta}
	\qty(
		\sum_{j'=J+1}^{j}f_{j'}\Delta y
	)\delta_h(y^{\mathcal{V}}_j-\eta)\Delta y=0
	\label{eq:cond_f_delta_dis_2}
\end{align}
for $y^\mathcal{V}_J\in\Omega_f\setminus\mathrm{supp} (f) $. By adding the divergence of the shear stress tensor $M\delta_h (\eta-y^\mathcal{V}_j) [\bm{e}_x (-\bm{e}_y) + (-\bm{e}_y)\bm{e}_x]$ distributed around the boundary $y=\eta$, the regularized boundary force $f$ is given by
\begin{align}
	f_j=F\delta_h(\eta-y^{\mathcal{F}_x}_j)-M(\Delta y)^{-1}(-\delta_h(\eta-y^{\mathcal{V}}_{j-1})+\delta_h(\eta-y^{\mathcal{V}}_j))
	\label{eq:Fd+Md'_dis_2}.
\end{align}
instead of Eq.~\eqref{eq:Fd+Md'_dis}. $M$ is determined to satisfy Eq.~\eqref{eq:cond_f_delta_dis_2}:
\begin{align}
	M=\qty[2\Delta y\sum_{y^{\mathcal{V}}_j\in \Omega_\delta}
	\qty(\sum_{j'=J+1}^{j}\delta_h(\eta-y^{\mathcal{F}_x}_{j'})\Delta y)
	\delta_h(y^{\mathcal{V}}_j-\eta)\Delta y]F.
	\label{eq:M-F_1D_2}
\end{align}
Note that the unified formulation is given for the two-dimensional case (Eqs.~\ref{eq:bf_FM_op} and \ref{eq:bf_F_op}), including explicitly the normal and tangential unit vectors of the boundary in the formulation.

\section{Derivation of the discretized consistency equation in two-dimension}
\label{app:derivation}
\subsection{Operator expression of the line integral}
\label{app:line_int}
\begin{figure}[t]
	\begin{center}
		\begin{overpic}[bb=0 0 160 120,width=6cm,clip,keepaspectratio]
				{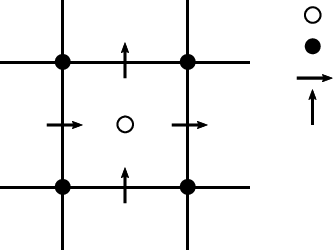}
				\put(60,60){$\psi_{i,j}$}
				\put(1,60){$\psi_{i-1,j}$}
				\put(-5,11){$\psi_{i-1,j-1}$}
				\put(60,11){$\psi_{i,j-1}$}
				\put(63,34.5){$(\hat{\partial}_y\psi)_{i,j}$}
				\put(-11,34.5){$(\hat{\partial}_y\psi)_{i-1,j}$}
				\put(30,65){$(\hat{\partial}_x\psi)_{i,j}$}
				\put(27,6){$(\hat{\partial}_x\psi)_{i,j-1}$}
				\put(100,68.5){$\mathcal{C}$}
				\put(100,59){$\mathcal{V}$}
				\put(100,49){$\mathcal{F}_x$}
				\put(100,41){$\mathcal{F}_y$}
		\end{overpic}
		\caption{Variable arrangements for a scalar variable $\psi$ defined on the vertices $\mathcal{V}$.}
	 \label{fig:V-cell}
	\end{center}
\end{figure}

\begin{figure}[t]
	\begin{center}
		\begin{overpic}[bb=0 0 360 241,width=8cm,clip,keepaspectratio]
				{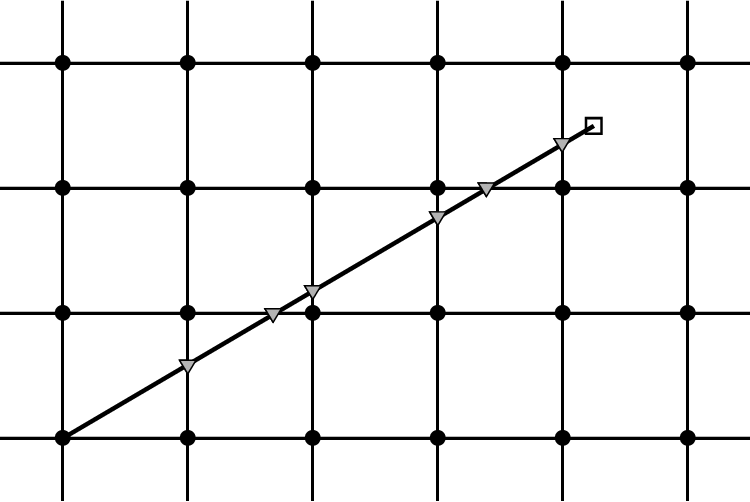}
				\put(1.5,3){\footnotesize $\bm{x}^\mathcal{V}$}
				\put(-8,-1.5){\footnotesize $(=\bm{x}^\mathrm{CP}_1)$}
				\put(75,52){\footnotesize $\bm{x}^\mathcal{V}+\nu\bm{n}$}
				\put(59,34){\footnotesize $\bm{x}^\mathrm{CP}_m$}
				\put(17,19){\footnotesize $\bm{x}^\mathrm{CP}_2$}
				\put(75,44){\footnotesize $\bm{x}^\mathrm{CP}_{m_\mathrm{max}}$}
		\end{overpic}
		\caption{Schematics of the line integral whose interval originates from a vertex. The intersections of the integration path with the cell boundaries are denoted by $\bm{x}^\mathrm{CP}$.}
	 \label{fig:integral_path}
	\end{center}
\end{figure}
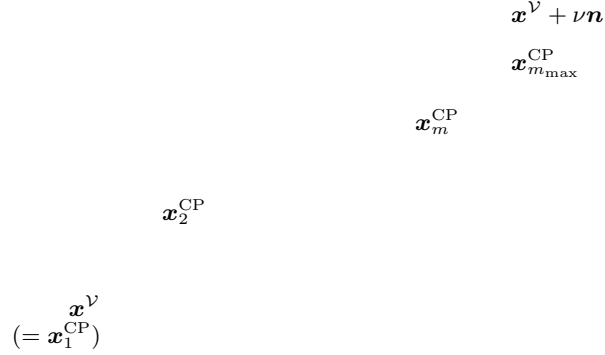

In this section, we first derive the operator expression of the line integral that is used in the discretization of Eq.~\eqref{eq:bf_cond}. Let us consider the line integral in the positive $n$ direction from $\bm{x}^\mathcal{V}$ to $\bm{x}^\mathcal{V}+\nu\bm{n}$ ($\nu >0$) of $\partial\psi/\partial n$ for a discrete variable $\psi\in\mathbb{R}^\mathcal{V}$ defined on $\mathcal{V}$ (for $\psi$ we have $\mathcal{V}$ part of $\bm{\tau}\cdot[\nabla\bm{u}+(\nabla\bm{u})^T]\cdot\bm{n}$ in Eq.~\ref{eq:st_dndt} in mind\footnote{The decomposition of $\bm{\tau}\cdot[\nabla\bm{u}+(\nabla\bm{u})^T]\cdot\bm{n}$ into $\mathcal{V}$ and $\mathcal{C}$ parts, see pp.~\pageref{pp:decomp}.}). The derivation for the line integral from $\bm{x}^\mathcal{C}$ to $\bm{x}^\mathcal{C}+\nu\bm{n}$ of a variable defined on $\mathcal{C}$ is the exact analogue. We require for the numerical integration scheme to preserve the property that the continuous form possesses: 
\begin{align}
\int_{\bm{x}^\mathcal{V}}^{\bm{x}^\mathcal{V}+\nu\bm{n}}\pd{\psi}{n}\dd{n}=
	\psi|_{\bm{x}^\mathcal{V}+\nu\bm{n}}-\psi|_{\bm{x}^\mathcal{V}}.
	\label{eq:line_int_V}
\end{align}
In this equation, only the last term on the RHS can be evaluated directly. For the RHS, $\psi|_{\bm{x}^\mathcal{V}+\nu\bm{n}}\notin\mathbb{R}^\mathcal{V}$ since $\bm{x}^\mathcal{V}+\nu\bm{n}\notin\mathcal{V}$ in general. For the LHS, the direct summation $\partial\psi/\partial n=n_x (\partial\psi/\partial x) +n_y (\partial\psi/\partial y)$ is not possible since $\hat{\partial}_x\psi$ and $\hat{\partial}_y\psi$ have different locations of definition, $\mathbb{R}^{\mathcal{F}_y}$ and $\mathbb{R}^{\mathcal{F}_x}$, respectively.
Therefore, Eq.~\eqref{eq:line_int_V} has to assume the distribution of $\psi$ that gives the value of $\psi$ to the points other than the points of definition. In the present method, we consider the cell whose vertices belong to $\mathcal{V}$ (Fig.~\ref{fig:V-cell}) as a bilinear quadrilateral element often employed in the finite element method, and the bilinear distribution of $\psi$ in the cell. Writing $\bm{x}^\mathcal{V}_{i, j}=[x^\mathcal{V}_i, y^\mathcal{V}_j]^T$ and $\psi$ values on the vertices $\bm{x}^\mathcal{V}_{i-1, j-1},\bm{x}^\mathcal{V}_{i, j-1},\bm{x}^\mathcal{V}_{i-1, j},\bm{x}^\mathcal{V}_{i, j}$ as $\psi_{i-1, j-1},\psi_{i, j-1},\psi_{i-1, j},\psi_{i, j}$, the value of $\psi$ at an arbitrary point $\bm{x}$ in the cell is given by
\begin{align}
	\psi|_{\bm{x}}&=\frac{x^\mathcal{V}_i-x}{\Delta x}\frac{y^\mathcal{V}_j-y}{\Delta y}\psi_{i-1,j-1}\notag\\
							&+\frac{x-x^\mathcal{V}_{i-1}}{\Delta x}\frac{y^\mathcal{V}_j-y}{\Delta y}\psi_{i,j-1}\notag\\
							&+\frac{x^\mathcal{V}_i-x}{\Delta x}\frac{y-y^\mathcal{V}_{j-1}}{\Delta y}\psi_{i-1,j}\notag\\
							&+\frac{x-x^\mathcal{V}_{i-1}}{\Delta x}\frac{y-y^\mathcal{V}_{j-1}}{\Delta y}\psi_{i,j}
	\label{eq:psi_bilin_V}.
\end{align}
At a lattice point, it coincides with the value of a discrete variable: $\psi|_{\bm{x}^\mathcal{V}_{i, j}}=\psi_{i, j}$. Since the value at a cell boundary is identical for the two cells sharing the boundary, $\psi$ computed by Eq.~\eqref{eq:psi_bilin_V} is a continuous function throughout the computational domain. By analytically differentiating the distribution (\ref{eq:psi_bilin_V}), the integrand in Eq.~\eqref{eq:line_int_V}, $\partial\psi/\partial n$, is given as
\begin{align}
	\left.\pd{\psi}{n}\right|_{\bm{x}}
	&=n_x\qty[\frac{y^\mathcal{V}_j-y}{\Delta y}(\hat{\partial}_x\psi)_{i,j-1}+\frac{y-y^\mathcal{V}_{j-1}}{\Delta y}(\hat{\partial}_x\psi)_{i,j}]\notag\\
	&+n_y\qty[\frac{x^\mathcal{V}_i-x}{\Delta x}(\hat{\partial}_y\psi)_{i-1,j}+\frac{x-x^\mathcal{V}_{i-1}}{\Delta x}(\hat{\partial}_y\psi)_{i,j}].
	\label{eq:psi_n_bilin_V}
\end{align}
Unlike $\psi$, $\partial\psi/\partial n$ is discontinuous at the cell boundaries, and its line integration should be done piecewise for each cell. The line integral from $\bm{x}_1$ to $\bm{x}_2$ belonging to the identical cell is calculated by 
\begin{align}
	\int_{\bm{x}_1}^{\bm{x}_2}\pd{\psi}{n}\dd{n}=\psi|_{\bm{x}_2}-\psi|_{\bm{x}_1},
	\label{eq:line_int_cell_V}
\end{align}
which is a consequence of analytically integrating Eq.~\eqref{eq:psi_n_bilin_V}. As shown in Fig.~\ref{fig:integral_path}, the integration path from any $\bm{x}^\mathcal{V}$ to $\bm{x}^\mathcal{V}+\nu\bm{n}$ can be decomposed into the piecewise linear paths whose end points are the intersections with the cell boundaries $\{\bm{x}^\mathrm{cp}_m\}_{m=1}^{m_\mathrm{max}}$ ($\bm{x}^\mathrm{cp}_1=\bm{x}^\mathcal{V}$): 
\begin{align}
	\int_{\bm{x}^\mathcal{V}}^{\bm{x}^\mathcal{V}+\nu\bm{n}}\pd{\psi}{n}\dd{n}
	&=\sum_{m=1}^{m_\mathrm{max}-1}\int_{\bm{x}^\mathrm{cp}_m}^{\bm{x}^\mathrm{cp}_{m+1}}\pd{\psi}{n}\dd{n}
	+\int_{\bm{x}^\mathrm{cp}_{m_\mathrm{max}}}^{\bm{x}^\mathcal{V}+\nu\bm{n}}\pd{\psi}{n}\dd{n} \notag\\
	&=\psi|_{\bm{x}^\mathcal{V}+\nu\bm{n}}-\psi|_{\bm{x}^\mathcal{V}},
	\label{eq:line_int_V_dis}
\end{align}
which recovers Eq.~\eqref{eq:line_int_V}.

Remembering that we have in mind the first term on the LHS of Eq.~\eqref{eq:st_dndt} for $\pdv*{\psi}{n}$, it is reasonable to assume the distribution given by Eq.~\eqref{eq:psi_n_bilin_V} also for the RHS of Eq.~\eqref{eq:st_dndt}. If we write the RHS of Eq.~\eqref{eq:st_dndt} symbolically as $\bm{c}\cdot\bm{q}$ ($q_x\in\mathbb{R}^{\mathcal{F}_x}$ and $q_y\in\mathbb{R}^{\mathcal{F}_y}$, and $\bm{c}$ is a constant vector), its distribution in a cell is given by
\begin{align}
	\bm{c}\cdot\bm{q}|_{\bm{x}}
	&=c_x\qty[\frac{x^\mathcal{V}_i-x}{\Delta x}(q_x)_{i-1,j}+\frac{x-x^\mathcal{V}_{i-1}}{\Delta x}(q_x)_{i,j}]\notag\\
	&+c_y\qty[\frac{y^\mathcal{V}_j-y}{\Delta y}(q_y)_{i,j-1}+\frac{y-y^\mathcal{V}_{j-1}}{\Delta y}(q_y)_{i,j}].
	\label{eq:cq_bilin_V}
\end{align}
Its line integral from $\bm{x}_1$ to $\bm{x}_2$ in the identical cell can be calculated as
\begin{align}
	\int_{\bm{x}_1}^{\bm{x}_2}\bm{c}\cdot\bm{q}\dd{n}=
\left.\bm{c}\cdot\bm{q}\right|_{\frac{\bm{x}_1+\bm{x}_2}{2}}\|\bm{x}_1-\bm{x}_2\|
	\label{eq:line_int_cell_V_Gauss}
\end{align}
since the distribution (\ref{eq:cq_bilin_V}) is linear. Finally the line integral from $\bm{x}^\mathcal{V}$ to $\bm{x}^\mathcal{V}+\nu\bm{n}$ is calculated by the cell-wise decomposition and can be written with a linear operator as 
\begin{align}
	\int_{\bm{x}^\mathcal{V}}^{\bm{x}^\mathcal{V}+\nu\bm{n}}
	\bm{c}\cdot\bm{q}\dd{n}
	=\hat{J}^\mathcal{V}\hat{c}\bm{q},
	\label{eq:line_int_V_dis_op_general}
\end{align}
where $\hat{c}\bm{q}=[(c_xq_x\in\mathbb{R}^{\mathcal{F}_x})^T, (c_yq_y\in\mathbb{R}^{\mathcal{F}_y})^T]^T \in\mathbb{R}^\mathcal{F}$.

In the present method, $\hat{J}^\mathcal{V}$ and $\hat{J}^\mathcal{C}$ are actually used only in the consistency condition for $\bm{f}$ (Eq.~\ref{eq:bf_cond_l_op}). In this expression, the line integral should be performed so that the non-zero distribution of $\bm{f}$ in the fluid domain, $\mathrm{supp} (\bm{f})\cap\Omega_f$, is completely covered. In the evaluation of Eq.~\eqref{eq:bf_cond_l_op}, for all $\bm{x}^\mathcal{C}$ and $\bm{x}^\mathcal{V}$ in $\mathrm{supp}(\delta_h (\bm{x}-\bm{\xi}_l))$, 
we choose $\nu>0$ such that both $\bm{x}^\mathcal{C}+\nu\bm{n}_l$ and $\bm{x}^\mathcal{V}+\nu\bm{n}_l$ are located outside $\mathrm{supp}(\bm{f})$.

\subsection{Derivation of the discretized consistency equation}
\label{app:decomp}
To derive the discretized form of Eq.~\eqref{eq:bf_cond}, the integrands in Eq.~\eqref{eq:st_dndt_int} have to be described in the discretized form first. Due to the difference in the locations of definition on the Eulerian mesh, $\bm{\tau}\cdot [\nabla\bm{u}+ (\nabla\bm{u}) ^T]\cdot\bm{n}$ is decomposed into $\sigma_n^\mathcal{C} \equiv 2\tau_x n_x(\hat{\partial}_x u)+2\tau_y n_y(\hat{\partial}_y v)\in\mathbb{R}^\mathcal{C}$ defined on $\mathcal{C}$ and $\sigma_n^\mathcal{V} \equiv (\tau_x n_y+n_x\tau_y)(\hat{\partial}_x v+\hat{\partial}_y u)\in\mathbb{R}^\mathcal{V}$ defined on $\mathcal{V}$, and their derivatives in the normal direction to the boundary we write as 
\begin{align}
\hat{\partial}_n\sigma_n^\mathcal{C}&=[(n_x\hat{\partial}_x\sigma_n^\mathcal{C})^T,(n_y\hat{\partial}_y\sigma_n^\mathcal{C})^T]^T\in\mathbb{R}^\mathcal{F}, \\
\hat{\partial}_n\sigma_n^\mathcal{V}&=[(n_y\hat{\partial}_y\sigma_n^\mathcal{V})^T,(n_x\hat{\partial}_x\sigma_n^\mathcal{V})^T]^T\in\mathbb{R}^\mathcal{F},
\end{align}
respectively. Similarly, $\bm{\tau}\cdot [\nabla\bm{u}+ (\nabla\bm{u}) ^T]\cdot\bm{\tau}$ is decomposed into $\sigma_{\tau}^\mathcal{C}=2\tau_x^2(\hat{\partial}_x u)+2\tau_y^2(\hat{\partial}_y v)\in\mathbb{R}^\mathcal{C}$ and $\sigma_{\tau}^\mathcal{V}=2\tau_x\tau_y(\hat{\partial}_x v+\hat{\partial}_y u)\in\mathbb{R}^\mathcal{V}$, and their derivatives in the tangential direction to the boundary are written as
\begin{align}
\hat{\partial}_{\tau}\sigma_{\tau}^\mathcal{C}&=[(\tau_x\hat{\partial}_x\sigma_{\tau}^\mathcal{C})^T,(\tau_y\hat{\partial}_y\sigma_{\tau}^\mathcal{C})^T]^T\in\mathbb{R}^\mathcal{F}, \\
\hat{\partial}_{\tau}\sigma_{\tau}^\mathcal{V}&=[(\tau_y\hat{\partial}_y\sigma_{\tau}^\mathcal{V})^T,(\tau_x\hat{\partial}_x\sigma_{\tau}^\mathcal{V})^T]^T\in\mathbb{R}^\mathcal{F},
\end{align}
respectively. The summation of these terms on the LHS in Eq.~\eqref{eq:st_dndt} is not arbitrary because the RHS should correspond to the viscous term in the Navier-Stokes equation by the construction of Eq.~\eqref{eq:st_dndt_int}: the combination should be $\hat{\partial}_n\sigma_n^\mathcal{C}+\hat{\partial}_{\tau}\sigma_{\tau}^\mathcal{V}$ and $\hat{\partial}_n\sigma_n^\mathcal{V}+\hat{\partial}_{\tau}\sigma_{\tau}^\mathcal{C}$. It can be shown as follows. If we denote by $\varphi$ the angle formed between the $x$-axis and the $\tau$-axis, $\tau_x=\cos\varphi$, $\tau_y=\sin\varphi$, $n_x=-\sin\varphi$, and $n_y=\cos\varphi$. Then under $\hat{D}\bm{u}=0$, it can be shown that 
\begin{align}
\hat{\partial}_n\sigma_n^\mathcal{C}+\hat{\partial}_{\tau}\sigma_{\tau}^\mathcal{V}&=
	\begin{bmatrix}
		\sin\varphi\sin{2\varphi}(\hat{L}u)\\
		\cos\varphi\sin{2\varphi}(\hat{L}v)
	\end{bmatrix}=:\hat{\tau}^\mathcal{C}\hat{L}\bm{u}
	\label{eq:st_Lu_C},\\
	\hat{\partial}_n\sigma_n^\mathcal{V}+\hat{\partial}_{\tau}\sigma_{\tau}^\mathcal{C}&=
	\begin{bmatrix}
		\cos\varphi\cos{2\varphi}(\hat{L}u)\\
		-\sin\varphi\cos{2\varphi}(\hat{L}v)
	\end{bmatrix}=:\hat{\tau}^\mathcal{V}\hat{L}\bm{u},
	\label{eq:st_Lu_V}
\end{align}
Eqs.~\eqref{eq:st_Lu_C} and \eqref{eq:st_Lu_V} are nothing but the discretized expressions of Eq.~\eqref{eq:st_dndt} since $\hat{\tau}^\mathcal{C}\hat{L}\bm{u}+\hat{\tau}^\mathcal{V}\hat{L}\bm{u}=[(\tau_x\hat{L}u) ^T, (\tau_y\hat{L}v) ^T]^T$. The origin of the integration interval and the corresponding line integral operator $\hat{J}$ are chosen according to the location where the LHS (that is $\sigma_n$) of Eq.~\eqref{eq:st_dndt_int} is defined, and we obtain Eqs.~\eqref{eq:st_dndt_int_C_op} and \eqref{eq:st_dndt_int_V_op}. Interpolating the third terms on the RHS of Eqs.~\eqref{eq:st_dndt_int_C_op} and \eqref{eq:st_dndt_int_V_op} to the boundary, we have the LHS of Eq.~\eqref{eq:bf_cond_l_op}, which is the discretized form of Eq.~\eqref{eq:bf_cond}.

\section{Invariance properties of the force regularization operator in two-dimension}
\label{app:consv}
We first show that the regularized force $\bm{f}$ given by Eq.~\eqref{eq:bf_FM_op} satisfies the force conservation (Eq.~\ref{eq:consv_force_dis}) and the torque conservation (Eq.~\ref{eq:consv_torque_dis}) regardless of the value of $M$. In this section, we use $i$ and $j$ to designate the lattice points on the Eulerian mesh as $\bm{x}_{i, j}=[x_i, y_j]^T$, and denote the discrete variable defined at $\bm{x}_{i, j}$ by $(\ )_{i,j}$. For example, the $x$-component of the regularized force $\bm{f}$ (Eq.~\ref{eq:bf_FM_op}) defined at $\bm{x}_{i, j}$ is written as
\begin{equation}\label{eq:f_index}
(f_x)_{i, j} = (\hat{H}F_x) _{i, j}+(\hat{\partial}_x m_{xx}+\hat{\partial}_y m_{yx})_{i, j}.
\end{equation}
For simplicity, we assume that $\Delta x=\Delta y=\Delta$, but the following discussion is equally valid for $\Delta x\neq\Delta y$. Noting that the following equations hold\footnote{Note that $\hat{\partial}_x\delta_h(\xi_l-x^\mathcal{C})$ and $\hat{\partial}_y\delta_h(\eta_l-y^\mathcal{V})$ are the quantities on $\mathcal{F}_x$.} for the discrete delta function $\delta_h$, 
\begin{align}
	[\hat{\partial}_x\delta_h(\xi_l-x^\mathcal{C})]_i
	&=\Delta^{-1}[-\delta_h(\xi_l-x^\mathcal{C}_i)+\delta_h(\xi_l-x^\mathcal{C}_{i+1})]\notag\\
	&=\Delta^{-1}[-\delta_h(\xi_l-(x^{\mathcal{F}_x}_i-\Delta/2))+\delta_h(\xi_l-(x^{\mathcal{F}_x}_i+\Delta/2))]
	\label{eq:dx_delta_C},\\
	[\hat{\partial}_y\delta_h(\eta_l-y^\mathcal{V})]_j
	&=\Delta^{-1}[-\delta_h(\eta_l-y^\mathcal{V}_{j-1})+\delta_h(\eta_l-y^\mathcal{V}_{j})]\notag\\
	&=\Delta^{-1}[-\delta_h(\eta_l-(y^{\mathcal{F}_x}_j-\Delta/2))+\delta_h(\eta_l-(y^{\mathcal{F}_xn}_j+\Delta/2))]
	\label{eq:dy_delta_V}
\end{align}
and denoting $\xi_l^{\pm}=\xi_l\pm\Delta/2$ and $\eta_l^{\pm}=\eta_l\pm\Delta/2$, the full index expression of Eq.~\eqref{eq:f_index} is given by
\begin{align}
	(f_x)_{i,j}
	&=\sum_l F_{x,l}\delta_h(\xi_l-x^{\mathcal{F}_x}_i)\delta_h(\eta_l-y^{\mathcal{F}_x}_j)\Delta s\notag\\
	&+\sum_l 2\tau_{x,l}n_{x,l}M_l
	[-\delta_h(\xi_l^{+}-x^{\mathcal{F}_x}_i)+\delta_h(\xi_l^{-}-x^{\mathcal{F}_x}_i)]\delta_h(\eta_l-y^{\mathcal{F}_x}_j)\Delta s\Delta^{-1}\notag\\
	&+\sum_l (\tau_{x,l}n_{y,l}+n_{x,l}\tau_{y,l})M_l
	\delta_h(\xi_l-x^{\mathcal{F}_x}_i)[-\delta_h(\eta_l^{+}-y^{\mathcal{F}_x}_j)+\delta_h(\eta_l^{-}-y^{\mathcal{F}_x}_j)]\Delta s\Delta^{-1}.
	\label{eq:fx_ij}
\end{align}
Similarly for the $y$-component of $\bm{f}$, we obtain 
\begin{align}
	(f_y)_{i,j}
	&=\sum_l F_{y,l}\delta_h(\xi_l-x^{\mathcal{F}_y}_i)\delta_h(\eta_l-y^{\mathcal{F}_y}_j)\Delta s\notag\\
	&+\sum_l (\tau_{x,l}n_{y,l}+n_{x,l}\tau_{y,l})M_l
	[-\delta_h(\xi_l^{+}-x^{\mathcal{F}_y}_i)+\delta_h(\xi_l^{-}-x^{\mathcal{F}_y}_i)]\delta_h(\eta_l-y^{\mathcal{F}_y}_j)\Delta s\Delta^{-1}\notag\\
	&+\sum_l 2\tau_{y,l}n_{y,l}M_l
	\delta_h(\xi_l-x^{\mathcal{F}_y}_i)[-\delta_h(\eta_l^{+}-y^{\mathcal{F}_y}_j)+\delta_h(\eta_l^{-}-y^{\mathcal{F}_y}_j)]\Delta s\Delta^{-1}.
	\label{eq:fy_ij}
\end{align}
Evaluating $\sum_i\sum_j (f_x) _{i, j}\Delta^2$ and $\sum_i\sum_j (f_y) _{i, j}\Delta^2$ with the basic property of $\delta_h$ (Eq.~\ref{eq:consv_force}), the terms containing $M$ cancel out, and the force conservation is shown: $\sum_i\sum_j (f_x)_{i,j}\Delta^2=\sum_l F_{x,l}\Delta s$ and $\sum_i\sum_j (f_y) _{i, j}\Delta^2=\sum_l F_{y, l}\Delta s$. For the torque, evaluating $\sum_i\sum_j x^{\mathcal{F}_y}_i (f_y) _{i, j}\Delta^2-\sum_i\sum_j y^{\mathcal{F}_y}_j (f_x) _{i, j}\Delta^2$ with the basic properties of $\delta_h$ (Eqs.~\ref{eq:consv_force} and \ref{eq:consv_torque}) shows its conservation: $\sum_i\sum_j x^{\mathcal{F}_y}_i(f_y)_{i,j}\Delta^2-\sum_i\sum_j y^{\mathcal{F}_y}_j(f_x)_{i,j}\Delta^2
=\sum_l(\xi_lF_{y, l}-\eta_lF_{x, l})\Delta s$, with the terms containing $M$ cancel out again. 

Next, we discuss the translational invariance, specifically in the weak sense that the regularized boundary force $\bm{f}_l$ interpolated back on the Lagrangian point $\bm{\xi}_l$, i.e.\ $\hat{E}_l\bm{f}_l$, is independent of the positional relation between $\mathcal{M}$ and $\Gamma$. The components of $\bm{f}_l$ in  $x$ and $y$ are related to the corresponding components of $\bm{f}$ as
\begin{equation}
\begin{aligned}
(f_x)_{i,j}&=\sum_l 
f_{x,l}, \\
(f_y)_{i,j}&=\sum_l 
f_{y,l}.
\end{aligned}\label{eq:f_l}
\end{equation}
Using the basic property of $\delta_h$ by Eq.~\eqref{eq:trns_invar}, we obtain
\begin{align}
	\hat{E}_lf_{x,l}
	&=\frac{1}{4}F_{x,l}\frac{\Delta s}{\Delta^2}\notag\\
	&+\frac{1}{2}\qty{
		2\tau_{x,l}n_{x,l}(M_l/\Delta)\sum_i\delta_h(x^{\mathcal{F}_x}_i-\xi_l)[-\delta_h(\xi_l^{+}-x^{\mathcal{F}_x}_i)+\delta_h(\xi_l^{-}-x^{\mathcal{F}_x}_i)]\Delta^2
	}\frac{\Delta s}{\Delta^2}\notag\\
	&+\frac{1}{2}\qty{
		(\tau_{x,l}n_{y,l}+n_{x,l}\tau_{y,l})(M_l/\Delta)\sum_j\delta_h(y^{\mathcal{F}_x}_j-\eta_l)[-\delta_h(\eta_l^{+}-y^{\mathcal{F}_x}_j)+\delta_h(\eta_l^{-}-y^{\mathcal{F}_x}_j)]\Delta^2
	}\frac{\Delta s}{\Delta^2}
	\label{eq:Efx_l},\\
	\hat{E}_lf_{y,l}
	&=\frac{1}{4}F_{y,l}\frac{\Delta s}{\Delta^2}\notag\\
	&+\frac{1}{2}\qty{
		(\tau_{x,l}n_{y,l}+n_{x,l}\tau_{y,l})(M_l/\Delta)\sum_i\delta_h(x^{\mathcal{F}_y}_i-\xi_l)[-\delta_h(\xi_l^{+}-x^{\mathcal{F}_y}_i)+\delta_h(\xi_l^{-}-x^{\mathcal{F}_y}_i)]\Delta^2
	}\frac{\Delta s}{\Delta^2}\notag\\
	&+\frac{1}{2}\qty{
		2\tau_{y,l}n_{y,l}(M_l/\Delta)\sum_j\delta_h(y^{\mathcal{F}_y}_j-\eta_l)[-\delta_h(\eta_l^{+}-y^{\mathcal{F}_y}_j)+\delta_h(\eta_l^{-}-y^{\mathcal{F}_y}_j)]\Delta^2
	}\frac{\Delta s}{\Delta^2}
	\label{eq:Efy_l}
\end{align}
The first terms on the RHS of Eqs.~\eqref{eq:Efx_l} and \eqref{eq:Efy_l} are independent of the positional relation between $\mathcal{M}$ and $\Gamma$, while the second and third terms on the RHS depend on $\sum_i\delta_h (x_i-\xi) [-\delta_h (\xi^{+}-x_i) +\delta_h (\xi^{-}-x_i)]\Delta^2$. Since $M$ is typically estimated as $M_l/\Delta\sim\bm{\tau}_l\cdot\bm{F}_l(=F_{\tau, l})$, $\tau_l$-$n_l$ components of $\hat{E}_l\bm{f}_l$ can be estimated as
\begin{equation}\label{eq:Efl}
\begin{aligned}
\tau_{x,l}\hat{E}_lf_{x,l}+\tau_{y,l}\hat{E}_lf_{y,l}&\sim\qty(\frac{1}{4}+\phi)F_{\tau,l}\frac{\Delta s}{\Delta^2}
	,\\
	n_{x,l}\hat{E}_lf_{x,l}+n_{y,l}\hat{E}_lf_{y,l}&\sim\frac{1}{4}F_{n,l}\frac{\Delta s}{\Delta^2}.
\end{aligned}
\end{equation}
Here, $\phi$ depends on the positional relation between $\mathcal{M}$ and $\Gamma$. Considering that $|\sum_i\delta_h (x_i-\xi) [-\delta_h (\xi^{+}-x_i) +\delta_h (\xi^{-}-x_i)]\Delta^2|<1/20$ holds for any $\xi$, we can show that $|\phi|<1/20$
\footnote{
In detail,
        $|\phi|
        <\frac{1}{40}[|\tau_{x,l}|(|2\tau_{x,l}n_{x,l}|+|\tau_{x,l}n_{y,l}+n_{x,l}\tau_{y,l}|)+|\tau_{y,l}|(|\tau_{x,l}n_{y,l}+n_{x,l}\tau_{y,l}|+|2\tau_{y,l}n_{y,l}|)] < \frac{1}{20}$
}.
Therefore the relative error in Eq.~\eqref{eq:Efl} is about $1/5$ at most, and the violation of the translational invariance is not significant.

\section{Spacial order of accuracy of the present method}
\label{app:PoiCtt}
The spacial error introduced by the present regularization scheme of the boundary force is theoretically analyzed for the steady Couette and Poiseuille flows. For the temporal error, the theoretical derivation is given in \ref{S:procedure}. To consider only the error by the boundary force regularization, the following discussion is carried out for the continuous formulation with the approximate delta function $\delta_\varepsilon$. In order to simplify the discussion, the value on the boundary $\Gamma$ is evaluated directly, not by the interpolation (The interpolation just introduces a higher order error and does not influence the order of accuracy estimation). 

First, we consider the steady Couette flow.
Let $y=\eta_1=-H/2$ and $y=\eta_2=H/2$ be the boundary $\Gamma$, $\Omega_f$ be the region of the interval $[\eta_1,\eta_2]$ and $\Omega_s$ be the other region. The governing equation in the IB formulation is
\begin{align}
	\pdd{u}{y}=-\mathrm{Re}f
	\label{eq:GE_ctt}
\end{align}
and the Navier BC 
\begin{align}
	u|_{\eta_1}-\mathcal{L}_s\left.\pd{u}{y}\right|_{\eta_1}&=0
	\label{eq:NBC_ctt_1},\\
	u|_{\eta_2}+\mathcal{L}_s\left.\pd{u}{y}\right|_{\eta_2}&=U
	\label{eq:NBC_ctt_2}
\end{align}
is imposed on $y=\eta_1$ and $y=\eta_2$. 
From the condition that $f$ should satisfy in the present method, $\int_{\eta_1}^Yf\dd{y}=0$ and $\int_Y^{\eta_2}f\dd{y}=0$ for any $Y\in\Omega_f\setminus\mathrm{supp}(f)$,
$f$ is given by
\begin{align}
	f(y)&=F_1\qty{\delta_\varepsilon(\eta_1-y)+\frac{1}{2\delta_\varepsilon(0)}\drv{}{y}[\delta_\varepsilon(\eta_1-y)]}\notag\\
			&+F_2\qty{\delta_\varepsilon(\eta_2-y)-\frac{1}{2\delta_\varepsilon(0)}\drv{}{y}[\delta_\varepsilon(\eta_2-y)]}.
	\label{eq:f_1D_poictt}
\end{align}
The general solution of Eq.~\eqref{eq:GE_ctt} is $u=-\mathrm{Re}\int_Y^{y}\dd{y'}\int_Y^{y'}f\dd{y''}+c_1y+c_2$, and by determining the integral constants $c_1$ and $c_2$ from Eqs.~\eqref{eq:NBC_ctt_1} and \eqref{eq:NBC_ctt_2} we obtain the velocity $u$ as
\begin{align}
	u=u_a
	&+\frac{\mathrm{Re}}{H+2\mathcal{L}_s}\qty{
		\int_Y^{\eta_2}\dd{y'}\int_Y^{y'} f\dd{y''}-\int_Y^{\eta_1}\dd{y'}\int_Y^{y'} f\dd{y''}
		+\mathcal{L}_s\qty[\int_Y^{\eta_2}f\dd{y'}+\int_Y^{\eta_1}f\dd{y'}]
	}y\notag\\
	&+\frac{\mathrm{Re}}{2}\qty{
		\int_Y^{\eta_2}\dd{y'}\int_Y^{y'} f\dd{y''}+\int_Y^{\eta_1}\dd{y'}\int_Y^{y'} f\dd{y''}
		+\mathcal{L}_s\qty[\int_Y^{\eta_2}f\dd{y'}-\int_Y^{\eta_1}f\dd{y'}]
	}\notag\\
	&-\mathrm{Re}\int_Y^y\dd{y'}\int_Y^{y'}f\dd{y''}.
	\label{eq:u_ctt}
\end{align}
Here, $u_a$ is the analytical solution of Eq.~\eqref{eq:GE_ctt} without the RHS (i.e. the Stokes equation), and the terms including $f$ are the errors. 

We estimate the order of magnitude of these error terms. If we consider $F$ as $\order{1}$, then $\int_{Y}^{y} f\dd{y'}=\order{1}$. Since $f$ is locally distributed in the region of order $\varepsilon$ near $\Gamma$, $f=\order{\varepsilon^{-1}}$. The order of the double integral $\int_{Y}^{y}\dd{y'}\int_{Y}^{y'} f\dd{y''}$ of $f$ is $\order{\varepsilon}$ in $\mathrm{supp}(f)$ due to the locality of the distribution of $\int_{Y}^{y} f\dd{y'}$ in $\Omega_f$. Therefore, in Eq.~\eqref{eq:u_ctt}, the leading error term is the term containing $\int f\dd{y}$, which is $\order{1}$. However, our method gives the $f$ distribution such that $\int_{\eta_1}^Yf\dd{y}=0 $ and $\int_Y^{\eta_2}f\dd{y}=0 $, and the error of $\order{1}$ vanishes. Therefore the leading error eventually comes from the double integral terms of $\order{\varepsilon}$. In the discrete expression, the grid width $\Delta y$ corresponds to $\varepsilon$. It shows that the leading error term is $\order{\Delta y}$ and the present method is first order accurate in space.
Substituting Eq.~\eqref{eq:f_1D_poictt} into Eq.~\eqref{eq:u_ctt} with $F_2=-F_1=F>0$ for the steady Couette flow, we furthur obtain
\begin{align}
	u=u_a
	&+\frac{2\mathrm{Re}}{H+2\mathcal{L}_s}F\qty[\rho_\varepsilon(0)-\frac{\theta_\varepsilon(0)}{2\delta_\varepsilon(0)}]y
	\label{eq:u_ctt_F}
\end{align}
for $y\in\Omega_f\setminus\mathrm{supp} (f)$, where $\rho_\varepsilon=\int\theta_\varepsilon\dd{y}$ is the approximate ramp function ($\delta_\varepsilon(0)=\order{\varepsilon^{-1}}$, $\theta_\varepsilon(0)=\order{1}$, and $\rho_\varepsilon(0)=\order{\varepsilon}$). 
From Eq.~\eqref{eq:u_ctt_F}, the error term is proportional to the boundary force $F$ and becomes zero at the center $y=0$. The error in $\partial u/\partial y$ is obtained by differentiating the error term of Eq.~\eqref{eq:u_ctt_F} and is constant of $y$.

Similarly, we consider the steady Poiseuille flow. The governing equation in the IB formulation is
\begin{align}
	\pdd{u}{y}=-\mathrm{Re}f_\mathrm{ext}-\mathrm{Re}f
	\label{eq:GE_poi}
\end{align}
and Navier BC 
\begin{align}
	u|_{\eta_1}-\mathcal{L}_s\left.\pd{u}{y}\right|_{\eta_1}&=0
	\label{eq:NBC_poi_1},\\
	u|_{\eta_2}+\mathcal{L}_s\left.\pd{u}{y}\right|_{\eta_2}&=0
	\label{eq:NBC_poi_2}
\end{align}
is imposed on $y=\eta_1$ and $y=\eta_2$. The solution of Eq.~\eqref{eq:GE_poi} is the same form as Eq.~\eqref{eq:u_ctt} with $u_a$ replaced by the analytic solution of the steady Poiseuille flow. Therefore the error terms are the same as that of the steady Couette flow, and again it is shown that the present method is first order accurate in space. 
By Substituting Eq.~\eqref{eq:f_1D_poictt} into Eq.~\eqref{eq:u_ctt} with $F_1=F_2=-F<0$ for the steady Poiseuille flow, we obtain
\begin{align}
	u=u_a
	-\mathrm{Re}F\qty[\rho_\varepsilon(0)-\frac{\theta_\varepsilon(0)}{2\delta_\varepsilon(0)}]
	\label{eq:u_poi_F}
\end{align}
for $y\in\Omega_f\setminus\mathrm{supp}(f)$. The error term is proportional to $F$ and constant of $y$. The error in $\partial u/\partial y$ is obtained by differentiating the error term of Eq.~\eqref{eq:u_poi_F} and is zero in $\Omega_f\setminus\mathrm{supp} (f)$.

\section{From the viewpoint of the optimization problem with constraints: the KKT condition}\label{App:KKT}
The IBPM formulation for the no-slip boundary, Eq.~\eqref{eq:Mtrx_GE_sym}, can be regarded as the Karush-Kuhn-Tucker (KKT) condition in the optimization problem with equality constraints \cite{Taira2007}: 
\begin{subequations}\label{eq:opt_noslip}
\begin{align}
	&\underset{\bm{u}^{n+1}}{\text{Minimize}}
	&&\frac{1}{2}(\bm{u}^{n+1})^T\hat{R}\bm{u}^{n+1}-(\bm{u}^{n+1})^T(\bm{r}_{NS}+bc_1),
	\label{eq:opt_noslip_obj}\\
	&\text{Subject to}
	&&\hat{D}\bm{u}^{n+1}=bc_2,
	\label{eq:opt_noslip_cnst1}\\
	&&&\hat{E}\bm{u}^{n+1}=\bm{U}^{n+1}.
	\label{eq:opt_noslip_cnst2}
\end{align}
\end{subequations}
The objective function written as Eq.~\eqref{eq:opt_noslip_obj} is a quantity similar to the kinetic energy, and $\tilde{P} $ and $\tilde{\bm{F}} $ are the Lagrange multipliers to satisfy the constraints \eqref{eq:opt_noslip_cnst1} and \eqref{eq:opt_noslip_cnst2}, respectively. 
Its essence is the symmetry between the operators appearing in the constraints (Eqs.~\ref{eq:opt_noslip_cnst1} and \ref{eq:opt_noslip_cnst2}) and the operators appearing in the constraint forces (Eq.~\ref{eq:Dis_NS_IB}). This physically means that the constraint forces do zero total work for the virtual velocity change $\delta\bm{u}$ not violating the constraint. In fact, for $\delta\bm{u} $ satisfying $\hat{D}\delta\bm{u}=0 $ and $\hat{E}\delta\bm{u}=0$, the work done by the constraint forces is zero: 
\begin{align}
	\sum_i(-\hat{G}P)_i\cdot\delta\bm{u}_i\Delta x\Delta y
	&=\sum_i P_i(\hat{D}\delta\bm{u})_i\Delta x\Delta y=0
	\label{eq:work_P_noslip},\\
	\sum_i(\hat{H}\bm{F})_i\cdot\delta\bm{u}_i\Delta x\Delta y
	&=\sum_l\bm{F}_l\cdot(\hat{E}\delta\bm{u})_l\Delta s=0,
	\label{eq:work_F_noslip}
\end{align}
considering $\hat{G}=-\hat{D}^T$ and $\hat{H}=\Delta s (\Delta x\Delta y) ^{-1}\hat{E}^T$.

On the other hand, in the present method, the operator $\hat{E}_{\tau n}-\hat{\mathcal{L}}_s\hat{G}_{\tau n}$ for the velocity (Eq.~\ref{eq:NBC+PBC_op}) and the operator $\hat{H}_{\tau n}+\hat{D}_{\tau n}\hat{K}_{\tau n}$ for the boundary force (Eq.~\ref{eq:bf_F_op}) are not in the symmetry. Therefore, Eq.~\eqref{eq:Mtrx_GE_sym_NBC} cannot be considered as the KKT condition of the following optimization problem with the equality constraints, which is the naive analogue of the no-slip problem: 
\begin{subequations}\label{eq:opt_NBC}
	\begin{align}
		&\underset{\bm{u}^{n+1}}{\text{Minimize}}
		&&\frac{1}{2}(\bm{u}^{n+1})^T\hat{R}\bm{u}^{n+1}-(\bm{u}^{n+1})^T(\bm{r}_{NS}+bc_1),
		\label{eq:opt_NBC_obj}\\
		&\text{Subject to}
		&&\hat{D}\bm{u}^{n+1}=bc_2,
		\label{eq:opt_NBC_cnst1}\\
		&&&(\hat{E}_{\tau n}-\hat{\mathcal{L}}_s\hat{G}_{\tau n})\bm{u}^{n+1}=\bm{U}_{\tau n}.
		\label{eq:opt_NBC_cnst2}
	\end{align}
\end{subequations} 
In the present method, the accurate prediction of the velocity gradient on the boundary is an important issue, and the additional constraint (Eq.~\ref{eq:bf_cond_op}) was introduced. To make the analysis simple, let us consider the case where the velocity gradinet on the boundary $(\nabla\bm{u}) _\Gamma$ is known and the corresponding constraint is written as 
\begin{align}
	\hat{G}_{\tau n}\bm{u}^{n+1}=(\hat{T}^T\hat{N}^T_2+\hat{N}^T\hat{T}^T_2)(\nabla\bm{u})_\Gamma^{n+1}
	\label{eq:gradcond_with_knowngrad}.
\end{align}
The velocity BC is then written as
\begin{align}
	\hat{E}_{\tau n}\bm{u}^{n+1}
	=\hat{\mathcal{L}}_s(\hat{T}^T\hat{N}^T_2+\hat{N}^T\hat{T}^T_2)(\nabla\bm{u})_\Gamma^{n+1}
	+\bm{U}_{\tau n}^{n+1}.
	\label{eq:NBC_with_knowngrad}
\end{align}
If we replace the constraint (\ref{eq:opt_NBC_cnst2}) by Eqs.~\eqref{eq:gradcond_with_knowngrad} and \eqref{eq:NBC_with_knowngrad}, the KKT condition to the optimization problem is
\begin{align}
	\begin{bmatrix}
		\hat{R} & (\hat{D}^\mathcal{F})^T & \hat{E}_{\tau n}^T & -\hat{G}_{\tau n}^T \\
		\hat{D}^\mathcal{F} & 0 											& 0 								 & 0 \\
		\hat{E}_{\tau n} 		& 0 											& 0 								 & 0 \\
		-\hat{G}_{\tau n} 	& 0 											& 0									 & 0 \\
	\end{bmatrix}
	\begin{bmatrix}
		\bm{u}^{n+1} \\
		\tilde{P} \\
		\tilde{\bm{F}}_{\tau n} \\
		\tilde{M} \\
	\end{bmatrix}
	=
	\begin{bmatrix}
		\bm{r}_{NS} \\
		0 \\
		\hat{\mathcal{L}}_s(\hat{T}^T\hat{N}^T_2+\hat{N}^T\hat{T}^T_2)(\nabla\bm{u})_\Gamma^{n+1}+\bm{U}_{\tau n}^{n+1} \\
		-(\hat{T}^T\hat{N}^T_2+\hat{N}^T\hat{T}^T_2)(\nabla\bm{u})_\Gamma^{n+1}
	\end{bmatrix}
	+
	\begin{bmatrix}
		bc_1 \\
		bc_2 \\
		0	\\
		0 \\
	\end{bmatrix}
	\label{eq:KKT_with_knowngrad}.
\end{align}
Here, $\tilde{M}$ is the Lagrange multiplier that corresponds to the forcing shear stress introduced in \ref{S:2DC}. To impose the constraints by Eqs.~\eqref{eq:gradcond_with_knowngrad} and \eqref{eq:NBC_with_knowngrad}, the boundary forcing $\bm{f}=\hat{E}_{\tau n}^T\tilde{\bm{F}}_{\tau n}-\hat{G}_{\tau n}^T\tilde{M}=\hat{H}_{\tau n}\bm{F}_{\tau n}
+\hat{D}_{\tau n}M$ is added to the Navier-Stokes equation, which is reminiscent of the boundary forcing (Eq.~\ref{eq:bf_FM_op}) in the present method. The total work done by the boundary forcing $\bm{f}$ is in fact zero to the virtual velocity change on the constrained surface. 

%





\bibliographystyle{elsarticle-num-names}
\bibliography{library.bib,omori.bib}

\begin{thebibliography}{21}
\expandafter\ifx\csname natexlab\endcsname\relax\def\natexlab#1{#1}\fi
\providecommand{\url}[1]{\texttt{#1}}
\providecommand{\href}[2]{#2}
\providecommand{\path}[1]{#1}
\providecommand{\DOIprefix}{doi:}
\providecommand{\ArXivprefix}{arXiv:}
\providecommand{\URLprefix}{URL: }
\providecommand{\Pubmedprefix}{pmid:}
\providecommand{\doi}[1]{\href{http://dx.doi.org/#1}{\path{#1}}}
\providecommand{\Pubmed}[1]{\href{pmid:#1}{\path{#1}}}
\providecommand{\bibinfo}[2]{#2}
\ifx\xfnm\relax \def\xfnm[#1]{\unskip,\space#1}\fi
\bibitem[{Peskin(1972)}]{Peskin1972}
\bibinfo{author}{C.~S. Peskin},
\newblock \bibinfo{title}{{Flow patterns around heart valves: A numerical
  method}},
\newblock \bibinfo{journal}{Journal of Computational Physics}
  \bibinfo{volume}{10} (\bibinfo{year}{1972}) \bibinfo{pages}{252--271}.
  \DOIprefix\doi{10.1016/0021-9991(72)90065-4}.
\bibitem[{Peskin(1982)}]{Peskin1982}
\bibinfo{author}{C.~S. Peskin},
\newblock \bibinfo{title}{The {Fluid} {Dynamics} of {Heart} {Valves}:
  {Experimental}, {Theoretical}, and {Computational} {Methods}},
\newblock \bibinfo{journal}{Annual Review of Fluid Mechanics}
  (\bibinfo{year}{1982}) \bibinfo{pages}{235--259}.
  \DOIprefix\doi{10.1146/annurev.fl.14.010182.001315}.
\bibitem[{Mittal and Iaccarino(2005)}]{Mittal2005a}
\bibinfo{author}{R.~Mittal}, \bibinfo{author}{G.~Iaccarino},
\newblock \bibinfo{title}{{IMMERSED BOUNDARY METHODS}},
\newblock \bibinfo{journal}{Annual Review of Fluid Mechanics}
  \bibinfo{volume}{37} (\bibinfo{year}{2005}) \bibinfo{pages}{239--261}.
  \URLprefix
  \url{http://www.annualreviews.org/doi/10.1146/annurev.fluid.37.061903.175743}.
  \DOIprefix\doi{10.1146/annurev.fluid.37.061903.175743}.
\bibitem[{Griffith and Patankar(2020)}]{Griffith2020}
\bibinfo{author}{B.~E. Griffith}, \bibinfo{author}{N.~A. Patankar},
\newblock \bibinfo{title}{{Immersed Methods for Fluid–Structure
  Interaction}},
\newblock \bibinfo{journal}{Annual Review of Fluid Mechanics}
  \bibinfo{volume}{52} (\bibinfo{year}{2020}) \bibinfo{pages}{421--448}.
  \URLprefix
  \url{https://www.annualreviews.org/content/journals/10.1146/annurev-fluid-010719-060228}.
  \DOIprefix\doi{https://doi.org/10.1146/annurev-fluid-010719-060228}.
\bibitem[{Kajishima and Taira(2017)}]{Kajishima2017a}
\bibinfo{author}{T.~Kajishima}, \bibinfo{author}{K.~Taira},
\newblock \bibinfo{title}{Immersed {Boundary} {Methods}},
\newblock in: \bibinfo{booktitle}{Computational {Fluid} {Dynamics}},
  \bibinfo{publisher}{Springer International Publishing}, \bibinfo{year}{2017},
  pp. \bibinfo{pages}{179--205}. \URLprefix
  \url{http://link.springer.com/10.1007/978-3-319-45304-0_5}.
  \DOIprefix\doi{10.1007/978-3-319-45304-0_5}.
\bibitem[{Taira and Colonius(2007)}]{Taira2007}
\bibinfo{author}{K.~Taira}, \bibinfo{author}{T.~Colonius},
\newblock \bibinfo{title}{{The immersed boundary method: A projection
  approach}},
\newblock \bibinfo{journal}{Journal of Computational Physics}
  \bibinfo{volume}{225} (\bibinfo{year}{2007}) \bibinfo{pages}{2118--2137}.
  \DOIprefix\doi{10.1016/j.jcp.2007.03.005}.
\bibitem[{Bocquet and Charlaix(2010)}]{Bocquet2010}
\bibinfo{author}{L.~Bocquet}, \bibinfo{author}{E.~Charlaix},
\newblock \bibinfo{title}{Nanofluidics, from bulk to interfaces},
\newblock \bibinfo{journal}{Chemical Society Reviews} \bibinfo{volume}{39}
  (\bibinfo{year}{2010}) \bibinfo{pages}{1073--1095}. \URLprefix
  \url{http://xlink.rsc.org/?DOI=B909366B}. \DOIprefix\doi{10.1039/B909366B}.
\bibitem[{Bocquet and Barrat(2007)}]{Bocquet2007}
\bibinfo{author}{L.~Bocquet}, \bibinfo{author}{J.~L. Barrat},
\newblock \bibinfo{title}{{Flow boundary conditions from nano- to
  micro-scales}},
\newblock \bibinfo{journal}{Soft Matter} \bibinfo{volume}{3}
  (\bibinfo{year}{2007}) \bibinfo{pages}{685--693}.
  \DOIprefix\doi{10.1039/b616490k}. \href{http://arxiv.org/abs/0612242}{{\tt
  arXiv:0612242}}.
\bibitem[{Matthews and Hill(2008)}]{Matthews2008}
\bibinfo{author}{M.~T. Matthews}, \bibinfo{author}{J.~M. Hill},
\newblock \bibinfo{title}{Nanofluidics and the {Navier} boundary condition},
\newblock \bibinfo{journal}{International Journal of Nanotechnology}
  \bibinfo{volume}{5} (\bibinfo{year}{2008}) \bibinfo{pages}{218--242}.
  \DOIprefix\doi{10.1504/IJNT.2008.016917}, \bibinfo{note}{iSBN: 1475-7435}.
\bibitem[{He et~al.(2018)He, Glowinski, and Wang}]{He2018}
\bibinfo{author}{Q.~He}, \bibinfo{author}{R.~Glowinski}, \bibinfo{author}{X.~P.
  Wang},
\newblock \bibinfo{title}{{A least-squares/fictitious domain method for
  incompressible viscous flow around obstacles with Navier slip boundary
  condition}},
\newblock \bibinfo{journal}{Journal of Computational Physics}
  \bibinfo{volume}{366} (\bibinfo{year}{2018}) \bibinfo{pages}{281--297}.
  \DOIprefix\doi{10.1016/J.JCP.2018.04.013}.
\bibitem[{Wang et~al.(2021)Wang, He, and Huang}]{WANG2021}
\bibinfo{author}{Z.~Wang}, \bibinfo{author}{Q.~He}, \bibinfo{author}{J.~Huang},
\newblock \bibinfo{title}{{The immersed boundary-lattice Boltzmann method for
  solving solid-fluid interaction problem with Navier-slip boundary
  condition}},
\newblock \bibinfo{journal}{Computers \& Fluids} \bibinfo{volume}{217}
  (\bibinfo{year}{2021}) \bibinfo{pages}{104839}. \URLprefix
  \url{https://www.sciencedirect.com/science/article/pii/S0045793021000050}.
  \DOIprefix\doi{https://doi.org/10.1016/j.compfluid.2021.104839}.
\bibitem[{Chen et~al.(2014)Chen, Zhang, and Koplik}]{Chen2014}
\bibinfo{author}{W.~Chen}, \bibinfo{author}{R.~Zhang},
  \bibinfo{author}{J.~Koplik},
\newblock \bibinfo{title}{{Velocity slip on curved surfaces}},
\newblock \bibinfo{journal}{Physical Review E - Statistical, Nonlinear, and
  Soft Matter Physics} \bibinfo{volume}{89} (\bibinfo{year}{2014}).
  \DOIprefix\doi{10.1103/PhysRevE.89.023005}.
  \href{http://arxiv.org/abs/1309.1423}{{\tt arXiv:1309.1423}}.
\bibitem[{Peskin(2002)}]{Peskin2002}
\bibinfo{author}{C.~S. Peskin},
\newblock \bibinfo{title}{{The immersed boundary method}},
\newblock \bibinfo{journal}{Acta Numerica} \bibinfo{volume}{11}
  (\bibinfo{year}{2002}) \bibinfo{pages}{479--517}. \URLprefix
  \url{http://www.journals.cambridge.org/abstract_S0962492902000077}.
  \DOIprefix\doi{10.1017/S0962492902000077}.
\bibitem[{Beyer and LeVeque(1992)}]{Beyer_LeVeque_1992}
\bibinfo{author}{R.~P. Beyer}, \bibinfo{author}{R.~J. LeVeque},
\newblock \bibinfo{title}{{Analysis of a One-Dimensional Model for the Immersed
  Boundary Method}},
\newblock \bibinfo{journal}{SIAM Journal on Numerical Analysis}
  \bibinfo{volume}{29} (\bibinfo{year}{1992}) \bibinfo{pages}{332--364}.
  \URLprefix \url{https://doi.org/10.1137/0729022}.
  \DOIprefix\doi{10.1137/0729022}.
\bibitem[{Roma et~al.(1999)Roma, Peskin, and Berger}]{Roma1999}
\bibinfo{author}{A.~M. Roma}, \bibinfo{author}{C.~S. Peskin},
  \bibinfo{author}{M.~J. Berger},
\newblock \bibinfo{title}{{An Adaptive Version of the Immersed Boundary
  Method}},
\newblock \bibinfo{journal}{Journal of Computational Physics}
  \bibinfo{volume}{153} (\bibinfo{year}{1999}) \bibinfo{pages}{509--534}.
  \DOIprefix\doi{10.1006/jcph.1999.6293}.
\bibitem[{Coutanceau and Bouard(1977)}]{Coutanceau_Bouard_1977}
\bibinfo{author}{M.~Coutanceau}, \bibinfo{author}{R.~Bouard},
\newblock \bibinfo{title}{{Experimental determination of the main features of
  the viscous flow in the wake of a circular cylinder in uniform translation.
  Part 1. Steady flow}},
\newblock \bibinfo{journal}{Journal of Fluid Mechanics} \bibinfo{volume}{79}
  (\bibinfo{year}{1977}) \bibinfo{pages}{231--256}.
  \DOIprefix\doi{10.1017/S0022112077000135}.
\bibitem[{Tritton(1959)}]{Tritton_1959}
\bibinfo{author}{D.~J. Tritton},
\newblock \bibinfo{title}{{Experiments on the flow past a circular cylinder at
  low Reynolds numbers}},
\newblock \bibinfo{journal}{Journal of Fluid Mechanics} \bibinfo{volume}{6}
  (\bibinfo{year}{1959}) \bibinfo{pages}{547--567}.
  \DOIprefix\doi{10.1017/S0022112059000829}.
\bibitem[{Dennis and Chang(1970)}]{Dennis_Chang_1970}
\bibinfo{author}{S.~C.~R. Dennis}, \bibinfo{author}{G.-Z. Chang},
\newblock \bibinfo{title}{{Numerical solutions for steady flow past a circular
  cylinder at Reynolds numbers up to 100}},
\newblock \bibinfo{journal}{Journal of Fluid Mechanics} \bibinfo{volume}{42}
  (\bibinfo{year}{1970}) \bibinfo{pages}{471--489}.
  \DOIprefix\doi{10.1017/S0022112070001428}.
\bibitem[{Linnick and Fasel(2005)}]{LINNICK2005157}
\bibinfo{author}{M.~N. Linnick}, \bibinfo{author}{H.~F. Fasel},
\newblock \bibinfo{title}{{A high-order immersed interface method for
  simulating unsteady incompressible flows on irregular domains}},
\newblock \bibinfo{journal}{Journal of Computational Physics}
  \bibinfo{volume}{204} (\bibinfo{year}{2005}) \bibinfo{pages}{157--192}.
  \URLprefix
  \url{https://www.sciencedirect.com/science/article/pii/S0021999104004127}.
  \DOIprefix\doi{https://doi.org/10.1016/j.jcp.2004.09.017}.
\bibitem[{Canuto and Taira(2015)}]{Canuto_Taira_2015}
\bibinfo{author}{D.~Canuto}, \bibinfo{author}{K.~Taira},
\newblock \bibinfo{title}{{Two-dimensional compressible viscous flow around a
  circular cylinder}},
\newblock \bibinfo{journal}{Journal of Fluid Mechanics} \bibinfo{volume}{785}
  (\bibinfo{year}{2015}) \bibinfo{pages}{349--371}.
  \DOIprefix\doi{10.1017/jfm.2015.635}.
\bibitem[{Legendre et~al.(2009)Legendre, Lauga, and
  Magnaudet}]{LEGENDRE_LAUGA_MAGNAUDET_2009}
\bibinfo{author}{D.~Legendre}, \bibinfo{author}{E.~Lauga},
  \bibinfo{author}{J.~Magnaudet},
\newblock \bibinfo{title}{{Influence of slip on the dynamics of two-dimensional
  wakes}},
\newblock \bibinfo{journal}{Journal of Fluid Mechanics} \bibinfo{volume}{633}
  (\bibinfo{year}{2009}) \bibinfo{pages}{437--447}.
  \DOIprefix\doi{10.1017/S0022112009008015}.

\end{thebibliography}

\end{document}